\documentclass[12pt]{article}
\usepackage{mathrsfs}
\usepackage{amsmath}
\usepackage{mathrsfs}
\usepackage{amssymb}
\usepackage{color}
\pdfoutput=1
\usepackage{graphicx}
\usepackage{xcolor}
\usepackage{hyperref}
\usepackage{color}
\usepackage[utf8]{inputenc}
\usepackage{setspace}
\usepackage{amsmath, amssymb, amsthm}
\numberwithin{equation}{section}
\hypersetup{colorlinks=false, linkcolor=blue, citecolor=red}
\newcommand{\be}{\begin{equation}}
\newcommand{\ee}{\end{equation}}
\def\g{\gamma}
\def\G{\Gamma}

\def\s{\sigma}
\def\m{\mu}
\def\n{\nu}
\def\a{\alpha}

\def\e{\epsilon}
\def\l{\lambda}
\def\k{\kappa}
\def\b{\beta}
\def\d{\delta}

\def\f{\phi}

\def\D{\Delta}

\def\ta{\tau}

\def\ra{\rangle}

\newcommand{\bea}{\begin{eqnarray}}
\newcommand{\eea}{\end{eqnarray}}

\renewcommand{\a}{\alpha}
\renewcommand{\b}{\beta}

\renewcommand{\d}{\delta}
\newcommand{\dsl}{\pa \kern-0.5em /}
\newcommand{\la}{\lambda}
\newcommand{\half}{\frac{1}{2}}
\newcommand{\pa}{\partial}

\newcommand{\nn}{\nonumber\\}

\begin{document}
	\title{\bf Mellin space bootstrap for global symmetry}
	\date{}
	
\author{Parijat Dey\footnote{parijat@chep.iisc.ernet.in},~ Apratim Kaviraj\footnote{apratim@chep.iisc.ernet.in} ~and Aninda Sinha\footnote{asinha@chep.iisc.ernet.in} \\ ~~~~\\
	\it Centre for High Energy Physics,
	\it Indian Institute of Science,\\ \it C.V. Raman Avenue, Bangalore 560012, India. \\}	
\maketitle
\abstract{We apply analytic conformal bootstrap ideas in Mellin space to conformal field theories with $O(N)$ symmetry and cubic anisotropy. We write down the conditions arising from the consistency between the operator product expansion and crossing symmetry in Mellin space.  We solve the constraint equations to compute the anomalous dimension and the OPE coefficients of all operators quadratic in the fields in the epsilon expansion. We reproduce known results and derive new results up to $O(\epsilon^3)$. For the $O(N)$ case, we also study the large $N$ limit in general dimensions and reproduce known results at the leading order in $1/N$.}
\tableofcontents

\onehalfspacing
\section{Introduction and summary of results }
Wilson's renormalization group approach to understanding critical phenomena \cite{wilsonkogut} has led to profound insights over many years. This approach relies on a Feynman diagram expansion which needs regularization of divergences and, as is usual in a perturbative approach, leads to asymptotic expansions. While this is well understood for renormalizable theories, it does not make use of the enhanced conformal symmetry at the fixed point. In the 1970s, \cite{Migdal:1972tk,Ferrara:1973vz, Polyakov} initiated the study of the conformal bootstrap approach in understanding critical phenomena. Unfortunately, the resulting equations proved very difficult to solve and not much progress was made. The work of \cite{bpz} in the 1980s made remarkable progress in understanding 2d CFTs. It would take another two decades before progress was made, starting with the work of \cite{rrtv} which made use of the development in understanding conformal blocks in \cite{dolanosborn,do2} in the bootstrap program in higher dimensions \cite{reviews,bootstrap,3dising,mostprecise,dlce,others,kss,spins}.

In the modern formalism of the conformal bootstrap, building on the work of \cite{rrtv}, one expands a four point function in a conformal field theory in terms of the conformal blocks of one of the channels (direct channel). Then one imposes crossing symmetry in the next step. This is a nontrivial constraint and forms the starting point for the powerful numerical approach to constraining conformal field theories. Analytic progress, with this as the starting point, has been limited \cite{dlce,kss}. In cases with weakly broken higher spin symmetry, some progress has been made in understanding the leading order anomalous dimensions \cite{ giombii, skvortt} for lower spin operators as well--however, it is not clear how to systematize this approach to get subleading orders. The double light cone limit of the bootstrap equations in the works of \cite{alday,Alday2,Alday4,Alday3} gives a systematic approach for the large spin limit. For low spin cases, the methods of \cite{Alday4} allow a resummation to finite values of the spin, including spin zero but the issue is subtle.\footnote{In some cases one can add solutions consistent with crossing and with finite support in the spin. In many interesting cases this appears not to be the case, we thank F. Alday for discussions on this.}. It is worth exploring other methods which do not require a resummation.
 In \cite{rychkovtan} it was shown how to make use of conformal symmetry of three point functions to get the leading order (in epsilon) anomalous dimension of a large class of scalar operators (see also \cite{rtothers}). This approach depended indirectly on the equations of motion that follows from a lagrangian and leads to the question: How does one recover these results using the bootstrap and go further? The modern incarnation of the bootstrap can be used to gain some insight into the epsilon expansion using numerics \cite{epsnum} but is not very efficient in getting analytic results. Hence it is desirable to seek a different starting point.

In \cite{Polyakov}, Polyakov considered a version of the conformal bootstrap that made use of crossing symmetric blocks from the beginning. Thus while crossing symmetry was in-built, consistency with the operator product expansion, for instance in the direct channel, was not guaranteed. There are spurious poles in the expansion which need to be cancelled. Demanding this consistency leads to an infinite number of constraints. This approach lay dormant for a long time. In \cite{sensinha},  this approach was revisited and it was pointed out that it could be made to work at the next nontrivial order in the epsilon expansion. In \cite{usprl,uslong}, it was realized that the full power of \cite{Polyakov} could be harnessed in Mellin space \cite{mack,pene,mig,dolanosborn2,fitzpatrick,joao,costa,joaoreview,aldaybissi2,rastelli,Aharony:2016dwx} where the systematics of non-zero spin exchange was both conceptually and calculationally simpler. Quite remarkably, the epsilon expansion results at three loops in the Feynman diagram approach were reproduced leading to agreement with existing results for anomalous dimensions as well as new results for OPE coefficients which have never been calculated, barring the stress tensor and conserved current exchanges (which are known upto two loop order in the epsilon expansion for the Ising case).

The reason why \cite{usprl, uslong} worked so efficiently relied on two key ingredients. First, the direct channel expression for the leading spurious pole naturally leads to an expansion in terms of a convenient orthonormal basis in terms of the continuous Hahn polynomials \cite{AAR,Luke}. Second, the crossed channels got contributions from only one scalar operator upto the first two or three (depending on the spin in the s-channel) subleading orders in epsilon. We will demonstrate that this approach works for the $O(N)$ case as well in the epsilon expansion. This will lead to reproducing known three loop results as well as new results for the OPE coefficients for various operators. Another reason for looking at the $O(N)$ case is that $1/N$ in the large-$N$ limit provides another expansion parameter for a fixed spacetime dimension $d$ and it is natural to ask what happens in this case. There is a large body of work using a bootstrap type approach and conformal symmetry to understand this very important case \cite{petkoulots, Lang:1992zw} which ties up with the AdS/CFT correspondence. As we will show that the single operator contribution in the crossed channel holds only upto leading order in $1/N$, enabling us to easily extract the leading $1/N$ terms. To go beyond these orders, will require a careful study of the systematics of all the spurious poles, not just the leading one, and also some mixed correlators. This will be taken up in the near future in a separate work.\\

Another important case of $N$-scalars that we will consider is the theory with cubic anisotropy \cite{phi4}. In the space of couplings there are four fixed points--the Gaussian fixed point corresponding to the free theory, the Ising fixed point corresponding to $N$ decoupled $\phi^4$ theories, the $O(N)$ fixed point arising from an interaction $g_1 (\sum_{i=1}^N \phi_i \phi_i)^2$ and the cubic fixed point corresponding to a continuum theory with the interaction $g_1 (\sum_{i=1}^N \phi_i \phi_i)^2+g_2 \sum_{i=1}^N (\phi_i)^4$--for this last case the discrete symmetry $\phi_i \leftrightarrow \phi_j, \phi_i\rightarrow -\phi_i$ is preserved. For a certain $N<N_c$, the $O(N)$ fixed point is the stable fixed point while for $N>N_c$ the cubic fixed point is the stable fixed point. The value of $N_c$ that follows from an epsilon expansion analysis is less than 3. To our knowledge, this value has not been determined using the modern numerical bootstrap and our analysis may be a useful starting point to address the same. The $N=3$ case is relevant for ferromagnets. We will set up the equations for this problem and derive anomalous dimensions and OPE coefficients for operators quadratic in the field. While the anomalous dimension of the singlet scalar and the fundamental scalar are known to five loop order \cite{phi4}, many of the results we will quote appear to be unknown in the literature (to the best of our knowledge). 

{\bf Assumptions:}\\
The essential assumptions that we will make in order to solve the bootstrap conditions, in addition to there being a $\mathbb{Z}_2$ symmetry in all cases are: 
\begin{enumerate}
\item There is a unique conserved stress tensor and a conserved spin-1 current. However, the conservation of spin-1 current does not hold for cubic anisotropy.
\item In the $\epsilon$-expansion the OPE coefficients of higher order operators like $(\phi_i\phi_i)^2$ begin at $O(\e)$. This is expected from the free theory, the only nontrivial bit in this assumption is that it begins at $O(\e)$ rather than say $O(\e^{1/2})$.
\end{enumerate}

\subsection{Summary of the results}

We summarize below the findings of the paper. We use the colour code of blue to indicate results that are new.

\subsubsection{$\e$-expansion}


We find the anomalous dimensions and OPE coefficients (squared) of operators for the critical $O(N)$ model in $d=4-\e$ at the Wilson-Fisher fized point, for general $N$. The results are obtained as an expansion in $\e$. The table below summarizes the operators, and the equations showing the corresponding corrections. The anomalous dimensions below agree with literature \cite{Braun,kleinert,gracey} while the OPE coefficients are all new results.\footnote{The symmetrization and antisymmetrization brackets are defined as $A_{(ab)}=(A_{ab}+A_{ba})/2$ and $A_{[ab]}=(A_{ab}-A_{ba})/2$.}  \footnote{We show only a schematic form of the operators. For example $\phi_i\partial^\ell\phi_i$ indicates a primary with symmetrized traceless combination of $\ell$ derivatives.}

\begin{center}
\begin{tabular}{|c|c|l|l|l|}
 \hline
 \text{operator} $O$ &\text{dimension} $\D_O$ & \text{OPE coefficient} $C_{\phi\phi O} \equiv C_{O}$ \\ \hline
 $\phi$ & \eqref{dphi} & - \\
  \hline   $\phi_i \phi_i$ & \eqref{sol3} & \textcolor{blue}{\eqref{sol}} \\
    \hline $\phi_{(i} \phi_{j)}-\frac{\d_{ij}}{N}\phi_k\phi_k$ & \eqref{sol4} & \textcolor{blue}{\eqref{sol2}} \\
        \hline
 $\phi_i \partial^\ell\phi_i$ & \eqref{higherspindim} & \textcolor{blue}{\eqref{opesing}} \\
     \hline $\phi_{(i} \partial^\ell\phi_{j)}-\frac{\d_{ij}}{N}\phi_k\partial^\ell\phi_k$ & \eqref{higherspindim} & \textcolor{blue}{\eqref{opetraceless}}\\
          \hline   $\phi_{[i} \partial^\ell\phi_{j]}$ & \eqref{higherspindim} & \textcolor{blue}{\eqref{opeanti}}\\
                    \hline       
\end{tabular}
\end{center} 

\subsubsection{Large $N$ expansion}
Our progress with the $1/N$ expansion is more modest and we do not report any new result. The leading order results that arise from our analysis are consistent with known results.  

\begin{center}
\begin{tabular}{|c|c|l|l|l|}
 \hline
 \text{operator} $O$ &\text{dimension} $\D_O$ & \text{OPE coefficient} $C_{\phi\phi O} \equiv C_{O}$ \\ \hline
 $\phi$ & \eqref{p1} & - \\
  \hline   $\s$ & \eqref{spin0dim},\eqref{zero} & \eqref{spin0ope},\eqref{zero} \\
    \hline $\phi_{(i} \phi_{j)}-\frac{\d_{ij}}{N}\phi_k\phi_k$ & \eqref{spin0dim},\eqref{zero} & \eqref{spin0ope},\eqref{zero} \\
        \hline
 $\phi_i \partial^\ell\phi_i$ & \eqref{largen1} & \eqref{largen3} \\
     \hline $\phi_{(i} \partial^\ell\phi_{j)}-\frac{\d_{ij}}{N}\phi_k\partial^\ell\phi_k$ & \eqref{largen2} & \eqref{largen4}\\
          \hline   $\phi_{[i} \partial^\ell\phi_{j]}$ & \eqref{largen2} & \eqref{largen4}\\
                    \hline       
\end{tabular}
\end{center} 
These results are in agreement with the results from \cite{Lang:1992zw, giombii}\,.

\subsubsection{Cubic anisotropy}
We also consider the special case of a broken $O(N)$ symmetry with an interacting term like $g_1 (\sum_i \phi_i \phi_i)^2+g_2\sum_i \phi_i^4$, in $d=4-\e$. In this model the symmetric traceless operators break into a diagonal part and an off-diagonal part. We have obtained the anomalous dimensions and OPE coefficients for a few operators, which are summarized below. The generalized $\d_{ijkl}$ notation is introduced in \eqref{geng}.
\begin{center}
\begin{tabular}{|c|c|l|l|l|}
 \hline
 \text{operator} $O$ &\text{dimension} $\D_O$ & \text{OPE coefficient} $C_{\phi\phi O} \equiv C_{O}$ \\ \hline
 $\phi$ & \eqref{Dphicubic} & - \\
  \hline   $\phi_i\phi_i$ & \eqref{Dim0} & \textcolor{blue}{\eqref{ope1}} \\
    \hline $\phi_{(i}\phi_{j)}-\d_{ijkl}\phi_k\phi_l$ & \textcolor{blue}{\eqref{dimT}} & \textcolor{blue}{\eqref{ope2}} \\
        \hline $\d_{ijkl}\phi_{k}\phi_{l}-\frac{\d_{ij}}{N}\phi_k\phi_k$ & \textcolor{blue}{\eqref{dimV}} & \textcolor{blue}{\eqref{Dimlast}} \\
         \hline  $\phi_i\partial^\ell\phi_i$ & \textcolor{blue}{\eqref{dimspinS}} & \textcolor{blue}{\eqref{cubicope},Table \ref{cS2},\ref{cS3}} \\ \hline       
 $\phi_{(i}\partial^\ell\phi_{j)}-\d_{ijkl}\phi_k\partial^\ell\phi_l$ & \textcolor{blue}{\eqref{dimspinT} } & \textcolor{blue}{\eqref{cubicope},Table \ref{cT2},\ref{cT3}}\\ \hline       
  $\d_{ijkl}\phi_k\partial^\ell\phi_l-\frac{\d_{ij}}{N}\phi_k\partial^\ell\phi_k$ & \textcolor{blue}{\eqref{dimspinV}}  & \textcolor{blue}{\eqref{cubicope},Table \ref{cV2},\ref{cV3}} \\ \hline $\phi_{[i}\partial^\ell\phi_{j]}$ & \textcolor{blue}{\eqref{dimspinV}} &  \textcolor{blue}{\eqref{cubicope},Table \ref{cA2},\ref{cA3}} \\ \hline
\end{tabular}
\end{center}

The paper is organised as follows: In section \ref{setup} we set up our equations for theories with a $O(N)$ symmetry. Section \ref{eExp} uses these equations to obtain the anomalous dimensions and OPE coefficients for $O(N)$ models in $d=4-\e$ at the Wilson-Fisher fixed point. In section \ref{largeN} we apply our method for large $N$ critical models. Section \ref{compare} makes some $d=3$ predictions  and also a large spin analysis to compare with known results.   In section \ref{anisotropy} we show how to modify our equations for a different kind of symmetry, which is the cubic anisotropic case. There are four appendices which give the essesntial formulas, alternate methods to verify our findings and some results which were too big for the main text.

\section{Mellin space bootstrap for $O(N)$}\label{setup}
 We begin by reviewing the analysis of \cite{usprl,uslong} and extend the ideas to theories with $O(N)$ symmetry. For the $O(N)$ case the spectrum contains operators that behave differently from one another under $O(N)$ transformations. A two point OPE can have the following operator content in the spectrum,
 \be
 \phi_i \times \phi_j \supset \{1,S,T_{(ij)},A_{[ij]}\}\,,
 \ee
 where $S$ denotes $O(N)$ singlets of even spin, $T$ denotes $O(N)$ symmetric traceless tensors of even spin, and $A$ denotes $O(N)$ antisymmetric tensors of odd spin.  These operators do not mix with each other under $O(N)$ rotations, and give rise to different symmetry structures in a four point OPE. More discussion for $O(N)$ models and their symmetries can be found in \cite{petkoulots,Kos:2013tga}. 
 
\subsection{$s$-channel}
A generic four point function of the fundamentals of the $O(N)$ in the $s$-channel can be written as
\bea\label{s}
\langle\phi_{i_1}(x_1)\phi_{i_2}(x_2)\phi_{i_3}(x_3)\phi_{i_4}(x_4)\rangle &=& \frac{\mathcal{A}(u,v)}{(x_{12}\,x_{34})^{2\D_\f}}
\eea
where\footnote{We will choose the normalization of blocks such that in small $u$ and $1-v$ limit $g_{\D,\ell}(u,v)\sim u^{(\D-\ell)/2}(1-v)^\ell$. }
\bea\label{s-ope}
\mathcal{A}(u,v) &=& \sum_{S^+}C_{\D,\ell}\,(\delta_{{i_1}{i_2}}\delta_{{i_3}{i_4}})\,g_{\D,\ell}(u,v)\nn &&
+\sum_{T^+}C_{\D,\ell}\,(\delta_{{i_1}{i_3}}\delta_{{i_2}{i_4}}+\delta_{{i_1}{i_4}}\delta_{{i_2}{i_3}}-\frac{2}{N}\,\delta_{{i_1}{i_2}}\delta_{{i_3}{i_4}})\,g_{\D,\ell}(u,v)
\nn && +\sum_{A^-}C_{\D,\ell}\,(\delta_{{i_1}{i_4}}\delta_{{i_2}{i_3}}-\delta_{{i_1}{i_3}}\delta_{{i_2}{i_4}})\,g_{\D,\ell}(u,v)\,.
\eea
The $+$ denotes even spins, and the $-$ denotes odd spins. Each sector has  different $C_{\D,\ell}$-s corresponding to exchanges in that sector. The sums run over primary operators of dimension $\D$ and spin $\ell$ and $C_{\D,\ell}$ is the square of the OPE coefficient of the operator carrying dimension $\D$ and spin $\ell$. We will sometimes loosely refer to $C_{\D,\ell}$ as the OPE coefficient.

Following the analysis of \cite{usprl,uslong} we will write the four point function in the  basis of Witten diagrams, as follows
\bea
\mathcal{A}(u, v)=\sum_{\D,\ell}c_{\D,\ell}(W^{(s)}_{\D,\ell}(u,v)+W^{(t)}_{\D,\ell}(u,v)+W^{(u)}_{\D,\ell}(u,v))\,.
\eea
Here the constants $c_{\D,\ell}$ are related to the OPE coefficients $C_{\D,\ell}$ via a normalization factor defined in \eqref{norm}.
We can write the Mellin representation of a Witten  diagram (for identical external scalars) as follows,
\bea\label{mellinform}
W^{(s)}_{\D,\ell}(u, v)=\int \frac{ds\, dt}{(2\pi i)^2}\, u^s\, v^t\, \G(-t)^2\,\G(s+t)^2\,\G(\D_\f-s)^2\, M_{\D,\ell}^{(s)}(s, t)
\eea
where $M^{(s)}_{\D,\ell}(s, t)$ is the Mellin amplitude of $W^{(s)}_{\D,\ell}(u, v)$ given by,
\begin{align}
\begin{split}\label{sunitrymell}
M^{(s)}_{\D,\ell}(s, t) = \int_{-i\infty}^{i\infty} d\nu\, \mu^{(s)}_{\Delta,\ell}(\nu)
\Omega_{\nu, \ell}^{(s)}(s)P^{(s)}_{\nu, \ell}(s,t)
\end{split}
\end{align}
where we have,
\be\label{specunitry}
\mu^{(s)}_{\Delta,\ell}(\nu)=\frac{\G^2(\frac{2\D_\phi-h+\ell+\nu}{2})\G^2(\frac{2\D_\phi-h+\ell-\nu}{2})}{2\pi i((\D-h)^2-\nu^2)\G(\nu)\G(-\nu)(h+\nu-1)_\ell(h-\nu-1)_\ell}\,.
\ee
and 
\be
\Omega_{\nu, \ell}^{(s)}(s)=\frac{\G(\lambda_2-s)\G(\bar{\lambda}_2-s)}{\G(\D_\phi-s)^2}\,.
\ee
Here we have $\l_2=(h+\nu -\ell)/2$,  \ $\bar{\l}_2  =(h- \nu -\ell)/2$ and $h=\frac{d}{2}$. The Mack polynomials $P^{(s)}_{\Delta-h,\ell}(s,t)$ are polynomials of degree $\ell$ in $s$ and $t$. Their form is shown explicitly in Appendix \ref{mack}.

For the four point function written in \eqref{s} carrying the $O(N)$ indices, the sum of Witten diagrams in a certain channel can be decomposed according to singlet, traceless symmetric and anti-symmetric operator exchanges. The $s$-channel  can be written as,
\bea
\sum_{\D,\ell}c_{\D,\ell}W^{(s)}_{\D,\ell}(u,v) &=& \int \frac{ds\, dt}{(2\pi i)^2}\, u^s\, v^t\, \G(-t)^2\,\G(s+t)^2\,\G(\D_\f-s)^2\nn &&\bigg( (\delta_{{i_1}{i_2}}\delta_{{i_3}{i_4}})\,M^{S, (s)}(s, t)+(\delta_{{i_1}{i_3}}\delta_{{i_2}{i_4}}+\delta_{{i_1}{i_4}}\delta_{{i_2}{i_3}}-\frac{2}{N}\,\delta_{{i_1}{i_2}}\delta_{{i_3}{i_4}})\,M^{T,(s)}(s, t)\nn &&+ (\delta_{{i_1}{i_4}}\delta_{{i_2}{i_3}}-\delta_{{i_1}{i_3}}\delta_{{i_2}{i_4}})\,M^{A, (s)}(s, t) \bigg)
\eea
with,
\bea\label{Mdef}
M^{i,(s)}(s, t) &=& \sum_{\D, \ell,i}\, \int d\nu \,c_{\D,\ell}^i\, M_{\D,\ell}^{(s)}(s,t)\,.
\eea
where $i$ stands for $S, T, A$ for singlet, symmetric traceless and antisymmetric operators respectively.

%

\subsection{$t$-channel}
The $t$-channel can be done in a similar manner by replacing
$ x_2 \leftrightarrow x_4$ , $i_2 \leftrightarrow i_4$ and  $u \leftrightarrow v$.  After this interchange we can bring the integral into the form \eqref{mellinform}, by relabelling $t+\D_\phi\rightarrow s$  and $s-\D_\phi \rightarrow t$.
Then the Witten diagrams in the $t$-channel can be written in Mellin space as,
\begin{align}
&\sum_{\D,\ell'}c_{\D,\ell'} W_{\D,\ell'}^{(t)}(u,v)= \int \frac{ds\, dt}{(2\pi i)^2}\, u^s\, v^t\, \G(-t)^2\,\G(s+t)^2\,\G(\D_\f-s)^2 \bigg((\delta_{{i_1}{i_4}}\delta_{{i_3}{i_2}})\, M^{S,(t)}(s,t)\nonumber\\&+(\delta_{{i_1}{i_2}}\delta_{{i_4}{i_3}}+\delta_{{i_1}{i_3}}\delta_{{i_4}{i_2}}-\frac{2}{N}\,\delta_{{i_1}{i_4}}\delta_{{i_3}{i_2}})\,M^{T,(t)}(s,t) +(\delta_{{i_1}{i_2}}\delta_{{i_4}{i_3}}-\delta_{{i_1}{i_3}}\delta_{{i_4}{i_2}})\,M^{A,(t)}(s,t)\bigg) 
\end{align}
where,
\be
M^{i,(t)}(s,t) = \sum_{\D,\ell^\prime, i}\, \int d\nu \, c_{\D, \ell^\prime}^i\, M^{(t)}_{\D,\ell'}(s,t) = \sum_{\D,\ell^\prime, i}c_{\D,\ell'}^i \int_{-i\infty}^{i\infty} d\nu\, \mu^{(t)}_{\Delta,\ell'}(\nu)\Omega_{\nu, \ell'}^{(t)}(t)
P^{(t)}_{\nu, \ell'}(s-\D_\phi,t+\D_\phi)\,.
\ee
Here we have,
\be
\Omega^{(t)}_{\nu, \ell'}(t)= \frac{\G(\frac{1}{2}(h+\nu-\ell')-t-\D_\phi)\G(\frac{1}{2}(h-\nu-\ell')-t-\D_\phi)}{\G^2(-t)} \, .
\ee
and $P^{(t)}_{\nu, \ell}(s,t)$ is obtained from $P^{(s)}_{\nu, \ell}(s,t)$ by interchanging $s \leftrightarrow t$. For identical scalars, $\mu^{(t)}_{\Delta,\ell'}(\nu)$ is the same as $\mu^{(s)}_{\Delta,\ell'}(\nu)$.

Similarly in $u$-channel we can write,
\begin{align}
&\sum_{\D,\ell'}c_{\D,\ell'}  W_{\D,\ell'}^{(u)}(u,v) = \int \frac{ds\, dt}{(2\pi i)^2}\, u^s\, v^t\, \G(-t)^2\,\G(s+t)^2\,\G(\D_\f-s)^2  \bigg((\delta_{{i_1}{i_3}}\delta_{{i_2}{i_4}})\, M^{S,(u)}(s,t)\nonumber\\ & +(\delta_{{i_1}{i_2}}\delta_{{i_3}{i_4}}  +\delta_{{i_1}{i_4}}\delta_{{i_3}{i_2}}-\frac{2}{N}\,\delta_{{i_1}{i_3}}\delta_{{i_2}{i_4}})\,M^{T,(u)}(s,t) +\, (\delta_{{i_1}{i_4}}\delta_{{i_2}{i_3}}-\delta_{{i_1}{i_2}}\delta_{{i_3}{i_4}})\,M^{A,(u)}(s,t)\bigg)
\end{align}
where, 
\be
M^{i,(u)}(s,t)  = \sum_{\D,\ell^\prime, i}\, \int d\nu \, c_{\D, \ell^\prime}^i\, M^{(u)}_{\D,\ell'}(s,t)= \sum_{\D,\ell^\prime, i}\, \int d\nu \, c_{\D, \ell^\prime}^i \mu^{(u)}_{\Delta,\ell'}(\nu)\Omega_{\nu, \ell'}^{(u)}(s+t) P^{(u)}_{\nu, \ell'}(s-\D_\phi,t ) \,.
\ee
Here we have,
\be
\Omega^{(u)}_{\nu, \ell'}(s+t)= \frac{\G(\frac{1}{2}(h+\nu-\ell')+s+t-\D_\phi)\G(\frac{1}{2}(h-\nu-\ell')+s+t-\D_\phi)}
{\G^2(s+t)} \, .
\ee
In $u$-channel the Mack polynomial $P^{(u)}_{\nu,\ell}(s,t)$ is obtained from $P^{(s)}_{\nu,\ell}(s,t)$ by transforming $s\to -s-t$ and $t\to t$. Once again for identical scalars, we have $\mu^{(u)}_{\Delta,\ell'}(\nu)$ is the same as $\mu^{(s)}_{\Delta,\ell'}(\nu)$.

\subsection{Total amplitude }
The total crossing symmetric amplitude is given by,
\bea\label{unitary}
&& \mathcal{A}(u,v) =  \int \frac{ds\, dt}{(2\pi i)^2}\, u^s\, v^t\, \G(-t)^2\,\G(s+t)^2\,\G(\D_\f-s)^2 \nn && \bigg[(\delta_{{i_1}{i_2}}\delta_{{i_3}{i_4}})\bigg(M^{S, (s)}(s, t)-\frac{2}{N}\,M^{T,(s)}(s, t)+M^{T,(t)}(s,t)+M^{T,(u)}(s,t)+M^{A,(t)}(s,t)-M^{A,(u)}(s,t)\bigg)\nn &&+(\delta_{{i_1}{i_4}}\delta_{{i_3}{i_2}}) \bigg( M^{S,(t)}(s,t)+M^{T,(s)}(s, t)-\frac{2}{N}M^{T,(t)}(s, t)+M^{T,(u)}(s, t)+M^{A, (s)}(s, t)+M^{A, (u)}(s, t)\bigg)\nn &&(\delta_{{i_1}{i_3}}\delta_{{i_2}{i_4}})\bigg( M^{S,(u)}(s,t)+M^{T,(s)}(s, t)+,M^{T,(t)}(s, t)-\frac{2}{N}M^{T,(u)}(s, t)-M^{A, (s)}(s, t)-M^{A, (t)}(s, t)\bigg)\bigg]\nn \,.
\eea
Now the Mellin integral \eqref{mellinform} in every component Witten diagram $W_{\D,\ell}(u,v)$ has poles in the Mellin variable $s$ at,
\be
2s=(\D-\ell)+2 n \text{  where  } n=0,1,2\cdots\,.
\ee
These poles correspond to operators present in the OPE (there are also shadow poles occuring at $(d-\D-\ell)+2 n$ but they can be eliminated by an appropriate choice of contour). They come from the Mellin amplitude $M_{\D,\ell}$ when we look at the simple pole at $\nu=\pm(\D-h)$ and the $\G$-functions in $\Omega_{\nu,\ell}^{(s)}(s)$. These poles then reproduce the $u^{(\D-\ell)/2+n}$ dependence that one expects in the OPE. We call these the physical poles. 

Now there are also poles in $s$ that do not correspond to operators present in the OPE. These poles occur at,
\be\label{unphyspoles}
s=\D_\phi+n \text{  where  } n=0,1,2\cdots\,.
\ee
These poles come from the $\G^2(\D_\phi-s)$ function measure in the Mellin integral, as well as from the  $\G(\bar{\lambda}_2-s)$ and  $\G^2(\frac{2\D_\phi-h+\ell-\nu}{2})$ combined in $M_{\D,\ell}(s,t)$. In the Mellin integral these poles give $u^{\D_\phi+n}\log u$ and $u^{\D_\phi+n}$  dependence which are spurious because they typically do not occur in the $s$-channel OPE\footnote{Except when the operator with dimension $2\D_\phi$ is protected in which case we will have to consider the contribution from these operators like the way we would treat the disconnected part.}. Since one already obtains $s$-channel OPE, which is the full $\mathcal{A}(u,v)$, from the physical poles, these other poles are called unpysical poles, and the spurious $u$-dependences as unphysical terms.

Let us look at unphysical terms, with the leading order in $u$. These occur at the pole $s=\D_\phi$ and the residues are simply given  by the individual Mellin apmlitudes evaluated at $s=\D_\phi$. They can be expanded in terms of the basis of the continuous Hahn polynomials $Q^{2s+\ell}_{\ell, 0}$. So let us write
\begin{align}\label{Mtoq}
M^{i,(s)}(s\to\D_\phi,t)=&  \sum_{\D, \ell, i}\, c_{\D,\ell}^iq^{i,(s)}_{\D, \ell} \, Q^{2\D_\phi+\ell}_{\ell, 0}(t)+\cdots\nonumber
\end{align}
\begin{align}
M^{i,(t)}(s\to \D_\phi,t) 
=&  \sum_{\D, \ell,\ell',i}\, c_{\D,\ell}^iq^{i,(t)}_{\D, \ell|\ell'} \, Q^{2\D_\phi+\ell}_{\ell, 0}(t)+\cdots\nonumber\\
M^{i,(u)}(s\to\D_\phi,t)
=&  \sum_{\D, \ell,\ell',i}\, c_{\D,\ell}^iq^{i,(u)}_{\D, \ell|\ell'} \, Q^{2\D_\phi+\ell}_{\ell, 0}(t)+\cdots
\end{align}
Here the $\cdots$ denote physical pole contributions, and other spurious poles. The polynomials $Q^{\Delta}_{\ell,0}(t)$ are given in terms of the Mack polynomaials $P_{\nu,\ell}^{(s)}(s,t)$ as
\be
Q^{\Delta}_{\ell,0}(t)= \frac{4^\ell }{(\Delta-1)_{\ell}(2h-\Delta-1)_{\ell}}P_{\Delta-h,\ell}(s=\frac{\Delta-\ell}{2},t) \,.
\ee
In the $s$-channel we have,
\begin{align}\label{ss1}
&  q^{i,(s)}_{\D, \ell}(s)= -4^{1-\ell}\,\frac{(2s+\ell-1)_{\ell}\,(2h-2s-\ell-1)_{\ell}\,\G(h-\ell-2s)\,\m^{(s)}_{\D,\ell}(\nu)}{\G(\D_\f-s)^2}\,.
\end{align}
Let us write this as,
\begin{align}\label{ss}
&q^{i,(s)}_{\D, \ell}(s)=q_{\D,\ell}^{(2,t)}+(s-\D_\phi)q^{(1,s)}_{\D,\ell}+O((s-\D_\phi)^2)\nonumber\\
&=-\frac{4^{1-\ell}\G(2\D_\phi+\ell-h)}{(\ell-\D+2\D_\phi)(\ell+\D+2\D_\phi-2h)}+(s-\D_\phi)\frac{4^{2-\ell}\G(2\D_\phi+\ell-h+1)}{(\ell-\D+2\D_\phi)^2(\ell+\D+2\D_\phi-2h)^2}\,,
\end{align}
where $i$ stands for $S, T, A$ for singlet, symmetric traceless and antisymmetric operators respectively. In the above equation the first term in the second line, is associated with the log term $u^{\D_\phi}\log u$\,, while the second term is part of the coefficient of the non-logarithmic term $u^{\D_\phi}$ (we will call this term the power law). We will need to sum up the coefficients of both log and power law terms from all the three channels and equate them to 0. For this purpose, only the two terms shown in \eqref{ss} are enough since once the log coefficients are 0, all that is left in the power law coefficient is the second term in \eqref{ss}.

The expansions in the $t$ and $u$ channels are possible because the continuous Hahn polynomials $Q_{\ell,0}^\D(t)$ are orthogonal polnomials. Their orthogonality property reads
\be
\frac{1}{2\pi i}\int_{-i\infty}^{i\infty}dt \ \G^2(s+t)\G^2(-t)Q^{2s+\ell}_{\ell,0}(t)Q^{2s+\ell '}_{\ell',0}(t)=\kappa_{\ell}(s)\d_{\ell,\ell'}\,,
\ee
where $\k_{\ell}(s)$ is defined in \eqref{kappadef}. The properties of continuous Hahn polynomials are detailed in Appendix \ref{A3} . Using this for the crossed channels we have,
\begin{align}\label{ts}
q^{i,(t)}_{\D, \ell|\ell'}(s)=& \frac{1}{\kappa_{\ell}(s)} \int \frac{dt}{2\pi i}\, \G(s+t)^2\,\, Q^{2s+\ell}_{\ell, 0}(t)  \int \, d\nu\, \, \G(\la_2-t-\D_\f)\,\G(\bar{\la}_2-t-\D_\f)\,\, \m^{(t)}_{\D,\ell}(\nu)\, P^{(t)}_{\nu,\ell'}(s-\D_\f, t+\D_\f)\\
\label{tu}
q^{i,(u)}_{\D, \ell}(s)=& \frac{1}{\kappa_{\ell}(s)}\, \int \frac{dt}{2\pi i}\, \G(s+t)^2\, \, Q^{2s+\ell^+}_{\ell^+, 0}(t)   \times  \int \, d\nu\, \, \G(\la_2-t-\D_\f)\,\G(\bar{\la}_2-t-\D_\f)\,\, \m^{(u)}_{\ell^\prime}(\nu)\, P^{(u)}_{\nu,\ell'}(s-\D_\f, t)\,.
\end{align}
So near the pole $s=\D_\phi$ the integrand would look like, 
\begin{align}\label{Atoq}
 \mathcal{A}(u,v) &=\int \frac{ds\, dt}{(2\pi i)^2}\, u^s\, v^t\, \G(-t)^2\,\G(s+t)^2\,\G(\D_\f-s)^2   \bigg[(\delta_{{i_1}{i_2}}\delta_{{i_3}{i_4}}) \bigg(\sum_{\D, \ell, S^+} \bigg(c_{\D,\ell}^S q^{S(s)}_{\D, \ell}+\sum_{\ell'}\frac{c_{\D,\ell'}^S}{N}(q^{S(t)}_{\D, \ell|\ell'}+q^{S(u)}_{\D,\ell| \ell'})\bigg)\nonumber\\ &+\sum_{\D, \ell,\ell', T^+}c_{\D,\ell'}^T\bigg(1+\frac{1}{N}-\frac{2}{N^2}\bigg) (q^{T(u)}_{\D, \ell|\ell'}+q^{T(t)}_{\D, \ell|\ell'}) + \sum_{\D, \ell, \ell',A^-}c_{\D,\ell'}^A(1-\frac{1}{N})\, (q^{A(t)}_{\D, \ell|\ell'}-q^{A(u)}_{\D, \ell|\ell'})\bigg)\nonumber\\ & +(\delta_{{i_1}{i_4}}\delta_{{i_3}{i_2}}+\delta_{{i_1}{i_3}}\delta_{{i_2}{i_4}}-\frac{2}{N}\delta_{{i_1}{i_2}}\delta_{{i_3}{i_4}}) \half \bigg(\sum_{\D, \ell,\ell', S^+}c_{\D,\ell'}^S(q^{S(t)}_{\D, \ell|\ell'} +q^{S(u)}_{\D, \ell|\ell'} )\nonumber\\ & + \sum_{\D, \ell, T^+}\bigg(2c_{\D,\ell}^T\,q^{T(s)}_{\D, \ell}+\sum_{\ell'}c_{\D,\ell'}^T(1-\frac{2}{N})\,(q^{T(t)}_{\D, \ell|\ell'}+\,q^{T(u)}_{\D, \ell|\ell'})\bigg) +\sum_{\D, \ell,\ell', A^-}c_{\D,\ell'}^A(q^{A(u)}_{\D, \ell|\ell'}-q^{A(t)}_{\D, \ell|\ell'})\bigg) \nonumber\\ & + (\delta_{{i_1}{i_4}}\delta_{{i_3}{i_2}}-\delta_{{i_1}{i_3}}\delta_{{i_2}{i_4}})\half \bigg(\sum_{\D, \ell, \ell',S^+}c_{\D,\ell'}^S(q^{S(t)}_{\D, \ell|\ell'}-q^{S(u)}_{\D, \ell|\ell'} ) +\sum_{\D, \ell,\ell', T^+}\bigg(1+\frac{2}{N}\bigg)c_{\D,\ell'}^T(q^{T(u)}_{\D, \ell|\ell'}-q^{T(t)}_{\D, \ell|\ell'})\nonumber\\ & +\sum_{\D, \ell, A^-} (2c_{\D,\ell}^A\,q^{A(s)}_{\D, \ell}+\sum_{\ell'}c_{\D,\ell'}^A(q^{A(u)}_{\D, \ell|\ell'}+q^{A(t)}_{\D, \ell|\ell'})) \bigg)\bigg]\,Q^{2s+\ell}_{\ell, 0}(t) \ + \ \cdots
 \end{align}
The explicit formulas for $q^{i,(t)}_{\D,\ell|\ell'}$ and $q^{i,(t)}_{\D,\ell|\ell'}$ are given in \eqref{qtgen}. As given in \eqref{ss} we can expand $q^{i,(t)}_{\D,\ell|\ell'}(s)$ and $q^{i,(u)}_{\D,\ell|\ell'}(s)$ around the point $s=\D_\phi$ to get,
\begin{align}
q_{\D,\ell|\ell'}^{i,(t)}(s)&\equiv \, \, q^{i,(2,t)}_{\D, \ell|\ell'} + (s-\D_\phi)q^{i,(1,t)}_{\D, \ell|\ell'}+ \ldots\nonumber\\
q_{\D,\ell|\ell'}^{i,(u)}(s)&\equiv \, \, q^{i,(2,u)}_{\D, \ell|\ell'} + (s-\D_\phi)q^{i,(1,u)}_{\D, \ell|\ell'}+ \ldots
\end{align}
The terms $q^{i,(2,s/t/u)}_{\D,\ell}$ give the $\log$ unphysical terms. Once we set the $\log$ terms to 0, the power law unphysical terms are given in terms of $q^{i,(1,s/t/u)}_{\D,\ell}$. In the crossed channel we will mostly need the expression for the $\ell'=0$ contributions which are given by,
\begin{align}\label{qtid}
q^{(2,t)}_{\D, \ell |\ell'=0}=&\int d\n  \frac{\m^{(t)}_{\D,0}(\n)\G(\l)\G(\bar{\l})2^\ell((\D_\phi)_\ell)^2}{\k_\ell(\D_\phi)(2\D_\phi+\ell-1)_\ell} \sum_{q=0}^\ell \frac{(-\ell)_q(2\D_\phi+\ell-1)_q\G(q+\l)\G(q+\bar{\l})}{((\D_\phi)_q)^2\ q!\G(q-2k+\l+\bar{\l})}  \nonumber\\
 &  \  \  \,,
\end{align}
and
\begin{align}\label{qt1}
\begin{split}
&q_{\D,\ell |\ell'=0}^{(1,t)}= \int d\n   \m^{(t)}_{\D,0}(\n)\partial_s\Bigg[\frac{2^\ell((s)_\ell)^2}{\k_\ell(s)(2s+\ell-1)_\ell}\G(s-\D_\phi+\l) \G(s-\D_\phi+\bar{\l})\\
&\times \sum_{q=0}^\ell\frac{(-\ell)_q(2s+\ell-1)_q}{((s)_q)^2\ q!} \frac{\G(q+s+\l-\D_\phi)\G(q+s+\bar{\l}-\D_\phi)}{\G(q+2s+\l+\bar{\l}-2\D_\phi)} \Bigg]_{s=\D_\phi} \ \ \,,
\end{split}
\end{align}
where $\l=(h+\nu)/2$ and $\bar{\l}=(h-\nu)/2$. Using properties of $Q_{\ell,0}(t)$ as given in Appendix \ref{A3} one can show that,
\bea\label{ueqtot}
q^{(t)}_{\D, \ell|\ell'}(s) &=& (-1)^{\ell+\ell'}\,q^{(u)}_{\D, \ell|\ell'}(s)\,.
\eea

 \subsection{Disconnected piece}

The above analysis does not include the case where the exchange operator is an identity operator which gives the disconnected part of the four point function. This is given by,
\be
\mathcal{A}_{dis}(u,v)=\d_{i_1 i_2}\d_{i_3 i_4}+\d_{i_1 i_4}\d_{i_2 i_3}\left(\frac{u}{v}\right)^{\D_\phi}+\d_{i_1 i_3}\d_{i_2 i_4}u^{\D_\phi}\,.
\ee
This has a Mellin transform that can be written as,
\be\label{mellindisc}
\mathcal{A}_{dis}(u,v)=\int \frac{ds\, dt}{(2\pi i)^2}\, u^s\, v^t\, \G(-t)^2\,\G(s+t)^2\,\G(\D_\f-s)^2 \left( \d_{i_1 i_2}\d_{i_3 i_4}M^{(s)}_{dis}+ \d_{i_1 i_4}\d_{i_2 i_3}M^{(t)}_{dis}+\d_{i_1 i_3}\d_{i_2 i_4}M^{(u)}_{dis}\right)\,,
\ee
where the Mellin amplitudes of the disconnected $s$, $t$ and $u$ channels are given by,
\begin{align}\label{mellinid}
\begin{split}
M^{(s)}_{dis}(s,t)=&\frac{(\G^2(\D_\phi-s)\G^2(-t)\G^2(s+t))^{-1}}{s t}\,,\\
M^{(t)}_{dis}(s,t)=&\frac{(\G^2(\D_\phi-s)\G^2(-t)\G^2(s+t))^{-1}}{(s-\D_\phi)(t+\D_\phi)}\,,\\
M^{(u)}_{dis}(s,t)=&\frac{(\G^2(\D_\phi-s)\G^2(-t)\G^2(s+t))^{-1}}{(\D_\phi-s-t)t}\,.\\
\end{split}
\end{align}
Note that, the Mellin amplitude of the identity piece is not well-defined. However, for our purposes it suffices to consider only the relevant poles, as in \eqref{mellinid}.

The equation \eqref{mellindisc} can be rearranged according to singlet, traceless symmetric and antisymmetric tensor combinations, as below,
 \begin{align}\label{AdisM}
 &\mathcal{A}_{dis} (u,v) = \int \frac{ds\, dt}{(2\pi i)^2}\, u^s\, v^t\, \G(-t)^2\,\G(s+t)^2\,\G(\D_\f-s)^2 \times \bigg[\d_{i_1 i_2}\,\d_{i_3 i_4}\,\big(M^{(s)}_{ dis}+ \frac{1}{N}\,(M^{(t)}_{ dis}+M^{(u)}_{ dis})\big) \nonumber \\& +\half\, \big(\delta_{{i_1}{i_4}}\delta_{{i_3}{i_2}}+\delta_{{i_1}{i_3}}\delta_{{i_2}{i_4}}-\frac{2}{N}\delta_{{i_1}{i_2}}\delta_{{i_3}{i_4}}\big)\big( M^{(t)}_{ dis}+M^{(u)}_{ dis}\big) + \half \big(\delta_{{i_1}{i_4}}\delta_{{i_3}{i_2}}-\delta_{{i_1}{i_3}}\delta_{{i_2}{i_4}}\big)\, \big(M^{(t) }_{dis}- M^{(u)}_{ dis}\big)\bigg]
 \end{align}
This integral also has spurious poles, but they only come from the simple poles of $M^{(t)}_{dis}(s,t)$ and $M^{(u)}_{dis}(s,t)$. Now the spurious terms come only from  $t$ and $u$ channel disconnected parts and there are no log terms from them. Since we have only power law $u^{\D_\phi}$ we will expand the Mellin amplitude in the basis of continuous Hahn Polynomials, as shown below,
 \be
 M^{(t)}_{ dis} (s\to\D_\phi, t) = \sum_{\ell} Q^{2s+ \ell}_{\ell, 0}(t)\, q^{(1,t)}_{\D=0,\ell|\ell'=0}\hspace{1cm} M^{(u)}_{ dis} (s\to\D_\phi, t)= \sum_{\ell} Q^{2s+ \ell}_{\ell, 0}(t)\, q^{(1,u)}_{\D=0,\ell|\ell'=0}
 \ee
where we have,
\begin{align}\label{qdisdef}
q^{(1,t)}_{\D=0,\ell|\ell'=0}(s)&=\frac{\k_\ell(s)^{-1}}{\G(\D_\phi-s)^2}\int \frac{dt}{2\pi i}\ M^{(t)}_{ dis} (s, t) Q^{2s+\ell}_{\ell,0}(t)\ = -\ \frac{\k_\ell(s)^{-1}(s-\D_\phi)}{\G(\D_\phi-s+1)^2}Q^{2s+\ell}_{\ell,0}(-\D_\phi)\nonumber\\
\text{and   \ \ } q^{(1,u)}_{\D=0,\ell|\ell'=0}(s)&=\frac{\k_\ell(s)^{-1}}{\G(\D_\phi-s)^2}\int \frac{dt}{2\pi i}\ M^{(u)}_{ dis} (s, t) Q^{2s+\ell}_{\ell,0}(t)\ = \ - \frac{\k_\ell(s)^{-1}(s-\D_\phi)}{\G(\D_\phi-s+1)^2}Q^{2s+\ell}_{\ell,0}(0)\,.
\end{align}
Since this term is associated with the power law term, the $q^{(1,t/u)}_{\D=0,\ell|\ell'=0}(s)$ will have to cancel with the $\partial_s((s-\D_\phi)q_{\D,\ell}')|_{s=\D_\phi}$ terms from \eqref{Mtoq}. In terms of $q^{(1,s/t/u)}_{\D,\ell}$, \eqref{AdisM} reads,
 \bea\label{Adistoq}
 \mathcal{A}_{dis}(u,v) &=& \int \frac{ds\, dt}{(2\pi i)^2}\, u^s\, v^t\, \G(-t)^2\,\G(s+t)^2\,\G(\D_\f-s)^2  \bigg[\d_{i_1 i_2}\,\d_{i_3 i_4}\,\sum_{\ell}\big(q^{(s) dis}_{\ell}+\frac{1}{N}\,(q^{(t) }_{\D=0,\ell|\ell'=0}+q^{(u) }_{\D=0,\ell|\ell'=0})\big)\nn && + \half\, \big(\delta_{{i_1}{i_4}}\delta_{{i_3}{i_2}}+\delta_{{i_1}{i_3}}\delta_{{i_2}{i_4}}-\frac{2}{N}\delta_{{i_1}{i_2}}\delta_{{i_3}{i_4}}\big)\sum_{\ell}\big( q^{(t)}_{\D=0,\ell|\ell'=0}+ q^{(u) }_{\D=0,\ell|\ell'=0}\big)\nn && + \half \big(\delta_{{i_1}{i_4}}\delta_{{i_3}{i_2}}-\delta_{{i_1}{i_3}}\delta_{{i_2}{i_4}}\big)\, \sum_{\ell}\big(q^{(t) }_{\D=0,\ell|\ell'=0}- q^{(u) }_{\D=0,\ell|\ell'=0}\big)\bigg]\,Q^{2s+\ell}_{\ell, 0}(t) \ + \ \cdots\nn
 \eea
Here the $\cdots$ indicate terms regular when $s\to\D_\phi$.

\subsection{Constraint Equations}
Let us now take the equations \eqref{Atoq} and \eqref{Adistoq} and equate the coefficients of logs and powers laws to 0. For the log terms we just put $s=\D_\phi$ in $q_{\D,\ell}(s)$ in all channels, and equate them to 0 for every value of $\ell$, and for each of the singlet, symmetric and antisymmetric sectors. This is because since the $Q^{2s+\ell}_{\ell}(t)$-s form a complete basis, each of them are independent, and so are each of the tensor structures. So we get six equations for every $\ell$, two corresponding to each sector. Using \eqref{ueqtot} the constraint equations undergo considerable simplifications. For even spin singlet exchange in the $s$-channel the constraints reduce to the following,
\be\label{es1}
\sum_{\D}\Bigg[c^S_{\D,\ell} q^{S(2,s)}_{\D, \ell|\ell'}+\frac{2}{N}\sum_{\ell'}c^S_{\D,\ell'}q^{S(2,t)}_{\D, \ell|\ell'}+ 2\,\bigg(1+\frac{1}{N}-\frac{2}{N^2}\bigg)\sum_{\ell'}c^T_{\D,\ell'}q^{T(2,t)}_{\D, \ell|\ell'}+ 2\left(1-\frac{1}{N}\right) \sum_{ \ell'}c_{\D,\ell'}^A\, q^{A(2, t)}_{\D, \ell|\ell'}\Bigg]=0\,.
\ee
\be\label{es3}
  \sum_{\D}\Bigg[c^S_{\D,\ell} q^{S(1,s)}_{\D, \ell|\ell'}+\frac{2}{N}\sum_{\ell'}c^S_{\D,\ell'}\, q^{S(1,t)}_{\D, \ell|\ell'}+ 2\,\bigg(1+\frac{1}{N}-\frac{2}{N^2}\bigg)\sum_{\ell'}c^T_{\D,\ell'} q^{T(1,t)}_{\D, \ell} + 2\left(1-\frac{1}{N}\right) \sum_{ \ell'}c_{\D,\ell'}^A\, q^{A(1, t)}_{\D, \ell|\ell'}+ \frac{2}{N}\,q^{(1,t)}_{0, \ell|0}\Bigg]=0\,.
\ee
For a symmetric traceless operator exchange in $s$-channel, we have,
\be\label{es2}
\sum_{\D}\Bigg[\sum_{\ell'}c^S_{\D,\ell}\,q^{S(2,t)}_{\D, \ell|\ell'} + c^T_{\D,\ell}q^{T(2,s)}_{\D, \ell}+\bigg(1-\frac{2}{N}\bigg)\,\sum_{\ell'}c^T_{\D,\ell'}q^{T(2,t)}_{\D, \ell|\ell'}-\sum_{\ell'}c_{\D,\ell'}^A q^{A(2, t)}_{\D, \ell|\ell'} \Bigg] =0\,.
\ee

\be\label{es4}
\sum_{\D}\Bigg[ \sum_{\ell'}c^S_{\D,\ell'}q^{S(1,t)}_{\D, \ell|\ell'}+  c^T_{\D,\ell}q^{T(1,s)}_{\D, \ell}+\bigg(1-\frac{2}{N}\bigg)\, \sum_{\ell'}c^T_{\D,\ell'}q^{T(1,t)}_{\D, \ell,\ell'}  -\sum_{\ell'}c_{\D,\ell'}^A q^{A(1, t)}_{\D, \ell|\ell'}+ q^{(1,t)}_{\D=0, \ell|\ell'=0}\Bigg]=0\,.
\ee
Similarly, the constraint equations for antisymmetric odd spin operator exchange in the $s$-channel are given by,
\be\label{os1}
\sum_{\D}\Bigg[\sum_{\ell'}c_{\D,\ell'}^S q^{S(2,t)}_{\D, \ell|\ell'}-\bigg(1+\frac{2}{N}\bigg)\sum_{\ell'}\,c_{\D,\ell'}^T q^{T(2,t)}_{\D, \ell|\ell'}+ c_{\D,\ell} ^A \,q^{A(2,s)}_{\D, \ell}+ \sum_{\ell'}c_{\D,\ell'}^A q^{A(2, t)}_{\D, \ell|\ell'}\Bigg]= 0 \,.
\ee
\be\label{os2}
\sum_{\D}\Bigg[\sum_{\ell'}c_{\D,\ell'}^S q^{S(1,t)}_{\D, \ell|\ell'}-\bigg(1+\frac{2}{N}\bigg)\sum_{\ell'}c_{\D,\ell'}^T q^{T(1,t)}_{\D, \ell|\ell'}+  c_{\D,\ell}^A \, q^{A(1,s)}_{\D, \ell} + \sum_{\ell'}c_{\D,\ell'}^A q^{A(1, t)}_{\D, \ell|\ell'}+  q^{(1,t)}_{0, \ell|0}\Bigg] =0 \,.
\ee
In writing the above equations we have used \eqref{ueqtot} and the fact that antisymmetric operators have odd spins and others have even spins.
\section{$\e$ -expansion from constraint equations} \label{eExp}

In this section we will use the above equations to get the dimensions and OPE coefficients of operators in $(\phi_i\phi^i)^2$ with an $O(N)$ global symmetry in $d=4-\e$ dimension at the Wilson-Fisher fixed point. The lagrangian in this theory is given by, 
\be\label{lagrangian}
S=\int d^d x\left[ \frac{(\partial \phi_i)^2}{2}+g(\phi_i\phi^i)^2\right]\,.
\ee
However we will not be using the explicit form of the lagrangian. Instead we will use the following assumptions:
\begin{itemize}
\item{There is a conserved stress tensor.}

\item{$Z_2$ symmetry ($\phi_i \leftrightarrow -\phi_i$) is present.}

\item{There are $N$ identical fundamental scalars.}

\item{The OPE coefficients of higher order operators which vanish in the free theory start at $O(\e)$ in the interacting theory so that the $C_{\D,\ell}$ of these operators are $O(\e^2)$.}

\end{itemize}
These assumptions will be enough, to determine the dimensions and OPE coefficients, from the equations above. Our starting point will be the conservation of stress tensor. This means we will use $\D_{\ell=2}=d$ as an input. 
Let us write the dimension of $\phi_i$ as $\D_\phi=1+\d_\phi^{(1)} \e+\d_\phi^{(2)} \e^2+\d_\phi^{(3)} \e^3+O(\e^4)$. It starts with 1 becuase in the free theory dimension of the fundamental scalar is $(d-2)/2=1+O(\e)$. For the stress tensor OPE coefficient we write $C_{2h, 2}=C_{S,2}^{(0)}+C_{S,2}^{(1)}\e+C_{S,2}^{(2)} \e^2+C_{S,2}^{(3)} \e^3 +O(\e^4)$. In the singlet equations \eqref{es1} and \eqref{es3}, we have,
\be\label{spin2-ep}
c_{2h,2}^Sq^{S(2,s)}_{2h,\ell=2}=-\frac{45}{4} \left(C_{S,2}^{(0)} \left(1+2 \delta^{(1)}_{\phi }\right)\right) \epsilon  \ + O(\e^2)\,.
\ee
Expansion of the derivative $q^{(1, s)}_{\ell=2}$ in $\e$ is given by,
\be\label{spin2der-ep}
c_{2h,2}^Sq^{S(1,s)}_{2h,\ell=2} =\frac{45 C_{S,2}^{(0)}}{2}+\frac{3}{2} \epsilon  \left(2 C_{S,2}^{(0)}+15 C_{S,2}^{(1)}+30 C_{S,2}^{(0)} \g_E \delta^{(1)}_{\phi }\right) +O(\e^2)\,.
\ee
Here $\g_E$ is the Euler-gamma. The disconnected part \eqref{qdisdef} for $\ell=0$ gives,
\be\label{kappa-ep}
q^{(1,t)}_{\D=0, \ell=2|\ell'=0}+q^{(1,u)}_{\D=0, \ell=2|\ell'=0}=-\frac{15}{2}-\frac{(47+60 \g_E) \delta^{(1)}_{\phi }}{4} \epsilon  \ + \ +O(\e^2)\,.
\ee
The $O(\e^2)$ and $O(\e^3)$ terms are too tedious and hence not written here. The crossed channel terms as shown below start from $O(\e^2)$ order. So solving \eqref{es1} and \eqref{es3} at $O(\e)$ we get,
\be
\d^{(1)}_\phi=-\frac{1}{2}\,,\hspace{1cm}C_{S,2}^{(0)}=\frac{1}{3 N}\hspace{1cm}\text{and}\hspace{1cm}C_{S,2}^{(1)}=-\frac{11}{36 N}\,.
\ee
The spin-0 singlet and traceless symmetric operators have the leading contributions in $t$-channel. We denote their dimensions as,
\bea
\D_{S,0} &=& 2+\d^{(1)}_{S,0} \e + \d_{S,0}^{(2)} \e^2 \ + \d_{S,0}^{(3)} \e^3+\, O(\e^4)\nn 
\D_{T,0} &=& 2+\d_{T,0}^{(1)} \e + \d_{T,0}^{(2)} \e^2 \ + \d_{T,0}^{(3)} \e^3+\, O(\e^4)
\eea
 and their respective OPE coefficients as,
\bea
C_{S,0} &=& C_{S,0}^{(0)} + C_{S,0}^{(1)}\,\e +C_{S,0}^{(2)}\,\e^2 \ +C_{S,0}^{(3)}\,\e^3 +\, O(\e^4)\nn 
C_{T,0} &=& C_{T,0}^{(0)}  + C_{T,0}^{(1)} \,\e + C_{T,0}^{(2)}\,\e^2+ C_{T,0}^{(3)}\,\e^3+\,O(\e^4)\,.
\eea
With this let us look at the singlet and traceless symmetric equations \eqref{es1}, \eqref{es3}, \eqref{es2} and \eqref{es4} for $\ell=0$. The  $q^{T(s)}_{\ell=0}$ and $q^{T(s)}_{\ell=0}$ for spin 0 are given by,
\begin{align}\label{phi21}
c_{\D_S,0}^Sq^{S(2,s)}_{\D_S,\ell=0}=& -C^{(0)}_{S,0} \d^{(1)}_{S,0} \left(1+\delta^{(1)}_{S,0}\right) \frac{\epsilon}{2} +\frac{\epsilon ^2}{2} \left(-\delta^{(1)}_{S,0} \left(1+\delta^{(1)}_{S,0}\right) \left(C_{S,0}^{(1)}+C_{S,0}^{(0)} \left(-\gamma_E +\delta^{(1)}_{S,0}\right)\right) \right.  \nonumber\\& \left. -C_{S,0}^{(0)} \left(1+2 \delta^{(1)}_{S,0}\right) \delta^{(2)}_{S,0}-2 C_{S,0}^{(0)} \left(1+2 \delta^{(1)}_{S,0} \left(1+\delta^{(1)}_{S,0}\right)\right) \delta^{(2)}_{\phi }\right) \nonumber\\
c^T_{\D_T,0}q^{T(2,s)}_{\D_T\ell=0}=& -C^{(0)}_{T,0} \d^{(1)}_{T,0} \left(1+\delta^{(1)}_{T,0}\right) \frac{\epsilon}{2} +\frac{\epsilon ^2}{2} \left(-\delta^{(1)}_{T,0} \left(1+\delta^{(1)}_{T,0}\right) \left(C_{T,0}^{(1)}+C_{T,0}^{(0)} \left(-\gamma_E +\delta^{(1)}_{T,0}\right)\right) \right.  \nonumber\\& \left. -C_{T,0}^{(0)} \left(1+2 \delta^{(1)}_{T,0}\right) \delta^{(2)}_{T,0}-2 C_{T,0}^{(0)} \left(1+2 \delta^{(1)}_{T,0} \left(1+\delta^{(1)}_{T,0}\right)\right) \delta^{(2)}_{\phi }\right)\,.
\end{align}
The derivatives are given by,
\begin{align}\label{phi22}
c^S_{\D_S,0}q^{S(1,s)}_{\D_S,\ell=0}=& C_{S,0}^{(0)} + \left(C_{S,0}^{(1)} +C_{S,0}^{(0)}  \left(-\gamma_E +\delta^{(1)}_{S,0}\right)\right) \epsilon \nonumber\\&
+ \left( C_{S,0}^{(2)} + C_{S,0}^{(1)}  \left(-\gamma_E +\delta^{(1)}_{S,0}\right)+\frac{C_{S,0}^{(0)}}{2}\text{  }\left(\gamma_E^2-2 \gamma_E  \delta^{(1)}_{S,0}+2 \delta^{(2)}_{S,0}\right)+2 C_{S,0}^{(0)} \gamma_E  \delta^{(2)}_{\phi }\right)\e^2 \nonumber\\
c^T_{\D_T,0}q^{T(1,s)}_{\D_T,\ell=0}=& C_{T,0}^{(0)} + \left(C_{T,0}^{(1)} +C_{T,0}^{(0)}  \left(-\gamma_E +\delta^{(1)}_{T,0}\right)\right) \epsilon \nonumber\\&
+ \left( C_{T,0}^{(2)} + C_{T,0}^{(1)}  \left(-\gamma_E +\delta^{(1)}_{T,0}\right)+\frac{C_{T,0}^{(0)}}{2}\text{  }\left(\gamma_E^2-2 \gamma_E  \delta^{(1)}_{T,0}+2 \delta^{(2)}_{T,0}\right)+2 C_{\D_T,0}^{(0)} \gamma_E  \delta^{(2)}_{\phi }\right)\e^2\,.
\end{align}
The spin 0 disconnected parts reads,
\be\label{phi23}
q^{(1,t)}_{\D=0, \ell=0|\ell'=0}+q^{(1,u)}_{\D=0, \ell=0|\ell'=0}=-2-4 (1+\gamma_E ) \epsilon  \delta_\phi^{(1)}-4 \epsilon ^2 \left(\gamma_E  (2+\gamma_E ) (\delta_\phi^{(1)})^2+(1+\gamma_E ) \delta_\phi^{(2)}\right)\,.
\ee
To determine the $C^{(0)}_{S,0}, C^{(0)}_{T,0}, \d^{(1)}_{S,0}, \d^{(1)}_{T,0}, \cdots$ we would also need the crosssed channels. In the $t(\text{or }u)$-channel for $q^{S,(t)}_{\ell=0}$ and $q^{T,(t)}_{\ell=0}$ only the $\ell'=0$ operators have the leading contributions. This is also true for $\ell>0$. This nice feature is discussed in detail later in this section. So we have,
\begin{align}\label{phi24}
&c_{\D_S,0}^Sq^{S(2,t)}_{\D_S,\ell=0|\ell'=0}=-C_{S,0}^{(0)}\left(1+\delta^{(1)}_{S,0}\right)^2 \frac{\epsilon}{2} \nonumber
\\ & +\frac{\epsilon^2}{108} \left(1+\delta^{(1)}_{S,0}\right)  \left(27 \left(\left(1+\delta^{(1)}_{S,0}\right) \left(C_{S,0}^{(0)}-2 C_{S,0}^{(1)}+2 C_{S,0}^{(0)} \gamma_E -2 C_{S,0}^{(0)} \delta^{(1)}_{S,0}\right)-4 C_{S,0}^{(0)} \delta^{(2)}_{S,0}\right)-2 C_{S,0}^{(0)} \delta^{(1)}_{S,0} \delta^{(2)}_{\phi }\right)\nonumber\\ & +O(\e^3) \,,
\end{align}
and the derivative,
\be\label{phi25}
c_{\D_S,0}^Sq^{S(1,t)}_{\D_S,\ell=0|\ell'=0}=-\frac{C_{S,0}^{(0)} \e^2}{18}+O(\e^3)+\cdots\,.
\ee
The correponding terms for $q^{T(t)}_{\ell=0}$ and its derivatives are simply given by replacing with the traceless symmetric scalar,
\be
q^{T(2,t)}_{\D_T,\ell=0|\ell'=0} \ = \  q^{S(2,t)}_{\ell=0} ( \ C^{(0)}_{S,0}\to C^{(0)}_{T,0},  \ C^{(1)}_{S,0}\to C^{(1)}_{T,0},  \ \d^{(1)}_{S,0} \to \d^{(1)}_{T,0} \  )
\ee
and
\be
q^{T(1,t)}_{\D_T,\ell=0|\ell'=0}\ = \  q^{T(1,t)}_{\D_S,\ell=0|\ell'=0} (C^{(0)}_{S,0}\to C^{(0)}_{T,0} )
\ee
Now using the above in \eqref{es1}, \eqref{es3}, \eqref{es2} and \eqref{es4} to get the solutions of $C^{(0)}_{S,0}, \  C^{(0)}_{T,0},  \ C^{(1)}_{S,0}$, \  $C^{(1)}_{T,0},   \ \d^{(1)}_{S,0}$ and  $\d^{(1)}_{T,0} $. These solutions are listed at the end of this subsection. But let us first use them to obtain the crossed channel terms with $\ell=2$. It is only the spin 0 operators that will contribute to the $q_{\ell=2}^{i(t)}$. They are given by,
\begin{align}
c^S_{\D_S,0}q^{S(2,t)}_{\D_S,\ell=2|0}=\frac{5}{32} C_{S,0}^{(0)} (\delta_{S,0}^{(1)} -2 \text{$\delta_\phi^{(1)} $})^2 \epsilon^2 \hspace{0.5cm}&\text{and}\hspace{0.5cm}c^T_{\D_T,0}q^{T(2,t)}_{\D_T,\ell=2|0}=\frac{5}{32} C_{T,0}^{(0)} (\delta_{T,0}^{(1)} -2 \text{$\delta_\phi^{(1)} $})^2 \epsilon^2\,.\nonumber\\
c^S_{\D_S,0}q^{S(1,t)}_{\D_S,\ell=2|0}=-\frac{21}{64} C_{S,0}^{(0)} (\text{$\delta_{S,0}^{(1)} $}-2 \text{$\delta_\phi^{(1)} $})^2 \epsilon^2 \hspace{0.5cm}&\text{and}\hspace{0.5cm} c^T_{\D_T,0}q^{T(1,t)}_{\D_T,\ell=2|0}=-\frac{21}{64} C_{T,0}^{(0)} (\text{$\delta_{T,0}^{(1)} $}-2 \text{$\delta_\phi^{(1)} $})^2 \epsilon^2\,.
\end{align}
This allows us to solve for \eqref{es1} and \eqref{es3} for $\ell=2$ at the $O(\e^2)$ order. Thus we get $\d_\phi^{(2)}$. With this information we go back to solving \eqref{es1}, \eqref{es3}, \eqref{es2} and \eqref{es4} for $\ell=0$ at the order of $O(\e^2)$. Then we return to $\ell=2$ and solve for $\ell=2$. That gives us $\D_\phi$ up to the $O(\e^3)$ order. Skipping the details of $O(\e^3)$ let us just give the results. The dimension of $\phi$ is obtained to be,
\be\label{dphi}
\D_\f = 1-\frac{\e}{2} +\frac{(N+2)}{4\,(N+8)^2}\, \e^2 -\frac{(2+N)\,(-272+N\,(N-56))}{16\,(8+N)^4}\,\e^3+\,O(\e^4)\,.
\ee
For the singlet and symmetric traceless scalars we get,
\begin{align}\label{sol}
 &  C_{S,0} = \frac{2}{N}-\frac{2(2+N)}{8N+N^2}\e -\frac{6\,(28+20\,N+3\,N^2)}{N\,(8+N)^3} \e^2 +C_{S,0}^{(3)}\e^3 \,, \\
& \label{sol2}C_{T,0}=1 -\frac{2 \ \e}{8+N}  -\frac{6\,(14+3\,N)}{(8+N)^3}\e^2 +C_{T,0}^{(3)}\e^3\,,\\  
&\label{sol3}\D_{S,0} = 2-\frac{6 \ \e}{8+N}+\frac{\e^2 \ (2+N)\,(44+13N)}{2\,(8+N)^3} \,, \\ &\label{sol4} \D_{T,0} =2 -\frac{6+N}{8+N}\e -\frac{(N-22)\,(4+N)}{2\,(8+N)^3}\e^2 \,.
\end{align}
The OPE coefficient of the stress tensor is given by,
\be
C_{S,2}=\frac{1}{3 N}-\frac{11 \epsilon }{36 N}+\frac{\left(514+145 N+7 N^2\right) \epsilon ^2}{108 N (8+N)^2}+\frac{\left(41824+27968 N+3462 N^2-193 N^3+N^4\right) \epsilon ^3}{1296 N (8+N)^4}+O(\e^4)\,.
\ee

This is how far one can go with $\ell=0$ and $\ell=2$. Obtaining the anomalous dimensions and OPE coefficients to next order of $\e$ becomes difficult since, as we discuss below, an infinite number of operators start contributing in both the channels at the next order. However for the terms $q_{\D,\ell=0}^{i,(1,s)}$ and $q_{\D,\ell=0,\ell'}^{i,(1,t)}$ at the $O(\e^3)$, only one operator contributes in each channel, even though for the analogous terms  $q_{\D,\ell=0}^{i,(2,s)}$ and $q_{\D,\ell=0,\ell'}^{i,(2,t)}$ an infinite number of terms contribute at $O(\e^3)$. This means even though we would not be able to compute the $O(\e^3)$ $\ell=0$ anomalous dimensions, $\d_{S,0}^{(3)}$ and $\d_{T,0}^{(3)}$, if we knew these quantites, we would be able to compute the $O(\e^3)$ OPE coefficients $C_{S,0}^{(3)}$ and $C_{T,0}^{(3)}$. Borrowing the $O(\e^3)$ anomalous dimensions from \cite{Braun, kleinert},
\begin{align}
\d_{S,0}^{(3)}&=-\frac{(2+N) \left(3 N^3+96 (8+N) (22+5 N) \zeta (3)-5312-2672 N-452 N^2\right)}{8 (8+N)^5}\,,\nonumber\\
\d_{T,0}^{(3)}&=\frac{10624+4192 N+56 N^2-134 N^3-5 N^4-192 (8+N) (22+5 N) \zeta (3)}{8 (8+N)^5}\,,
\end{align}
we can solve \eqref{es1} and \eqref{es2} for $\ell=0$ at $O(\e^3)$ order, to get,
\begin{align}
C_{S,0}^{(3)}&=\frac{(2+N) \left(-4256-1216 N-46 N^2+19 N^3+(8+N) (1504+N (344+N (14+N))) \zeta (3)\right)}{2 N (8+N)^5}\,,\nonumber\\
C_{T,0}^{(3)}&=\frac{27 N^3+N^4+2 (8+N) (752+N (204+7 N)) \zeta (3)-4256-1728 N-110 N^2}{2 (8+N)^5}\,.
\end{align}

\subsection{$\e$-expansion for higher spin exchange}
We now proceed to study the higher spin operators using the constraint equations.
We use the constraint equations\eqref{es1}-\eqref{os2} to find the OPE coefficient and anomalous dimension for the  spin $\ell$ singlet, symmetric traceless and antisymmetric operator exchange in the $s$-channel. The higher spin conformal dimension are of the order $\D =2\,\D_\f+\,\ell+\,O(\e^2)$. Let us denote their dimensions and OPE coefficients as,
\bea\label{unknowns}
\D_{i,\ell} &=& 2-\e+ \ell + \d_{i,\ell}^{(2)}\,\e^2+\,\d_{i,\ell}^{(3)}\,\e^3 +\, O(\e^4)\nn
C_{i,\ell} &=& C_{i,\ell}^{(0)} + C_{i,\ell}^{(1)}\,\e + C_{i,\ell}^{(2)}\, \e^2++ C_{i,\ell}^{(3)}\, \e^3\,+ O(\e^4)
\eea
where the subscript $i$ stands for  singlet($S$), symmetric traceless ($T$) and antisymmetric ($A$) exchange respectively. Here we use the fact that the singlet and symmetric traceless operators exist for even spins only, and the antisymmetric ones for odd spins. Even for the general $\ell$ cases, we find only the spin 0 singlet and symmetric traceless scalars contributing to the $t$ and $u$ channels, under the $\e$-expansion. The higher spin operators do not contribute to the crossed channels upto $O(\e^3)$. However, they will contribute to the $O(\e^4)$ order. This is discussed at the end of this section. 

To find the unknowns we solve \eqref{es1}-\eqref{os2}, order by order. The steps are exactly similar as chalked out for spin $2$. So we skip the details and give the solutions directly. The conformal dimensions in the three sectors are given by,
\bea\label{higherspindim}
\D_{S,\ell} &=& 2-\e+ \ell +\,\e^2\,\frac{(N+2)}{2\,(N+8)^2}\bigg(1-\frac{6}{\ell\,(\ell+1)}\bigg)\nn&&+\,\e^3\,\frac{(2+N)}{8\,\ell^2\,(\ell+1)^2\,(8+N)^4}\Big[8 N (\ell -1) \ell  (7 \ell  (\ell +3)+54)-N^2 \left((\ell  (\ell +2)-39) \ell ^2+28\right)\nn && -16 (N+8)^2 \ell  (\ell +1) H_{\ell -1}-448 (N+4)+16 \ell  (17 \ell  (\ell  (\ell +2)+3)-126)\Big]+\,O(\e^4) \nn
\D_{T,\ell} &=& \! 2-\e+\ell  +\, \e^2 \, \frac{(N+2)}{2\,(N+8)^2}\,\bigg(1-\frac{2\,(6+N)}{(N+2)\,\ell\,(\ell+1)}\bigg) \ -\,\frac{\e^3 \ell^{-2}\,(\ell+1)^{-2}}{8\,\,(8+N)^4}\nn && \times \big[  4(8+N) (112+N (30+N))+48 (6+N) (14+3 N) \ell -(4+N) (408+N (122+7 N)) \ell ^2 -272 \nn && +2 (2+N) ((N-56) N) \ell ^3+(2+N) ((N-56) N-272) \ell ^4+64 (4+N) (8+N) \ell  (1+\ell ) H_{\ell -1} \big] +\,O(\e^4)\nn
\D_{A,\ell} &=& 2-\e+ \ell +\, \e^2 \, \frac{(N+2)}{2\,(N+8)^2}\bigg(1-\frac{2}{\ell\,(\ell+1)}\bigg)  \nn&&+\,\e^3\,\frac{(2+N)}{8\,\ell^2\,(\ell+1)^2\,(8+N)^4}\big[16\,(\ell-1)(32+\ell\,(74+17\ell^2+51\,\ell))\nn&&+8\,N(\ell-1)(12+\ell(30+21\,\ell+7\ell^2))-N^2\,(4+\ell^2\,(-7+\ell^2+2\ell))-32\ell(\ell+1)(8+N)H_{\ell-1}\big]\nn&&+\,O(\e^4)\,,
\eea
where  $H_\ell$ is a harmonic number of order $\ell$.

It should be noted that for $\ell=1$ the anomalous dimension $\D_{A,\ell}$ vanishes as it is the conserved current. Also for $\ell=2$ we have vanishing anomalous dimension for singlet representation which is the conserved stress-tensor. However,  for $\ell=2$ symmetric traceless operators acquire an anomalous dimension. We can compute the the higher spin OPE coefficients for any given spin $\ell$ upto $O(\e^3)$. The explicit expressions for individual $N$ and $\ell$ up to the $O(\e^2)$ are given below. Their $O(\e^3)$ part have been obtained for individual $\ell$-s, and can be automated for any value of $\ell$--a general formula can be obtained using a different method \cite{RGAS}. First few have been listed here. 
\begin{align}\label{opesing}
& \frac{C_{S,\ell}}{C_{S,\ell}^{free}}=1 \ +  \e^2\Bigg[\frac{(2+N)\left(6+2 \left(-3-2 \ell +2 \ell ^2+\ell ^3\right) H_{\ell}-\left(-6-5 \ell +2 \ell ^2+\ell ^3\right) H_{2 \ell }\right)}{2 (8+N)^2 \ell  (1+\ell )^2} \Bigg]  \ + \ c^{(3)}_{S,\ell} \e^3 
\end{align}
\begin{align}\label{opetraceless}
& \frac{C_{T,\ell}}{C_{T,\ell}^{free}}=1 + \ \nonumber \\ &\e^2 \Bigg[ \frac{(2 (6+N)+2 (1+\ell ) (-6-N+(2+N) \ell  (1+\ell )) H_{\ell }-(1+\ell ) (-2 (6+N)+(2+N) \ell  (1+\ell )) H_{2 \ell }}{2 (8+N)^2 \ell  (1+\ell )^2}\Bigg] \ \nonumber \\ & + \ c^{(3)}_{T,\ell} \e^3 
\end{align}
\begin{align}\label{opeanti}
\frac{C_{A,\ell}}{C_{A,\ell}^{free}}=1 \ + \e^2 \Bigg[ \frac{(2+N) \left(2+2 \left(-1+2 \ell ^2+\ell ^3\right) H_{\ell }+\left(2+\ell -2 \ell ^2-\ell ^3\right) H_{2 \ell }\right)}{2 (8+N)^2 \ell  (1+\ell )^2} \Bigg] \ + \ c^{(3)}_{T,\ell} \e^3 
\end{align}

The $O(\e^3)$ terms $c_{i,\ell}^{(3)}$ can be computed for any given spin. They also obey a general $\ell$ formula. These are given in Appendix \ref{e3}\,.

The anomalous dimension corrections are in good agreement with the known results \cite{Braun}. The OPE ccoefficient corrections are new, except for $\ell=1$ and 2, which correspond to $c_J$ and $c_T$, which were calculated till $O(\e^2)$ \cite{petkou}, and agree with our results. In Appendix \ref{hugh} we give  an alternative computation of $c_T$ from symmetry arguments that agrees with our result \eqref{cT}, up to the $O(\e^3)$ order.

\subsection{Simplification under $\e$-expansion}\label{simpli}

While evaluating the equations \eqref{es1}-\eqref{os2}, we could get away with just one operator in the $s$-channel and two operators (the singlet and traceless symmetric scalars) in the crossed channels. This is because the other operators in the sum contribute from a subleading ($O(\e^4)$) order in $\e$. Let us see how the other operators are suppressed, in $\e$ in all the channels.

\subsubsection{$s$-channel}\label{heavier}

For a spin $\ell$ in the $s$-channel one can have operators with dimension  $\D_{2m,\ell}=\ell+2+2m +\d_m\e+O(\e^2)$. Such operators have the form $O_{2m,\ell}\sim\phi_i \partial^a \phi \partial^b \phi \partial^c \phi$, where among $a+b+c=2m+\ell$ derivatives, $2m$ derivatives are contracted and $\ell$ derivatives carry indices. For them, the $q^{(2,s)}_{\D,\ell}$ is given by,
\begin{align}
c_{\D,\ell}q_{\D,\ell}^{i(2,s)}=&\frac{C_{2m,\ell }\epsilon ^2(-1)^{2 m+\ell } 2^{-2+4 m+3 \ell }\text{  }m (m+\ell ) (1+2 m+2 \ell ) \left(\delta _m+1\right){}^2  \Gamma ^2(m) \Gamma (\ell ) \Gamma ^2\left(\frac{1}{2}+m+\ell \right)}{\pi  \Gamma ^4(1+m+\ell )}\nonumber\\ &+O\left(\epsilon ^3\right)\,.
\end{align}
Here $C_{2m,\ell}$ is the OPE coefficient of the operator. Since the operator is made of four $\phi$-s, it does not exist in the free theory, and in the interacting theory the generic three point function goes as $\langle \phi_i \phi_j O_{2m,\ell}\rangle \sim \e$. Since the OPE coefficient is the square of the three point function, we have $C_{2m,\ell}\sim \e^2$ and the whole expression above contributes at $O(\e^4)$. Now, there can be other operators too, with higher number of $\phi$-s. But they contribute at a more subleading order because their OPE coefficients are even further suppressed in $\e$.

\subsubsection{$t$-channel}\label{tch-cancel} 
In the crossed channels the simplifications happen due to cancellation of residues of various poles among one another, under $\e$ expansion--the discussion is similar to the one in \cite{uslong}. The cancellations are such that all the operators in the crossed channels start contributing from $O(\e^4)$. The only operators that can contribute at a more leading order are the lowest dimension scalars. Even for these operators there are only two poles whose residues contribute at a leading order, and all other residues cancel among one another to start from $O(\e^4)$ or more.

For $\ell'=0$ the expression \eqref{ts} or \eqref{qtgen} can have poles at (after substituting $s=\D_\phi$),
\begin{itemize}
\item{I. \ \ $\nu= (\D-h)$ }

\item{II. \  $\nu= (2\D_\phi-h+2n)$ }

\item{III. $\nu= (h+2n)$ }\,.

\end{itemize}
Here $n$ is a positive integer. For the lowest dimension operators, i.e. those with dimension $\D=2+\d_0^{(1)}\e+O(\e^2)$, we have the following residues cancelling each other,
\be\label{cancel1}
Res_{\n=h+2n-2}+Res_{\n=2\D_\phi-h+2n}=O(\e^4)\,.
\ee
So only the poles I and II (for $n_1=0$) contribute to our computations.

Heavier scalars having dimensions of the form $\D_{2m,0}=2+2m+\d_m \e+O(\e)$ contribute only from $O(\e^4)$. Tis is due to the following cancellations of residues:
\begin{align}
Res_{\nu=2\D_\phi-h}&=O(\e^5)\nonumber\\
Res_{\nu=\D-h}+Res_{\nu=h-2\D_\phi+2s+2m-2}+Res_{\nu=2\D_\phi-h+2m}&=O(\e^4)\nonumber\\
Res_{\nu=h-2\D_\phi+2s+2n-2}+Res_{\nu=2\D_\phi-h+n}&=O(\e^4) \text{ \ \ with \ \ }n\ne m\,. 
\end{align}

For spin $\ell'>0$ we have the following poles in the crossed channels:
\begin{itemize}
\item{I. \ \ $\nu= (\D-h)$}

\item{II. \  $\nu= (h+\ell'+2n)$}

\item{III. $\nu= (2\D_\phi+\ell'-h+2n)$ }

\item{IV.   $\nu= (h-1), (h-2), \cdots , (h-2+\ell')$ }

\end{itemize}
Here we observe two different cases:\\
{\bf Lowest dimension spin $\ell'$:}\\ 
These are operators of the form $\D_{\ell'}=2-\e+\ell'+O(\e^2)$, whose dimensions we computed in the paper\,. We observe the following cancellations for them,
\begin{align}
Res_{\nu=\D-h}+Res_{\nu=h-2+\ell'}+Res_{\nu=2\D_\phi+\ell'-h}&=O(\e^4)\nonumber\\
Res_{\nu=h+\ell'+2n}+Res_{\nu=2\D_\phi-h+\ell'+2n+2}&=O(\e^4) \text{\ \ where \ \ } n=0,1,2,\cdots\nonumber\\
Res_{\nu=h-1}\sim Res_{\nu=h-2}\sim \cdots Res_{\nu=h+\ell'-3}&=O(\e^4)\,.
\end{align}
{\bf Higher dimensional spin $\ell'$ operators:} \\
These are the operators $O_{2m,\ell'}$ we discussed in \ref{heavier} having the dimensions $\D_{2m,\ell}=\ell+2+2m +\d_m\e+O(\e^2)$\,. Their OPE coefficients go like $C_{2m,\ell'}\sim O(\e^2)$ or higher. Accounting for this suppression we have the following cancellations,
\begin{align}
Res_{\nu=h-2+\ell'}+Res_{\nu=2\D_\phi+\ell'-h}&=O(\e^4)\nonumber\\
Res_{\nu=\D-h}+Res_{\nu=h+\ell'+2m-2}+Res_{\nu=2\D_\phi-h+\ell'+2m}&=O(\e^4) \nonumber\\
Res_{\nu=h+\ell'+2n}+Res_{\nu=2\D_\phi-h+\ell'+2n+2}&=O(\e^4) \text{\ \ where \ \ } n\ne m-1\nonumber\\
Res_{\nu=h-1}\sim Res_{\nu=h-2}\sim \cdots Res_{\nu=h+\ell'-3}&=O(\e^4)\,.
\end{align}
Thus the crossed channels get contributions from only a finite number of operators. We refer the readers to Appendix F of \cite{uslong} for a more detailed discussion of these simplifcations.

\section{Large $N$ critical $O(N)$ model }\label{largeN}
In this section we focus on the constraint equations to fix
 the OPE coefficients and anomalous dimensions of operators appearing in the large $N$ expansion of the $\f^4$ theory in $d$ ($\equiv 2h$)dimension.
 Here we start with the conserved current which suggests that we will use $\D_{\ell=1}=d-1$ as the input. Let us write the dimension of $\phi$ as $\D_\f =  h-1+ \frac{1}{N}\,\d_\phi^{(1)} + O (1/N^2)$.  It starts with $h-1$ because the dimension of the fundamental scalar in free theory is $(d-2)/2= h-1$. We write the spin one conserved current OPE coefficient as $C_{2h-1, 1}=C_{A,1}^{(0)}+\frac{1}{N}C_{A, 1}^{(1)}+O(1/N^2) $. In the equations \eqref{os1} and \eqref{os2}, we have,
 
 \bea
 c_{2h-1,1}^Aq^{A(2,s)}_{2h-1,\ell=1}&=& -\frac{1}{N}\frac{2^{-1+2 h} C_{A,1}^{(0)} \delta _{\phi }^{(1)} \Gamma \left(\frac{1}{2}+h\right)}{ \sqrt{\pi } \Gamma (h-1) \Gamma [h]^2} + \, O(1/N^2)\nn
 c_{\D_S,0}^Sq^{S(2,t)}_{\D_S, \ell=1|\ell'=0} &=& -\frac{1}{N}\frac{C_{S,0}^{(1)}(h-2)^2 (-1+h) \Gamma (2 h)}{2 \Gamma ^3(h) \Gamma (1+h)}+\,O(1/N^2)\,.
 \eea
 In $q^{S,(2,t)}_{\D_S,\ell=1|\ell'}$ only $\ell'=0$ has contributed. Other operators contribute from $O(1/N^2)$ because of similar reasons as we explained in section \ref{simpli}. The $q^{T(t)}_{ \ell=1} $ also starts from $O(1/N^2)$.
 The large $N$ expansion of the derivative $q^{ A(1,s)}_{\ell=1}$ is given by,
 \bea
 c_{2h-1,1}^Aq^{ A(1,s)}_{2h-1,\ell=1}&=& \frac{2^{-1+2 h} C_{A,1}^{(0)}\Gamma \left(\frac{1}{2}+h\right)}{\sqrt{\pi } \Gamma (h-1) \Gamma ^2(h)} +\,O(1/N)\,.
 \eea
The disconnected part for $\ell=1$ is given by,
\be
q^{(1,t)}_{\D=0,\ell=1|\ell'=0}=-\frac{\Gamma (2 h)}{2 \Gamma ^2(h-1) \Gamma ^2(h)}\,.
\ee
Solving the constraints we get,
\be
 C_{A,1}^{(0)} =\frac{h-1}{2},\qquad  \d_{\phi}^{(1)} =\frac{2\,C_{S, 0}^{(1)}\,(h-2)^2}{h\,(h-1)}\,.
 \ee
 We find $C^{(1)}_{S,0}$ in the next subsection. Using that result given in \eqref{zero} we can fix $\d_\phi^{(1)} $.
%

\be\label{p1}
\d_\phi^{(1)} = \frac{4^{h-1}\,(h-2)\,\G(h-\half)\,\sin(\pi\,h)}{\pi^{3/2}\,\G(1+h)}
\ee
This agrees with eqn. 4.6 of \cite{giombii}.

\subsection{Spin $\ell=0$ in the $s$-channel}
We will first discuss the case of spin $0$ operators exchange.
Let us denote the anomalous dimensions of the spin-$0$ singlet and symmetric traceless operators as,
\bea\label{spin0dim}
\D_{S,0} &=& 2+\frac{1}{N}\, \d_{S,0}^{(1)} + O (1/N^2)\nn
\D_{T,0} &=& 2(h-1)+\frac{1}{N}\,\d_{T,0}^{(1)}+ O (1/N^2)
\eea
and their respective OPE coefficients as,
\bea\label{spin0ope}
C_{S,0} &=& C_{S,0}^{(0)}+\frac{1}{N}\, C_{S,0}^{(1)} + O (1/N^2)\nn
C_{T,0} &=& C_{T,0}^{(0)}+\frac{1}{N}\, C_{T,0}^{(1)}+ O (1/N^2)\,.
\eea
We have verified that it is essential to have $\D_{S,0}$  begin with 2, instead of $2h-1$ in order to have consistency of the equations. This is consistent with the fact that the Lagrange multiplier field is the shadow of $\phi_i\phi_i$.

We focus on the singlet and symmetric traceless equations \eqref{es1}-\eqref{es4} for $\ell=0$. 
In the large $N$ limit the $q^{S(s)}_{\ell=0}$ and $q^{T(s)}_{\ell=0}$ -channel has the following expansion,
\bea
c_{\D_S,0}^Sq^{S(2,s)}_{\D_{S},\ell=0} &=&-\frac{1}{N}\frac{C_{S,0}^{(0)} \left(\delta_{S,0}^{(1)}+2 \delta _{\phi }^{(1)}\right) \Gamma (3-h)}{2\text{  }\Gamma (h-1)}+\,O(1/N^2)\nn
c_{\D_T,0}^Tq^{T(2,s)}_{\D_T,\ell= 0}&=&\frac{1}{N}\frac{2^{2 h-4} C_{T,0}^{(0)} \left(\delta _{T,0}^{(1)}-2 \delta _{\phi }^{(1)}\right) \Gamma \left(h-\frac{1}{2}\right)}{N\sqrt{\pi } \Gamma ^3(h-1)}+\,O(1/N^2)
\eea
The derivatives are given by,
\bea
c_{\D_S,0}^Sq^{S(1,s)}_{\D_S,\ell=0} &=&\frac{C_{S,0} \Gamma (3-h)}{\Gamma (h-1)}+\frac{\left(C_{S,0}^{(1)}+C_{S,0}^{(0)} \delta _{S,0}^{(1)}\left(H_{2-h}+H_{h-2}\right)+C_{S,0}^{(0)} \left(\delta _{S,0}^{(1)}+2 \delta _{\phi }^{(1)} \gamma _E\right)\right) \Gamma (3-h)}{N \Gamma (h-1)}\nn && +\,O(1/N^2)\nn
c_{\D_T,0}^Tq^{ T(1,s)}_{\D_T,\ell=0}&=&\frac{2^{2h-3} C_{T,0}^{(0)} \Gamma \left(h-\frac{1}{2}\right)}{\sqrt{\pi } \Gamma (h-1)}+\frac{1}{N}\frac{\Gamma ^2\left(h-\frac{3}{2}\right)}{ \pi  \Gamma ^2(h-1) \Gamma (2 h-3)}\Big(4^{2 h-4} \left(C_{T,0}^{(1)} (2 h-3)+C_{T,0}^{(0)} (2 h-3) \delta _{T,0}^{(1)}\right.\nn&& \left.\left(\text{log}\ 4-H_{h-2}+H_{h-\frac{5}{2}}-H_{2 h-4}\right)+C_{T,0}^{(0)}\left(\delta _{T,0}^{(1)}+2 (2 h-3) \delta _{\phi }^{(1)} \gamma _E\right)\right) \Gamma ^2\big(h-\frac{3}{2}\big)\Big)+\,O(\frac{1}{N^2})\,.\nn 
\eea
The disconnected part is given by,
\be
q_{0,0|0}^{(1,t)}=-\frac{2 \Gamma (2 h-2)}{\Gamma ^4(h-1)}+\frac{4 \delta _{\phi }^{(1)} \left(2 H_{h-2}-H_{2 h-3}-\gamma_E\right) \Gamma (2 h-2)}{N \Gamma ^4(h-1)}\,.
\ee
In the crossed ($t$ or $u$) channel $q^{S,(t)}_{\ell=0}$ and $q^{T,(t)}_{\ell=0}$ only the $\ell^{\prime}=0$ operators have the leading contributions as discussed in section \ref{simpli}\,. The crossed channel and the derivatives have the following large $N$ expansion,
\bea
c_{\D_S,0}^Sq^{S(2,t)}_{\D_S,\ell=0|\ell'=0} &=&\frac{4^{h-2}\,C_{S,0}^{(0)}\,(2h-2)\,\G(h-\half)}{\sqrt{\pi}\,\G(h-1)^2\,\G(h)}+O(1/N)\nn
c_{\D_S,0}^Sq^{S(1,t)}_{\D_S,\ell=0|\ell'=0} &=&\frac{4^{h-2} C_{S,0}^{(1)} (h-2) \left(6 h-8-4 (h-1) (2 h-3) H_{h-1}+2 (h-1) (-3+2 h) H_{2 h-4}\right) \Gamma \left(h-\frac{3}{2}\right)}{(-1+h) N \sqrt{\pi } \Gamma ^2(h-1) \Gamma (h)} \nn  && +\,O(1/N)\,.
\eea
The $O(1/N)$ term of $q^{\prime S(t)}_{\ell=0}$ is too ugly to write here. 
The crossed channel $q^{T(t)}_{\ell=0}$ and the derivative start from $O(1/N^2)$.
We use the above in the constraint equations \eqref{es1}-\eqref{es4} to get the solutions for $C_{S,0}^{(0)}$, $C_{T,0}^{(0)}$, $C_{S,0}^{(1)}$, $C_{T,0}^{(1)}$ , $\d_{T,0}^{(1)}$. They are given by,
\be\label{zero}
C_{S,0}^{(0)}= 0, \qquad
C_{T,0}^{(0)}= 1,\qquad
C_{S,0}^{(1)} = \frac{2\G(2h-2)\,\sin(\pi\,h)}{\pi\,(h-2)\,\G(h-1)^2},\qquad
C_{T,0}^{(1)} = \frac{2^{2h-1} \sin (\pi  h) \Gamma \left(h-\frac{1}{2}\right)}{\pi ^{3/2} (h-1) \Gamma (h)}\nonumber
\ee
\be
\d_{T,0}^{(1)} 
= -\frac{4^h \sin (\pi  h) \Gamma \left(h-\frac{1}{2}\right)}{\pi ^{3/2} \Gamma (h+1)}\,.
\ee
The anomalous dimension of spin zero symmetric operators match with known results \cite{giombii}. An important point is that using the constraint equations we could not fix the anomalous dimension of the singlet as it requires the solution at $O(1/N^2)$. At this order the higher spin operators start contributing to the $t$-channel. This is a similar problem to the one we faced in $\e$-expansion, as dicussed in section \ref{simpli}, that prevented us from going beyond a certain order in $\e$.

\subsection{Higher spin in the $s$-channel}
Now we will use the spin zero results obtained above to determine the anomalous dimensions and OPE coefficients of operators with spin $\ell$. Let us denote the higher spin singlet, symmetric traceless and antisymmetric operators and the corresponding OPE coefficients as,
\bea
\D_{i,\ell} &=& 2\,h-2+\ell +\frac{1}{N}\,\d_{i,\ell}^{(1)} + O (1/N^2) \nn
C_{\D_i,\ell} &\equiv& C_{i,\ell}= C_{i,\ell}^{(0)} +\frac{1}{N}\,  C_{i,\ell}^{(1)} + O (1/N^2) 
\eea
where the subscipt $i$ is the shorthand $S, T, A$  for singlet, symmetric traceless and antisymmetric exchange respectively.
Now we  use the constraint equations \eqref{es1}-\eqref{os2} to extract the singlet, symmetric traceless and antisymmetric anomalous dimensions and OPE coefficients. The $q^{i(s)}_{\ell} , q^{i(t)}_{\ell}$ and their  derivatives have the following large $N$ expansion,
\bea
c^i_{\D_i,\ell}q^{i(2,s)}_{\D_i,\ell} &=& -\frac{1}{N}\,\frac{2^{-\ell-1}\,C_{i,\ell}^{(0)} \,(2\ell+2h-3)\,(2\,\d_\phi^{(1)}-\d_{i,\ell}^{(1)} )\,\G(2\ell+2h-3)^2}{\G(\ell+h-1)^4\,\G(\ell+2h-3)} + \, O(1/N^2)\nn
c^S_{\D_S,\ell}q^{S(2,t)}_{\D_S,\ell|\ell'=0}&=& \frac{1}{N}\frac{(-1)^\ell\,2^{\ell+4h-5}\,(h-2)\,\G(h-\half)\,\G(\ell+h-\half)\,\sin(h\,\pi)}{\pi^2\,(\ell+h-2)\,\G(h-1)^3\,\G(1+\ell)\,\G(\ell+h)}+\,O(1/N^2)\,.
\eea
The crossed channel $q^{T(t)}_{\ell}$ and its derivative start at $O(1/N^2)$.
The derivative of $q^{i(s)}_{\D_i,\ell} $ is given by,
\begin{align}
c^i_{\D_i,\ell}q^{i(1,s)}_{\D_i,\ell}  &= \frac{(-1)^{\ell } 2^{-8+4 h+3 \ell } C_{i,\ell }^{(0)} (2 h+2 \ell -3) \Gamma ^2\left(h+\ell -\frac{3}{2}\right)}{\pi  \Gamma ^2(h+\ell -1) \Gamma (2 h+\ell -3)}  +\,O(1/N)\nonumber\\
c^S_{\D_i,\ell}q^{S(1,t)}_{\D_i,\ell|\ell'=0}&=-\frac{1}{N}\frac{2^{1-\ell } C_{S,0}^{(1)} \left(H_{h+\ell -2}+H_{h+\ell -1}-H_{2 h+2 \ell -3}\right) \Gamma (h+\ell -2) \Gamma (2 (-1+h+\ell ))}{\ell ! \Gamma ^2(h-2) \Gamma ^2(h+\ell -1) \Gamma (h+\ell )} \ + \ O(1/N^2)\,.
\end{align}
The disconnected piece for spin $\ell$ has the following large $N$ expansion,
\be
q^{(1,t)}_{\D=0,\ell|\ell'=0} = -\frac{2^{1-\ell } \Gamma (-2+2h+2\ell )}{\ell ! \Gamma ^2(h-1) \Gamma ^2(h+\ell -1)}+\frac{1}{N}\frac{2^{2-\ell } \delta _{\phi }^{(1)}\text{  }\Gamma (2h-2+2\ell ) \left(H_{h-2}+H_{h-2+\ell }-H_{2 h-3+2 \ell }-\gamma _{E}\right)}{\ell ! \Gamma ^2(h-1) \Gamma ^2(h+\ell -1)}\,.
\ee
Now we solve the constraint equations \eqref{es1}-\eqref{os2} to compute the anomalous dimension and OPE coefficient of spin $\ell$ operators.
The solutions are given by,
\begin{align}\label{largen1}
&\D_{S,\ell} = 2h- 2+ \ell+ O(1/N)\\
& \label{largen2}\D_{T/A,\ell} = 2h-2+\ell+\frac{1}{N}\frac{2\, \d_\phi^{(1)}\, ((\ell-1) (2 h+\ell-2))}{(h+\ell-2) (h+\ell-1)}+ O(1/N^2)\\
& \label{largen3}C_{S,\ell} = \frac{1}{N}\,\frac{\sqrt{\pi } 2^{5-2 h-2 \ell } \Gamma (h+\ell-1) \Gamma (2 h+\ell-3)}{\Gamma (h-1)^2 \Gamma (\ell+1) \Gamma \left(h+\ell-\frac{3}{2}\right)} + O(1/N^2)\\
& \label{largen4} C_{T/A,\ell}=C_{T/A,\ell}^{(0)}+\frac{C_{T/A,\ell}^{(1)}}{N}+O(1/N^2)\\
&C^{(0)}_{{T/A},\ell} = \frac{\sqrt{\pi } 4^{2-h-\ell } \Gamma (h+\ell-1) \Gamma (2 h+\ell-3)}{\Gamma (h-1)^2 \Gamma (\ell+1) \Gamma \left(h+\ell-\frac{3}{2}\right)}\nonumber\\
&C_{{T/A}}^{(1)}=\frac{2^{3-2\ell} (h-2) \text{sin}(h \pi ) \Gamma \left(h-\frac{1}{2}\right) \Gamma (h+\ell-1) \Gamma (2 h+\ell-3)}{\pi  h(h-1)\Gamma ^3(h-1) \Gamma (1+\ell) \Gamma \left(h+\ell-\frac{3}{2}\right)(h+\ell-2) (h+\ell-1)}\Big[h(h-1)H_{h+\ell-2}\nonumber\\&
+\frac{2 h^4+(\ell-2) (\ell-1)^2 (2 \ell-3)+h^3 (8 \ell-13)+h^2 (27+\ell (6 \ell-35))+h (\ell (46+\ell (8 \ell-33))-22)}{(h+\ell-2) (h+\ell-1) (2 h+2 \ell-3)}\nonumber\\&
-H_{h-2}(h+\ell-2) (h+\ell-1)+(\ell-1) (2 h+\ell-2) H_{h+\ell-3}+(\ell-1) (2 h+\ell-2)\left( H_{2 h+\ell-4} - H_{2 (h+\ell-2)}\right)\Big]\,.
\end{align}
Note that the anomalous dimension vanishes for $\ell=2$ singlet and $\ell=1$ antisymmetric operators as it should, corresponding to the conservation of the stress-tensor and current. These results are in agreement with known results \cite{Lang:1991kp,Lang:1992zw,giombii}. In Appendix \ref{largeNexact} we show how to extract the OPE coefficients and anomlaous dimension corrections from the exact form of the correlator at $1/N$.

\section{Comparison with known results}\label{compare}
\subsection{\text{Pad$\acute{\text{e}}$} approximations} 
We can construct the \text{Pad$\acute{\text{e}}$} approximant for the $O(N)$ models
\be\label{pade}
 \text{Pad$\acute{\text{e}}$}_{[m, n]} (d) =\frac{A_0+\,A_1\,d+\,A_2\, d^2+\cdots+A_m\,d^m}{1+\,B_1\,d+\,B_2\, d^2+\cdots+B_n\,d^n}
\ee
for any given physical quantity known in the $d=4-\,\e$ and $d=2+\,\e$ expansions upto a given order--we will closely follow the discussion in \cite{klebanov2}. The coefficients in \eqref{pade} are fixed by demanding that the expansion \eqref{pade} agrees with the known perturbative expansions in $d= 4-\e$ and $d=2+\e$ at each order.

We can use the following $d=4-\e$ expansion results for ${c_J/c_J}_{free}$ and ${c_T/c_T}_{free}$ to order $\e^3$ given in section \ref{eExp} as well as the $d=2+\e$ results given in \cite{klebanov2} \footnote{These are given by $c_J/c_J^{free}=\frac{C_{A,1}^{free}}{C_{A,1}}$ and $c_T/c_T^{free}=\frac{ 4\Delta_\phi ^2 C_{S,2}^{free}}{(d-2)^2C_{S,2}}$.},
\bea
{c_J/c_J}_{free} &=& 1-\frac{3 \,(N+2)}{4\,(N+8)^2}\,\e^2 -\frac{(N+2)(N^2+132N+632)}{8\,(N+8)^4}\,\e^3+\,O(\e^4) \quad {\rm{in}} \quad d=4-\e  \nn
&=& \frac{N-2}{N}+\,\frac{\e}{N} +\,O(\e^2) \qquad {\rm{in}} \quad d=2+\e
\eea
\bea\label{cT}
{c_T/c_T}_{free} &=& 1-\frac{5\,(N+2)}{12\,(N+8)^2}\,\e^2-\,\frac{(N+2)(7N^2+382N+1708)}{36\,(N+8)^4}\,\e^3 +\,O(\e^4)\quad {\rm{in}} \quad d=4-\e  \nn
&=& 1-\frac{1}{N} +\frac{3\,(N-1)}{4\,N\,(N-2)}\,\e^2+\,O(\e^3) \qquad {\rm{in}} \quad d=2+\e
\eea
We construct the approximant $\text{Pad$\acute{\text{e}}$}_{[3,2]}$ and $\text{Pad$\acute{\text{e}}$}_{[4, 2]}$ for  ${c_J/c_J}_{free}$ and ${c_T/c_T}_{free}$ respectively. These approximants are well-behaved and in good agreement with large $N$ results in $2 < d < 4$. The results of Pade approximant for ${c_J/c_J}_{free}$ and ${c_T/c_T}_{free}$ for several values of $N$ are listed in the tables \ref{Pade1} and \ref{Pade2}.

\begin{table} 
	\begin{center}
		\begin{tabular}{ | c c  c c c|}
			\hline 
			\hspace{0.1cm}	$N$ \hspace{0.3cm}& \hspace{0.2cm} $\epsilon^3$ results \hspace{0.3cm}&\hspace{0.2cm} $\text{Pad$\acute{\text{e}}$}_{[4, 2]}$ & \hspace{0.7cm} Large $N$ results  & \hspace{0.7cm} Numerical results \\ 
			\hline 
			
			1 & 0.957933& 0.778495  & 0.549684& $0.946600$ \\ 
			2 & 0.955556 & 0.94661  & 0.774842& $0.94365$ \\ 
			3 & 0.955111 & 0.933861  & 0.849895 & $0.94418$ \\
			4 & 0.955729& 0.938871  & 0.887421 & $0.94581$\\
			5 & 0.956919& 0.944714 & 0.909937 & $0.9520$ \\
			6  & 0.958397& 0.949952  & 0.924947& $0.9547$\\
			10 & 0.964792& 0.964425  & 0.954968& 0.96394 \\
			20 & 0.97623& 0.979761  & 0.977484& 0.97936 \\
			\hline
		\end{tabular}
		
		\caption{The values of $\epsilon^3$, \text{Pad$\acute{\text{e}}$} approximants $\text{Pad$\acute{\text{e}}$}_{[4, 2]}$,  large $N$  and  numerical bootstrap results \cite{Kos:2013tga} for ${c_T/c_T}_{free}$ in $d=3$ for the $O(N)$ model. }\label{Pade1}
		
	\end{center}
\end{table}

\begin{table}
	\begin{center}
		\begin{tabular}{ | c c  c c c|}
			\hline 
			\hspace{0.1cm}	$N$ \hspace{0.3cm}& \hspace{0.2cm} $\epsilon^3$ results \hspace{0.3cm}&\hspace{0.2cm} $\text{Pad$\acute{\text{e}}$}_{[3,2]}$ & \hspace{0.7cm} Large $N$ results & \hspace{0.7cm} Numerical results   \\ 
			\hline 
			
			2 & 0.925 & 0.863795& 0.639747 & 0.9050(16)\\
			3 & 0.92474& 0.879552 & 0.759831 & 0.9065(27) \\
			4 & 0.926215 & 0.893426&   0.819873& -  \\
			5 & 0.928587& 0.905072&  0.855899& - \\
			6 & 0.931383& 0.914751&  0.879916 & -  \\
			10 & 0.942901& 0.940299&  0.927949 & -  \\
			20 & 0.962525 & 0.96654 & 0.963975  & 0.9674(8) \\
			\hline
		\end{tabular}
		\caption{The values of $\epsilon^3$, \text{Pad$\acute{\text{e}}$} approximants $\text{Pad$\acute{\text{e}}$}_{[3, 2]}$,  large $N$ and  numerical bootstrap  results \cite{Kosarchie} for ${c_J/c_J}_{free}$ in $d=3$ for the $O(N)$ model. (Only numerical results which were precisely computed in \cite{Kosarchie} have been presented.) For $N=2$ quantum Monte Carlo \cite{Witczak-Krempa:2013nua}  results quote the value 0.917 or 0.904 \cite{prlcj} depending on the extrapolation scheme used. }\label{Pade2} 
	\end{center}
\end{table}

\subsection{Large spin analysis}
In \cite{uslong}  it is shown how the  $s$, $t$ and $u$ channels simplify when we consider large spin in the $s$-channel. Let us assume a weakly coupled theory, which means a CFT where we have a certain suitable small parameter $g$, in terms of which the anomalous dimension can be expanded.  The large spin analysis predicts the behavior of large spin operators as an expansion in a small parameter, at large $\ell$, which matches with the prediction of \cite{alday}. In this section we will briefly review that analysis for $O(N)$ models with large $N$. So we will have $g=1/N$, and demonstrate how it can correctly reproduce known results for large spin dimensions in $4<d<6$ dimension.

We will start with the correlator $\langle \phi_i \phi_j \phi_k \phi_l \rangle$. The external fields have the dimension,
\be
\D=\D_\phi=\frac{d-2}{2}+\g_\phi\,.
\ee
In the $s$-channel we have the large spin operators having dimensions of the form 
\be
\D_{i,\ell}=d-2+\ell+\g_{i,\ell}\,.
\ee 
Both $\g_{\phi}$ and $\g_{\ell}^i$ allow expansions in $N^{-1}$ as $\g_\phi=\d_{\phi}^{(1)}/N+O(N^{-2})$ and $\g_{i,\ell}=\d_{i,\ell}^{(1)}/N+O(N^{-2})$. From \eqref{ss}, we can use $\ell \gg 1$ to obtain,
\be\label{weakgammalhs}
c_{\D,\ell}q^{i,(2,s)}_{\D,\ell}=\frac{P_i 2^{2\ell +d-\frac{5}{2}} e^{\ell } \ell ^{-\ell } (\gamma_{i,\ell} -2\g_\phi)}{\pi  \Gamma \left(\frac{d-2}{2}\right)^2}\,.
\ee
In the above equation, $P_S=1/N$, $P_T=P_A=1/2$. 

To evaluate the $t$-channel, we will use the following approximation of $Q_{\ell,0}^{2s+\ell}$ \cite{uslong},
\be\label{Qapprox}
Q^{2s+\ell}_{\ell,0}(t)\stackrel{\ell \gg s,t}{=} \ \frac{2^{\ell } \ell ^{-s-t} \Gamma (s+\ell )^2 \Gamma (-1+s-t+\ell )}{\Gamma (-t)^2 \Gamma (-1+2 s+2 \ell )}\,.
\ee
We use \eqref{ts} to evaluate $q_{\D,\ell|\ell'}^{(t)}$. Both $t$ and $\nu$ contours are determined from the power of $\ell$ in the integrand. With the above approximation in the integrand, if we do the $t$ integral first, we will have poles at $t=\l_2-\D_\phi$ and $t=\bar{\l}_2-\D_\phi$. All other poles in $t$ have residues suppressed in $\ell$ or lie out of the contour. Similarly in the $\nu$ integral we will only have poles at $\nu=\pm(\D_i-h)$ (signs depend on the pole of $t$ considered before). Other $\nu$ poles have residues suppressed in $\ell$, or are out of the contour. Writing $\D_i=\ta_i-\ell'$, we arrive at,
\be\label{afternuint}
c_{\D_i,\ell}q^{(2,t)}_{\D_i,\ell|\ell'}= -\frac{C_{\D_i,\ell'} \mathfrak{N}_{\ta_i+\ell',\ell'} 2^{-\frac{1}{2}+\ell -3 \ell'+2 \Delta _{\phi }-\tau_i} e^{\ell }\ell^{-\ell -\tau_i} \Gamma \left(\ell'+\frac{\tau_i}{2}\right) \Gamma \left(\ell'+\Delta_{\phi }+\frac{\tau_i}{2}-h\right)^2 \Gamma \left(\ell'+\tau_i-1\right)}{\sqrt{\pi } \Gamma \left(1-h+\ell'+\tau_i\right) \Gamma \left(\frac{2 \ell'+\tau_i-1}{2}\right)}\,.
\ee
The above formula gives the contributions from different $O(N)$ sectors in the $t$-channel. Since this is the $t$ channel we have a sum over $\ta_i$ and $\ell'$. However one can see from the large $\ell$ dependence, that only operators with small twists dominate the sum.  

Here we assume the presence of a singlet scalar of dimension 
\be\label{leading}
\D_S=\ta_S=2+O(g) =2+\frac{1}{N}\, \d_{S,0}^{(1)} + O (1/N^2)
\ee 
in the spectrum.  This operator is the lagrange multiplier field present in the large $N$ critical theory. It becomes significant in $4<d<6$ dimensions because it is the leading operator at large $\ell$  in the $t$-channel. Then the sum in the $t$-channel goes away and we use \eqref{leading} to expand \eqref{afternuint}. Finally using the contraint equations \eqref{es1}. \eqref{es2} and \eqref{os1}, we get,
\be\label{gammaellweak1}
\g_{i,\ell}-2\g_\phi=\frac{\a_0(N^{-1}) +\a_1(N^{-1}) \log \ell+\a_2(N^{-1}) 
(\log \ell)^2+\cdots}{\ell^{2}}\,.
\ee
Here $\a_p(N^{-1})$ are given by,
\begin{align}\label{alphaidim2}
\a_p(N^{-1})= & \frac{C_{S,0}^{(1)}}{N} \sum_{q=0}^{2}\frac{(-1)^{p+q+1}}{2 p!}  \left(\delta _{0}^{(1)}\right)^{p+q}(\delta _{0}^{(1)}-2\d_\phi^{(1)})^{q} \binom{2}{q} (d-4)^{2-q}\left(\frac{1}{N}\right)^{p+q} \ \nonumber\\&+O(\left(\frac{1}{N}\right)^{p-2} (d-4)^3,\left(\frac{1}{N}\right)^{p-1}(d-4)^2,\left(\frac{1}{N}\right)^{p}(d-4),\left(\frac{1}{N}\right)^{p+1})\,.
\end{align}
In general dimension the second line of \eqref{alphaidim2} is not significant, and we can just take $q=0$. For large $N$ critical model we have,
\be
C_{S,0}^{(1)} = \frac{2\G(2h-2)\,\sin(\pi\,h)}{\pi\,(h-2)\,\G(h-1)^2}\,.
\ee
Plugging this in, we get the correct $1/\ell^2$ dependence for the large spin currents in all the sectors as given in \cite{giombii}  for $p=0$,
\be
\a_0=\frac{2h\G(2h-1)\,\sin(\pi\,h)}{N\pi\,(h-2)\,\G(h+1)}\,.
\ee
One can also compute the leading $\log \ell$ term, given by,
\be
\a_1=\frac{1}{N^2}\frac{2^{-1+4 h}\text{  }(h-2) (2 h-1) \Gamma ^2\left(h-\frac{1}{2}\right) \sin ^2(h \pi )}{h \pi ^3 \Gamma ^2(h-1)}\,.
\ee
This matches with the expected $\log$ term at $O(1/N^2)$  \cite{Alday2}.

\section{Cubic anisotropy}\label{anisotropy}
The $\phi^4$ interaction of \eqref{lagrangian} can be extended to the case of cubic anisotropy, whose lagrangian can be written as,
\be
S=\int d^d x\left[ \frac{(\partial \phi_i)^2}{2}+g_{ijkl} \phi_i\phi_j\phi_k\phi_l \right]\,,
\ee
where
\be\label{geng}
g_{ijkl}=\frac{g_1}{3}(\d_{ij}\d_{kl}+\d_{il}\d_{jk}+\d_{ik}\d_{jl})+g_2 \d_{ijkl}\,.
\ee
Here we have introduced the generalized $\d_{ijkl}$-function. It is defined by
\begin{equation}
\d_{ijkl} = 
\begin{cases}
1, & \text{when $i=j=k=l$},\\
0, & \text{otherwise}\,.
\end{cases} 
\end{equation}
Also $\d_{ijkl}\d_{kl}=\d_{ij}$. The interaction term then looks like $g_1(\phi_i\phi_i)^2+g_2\sum_i\phi_i^4$\,. This action breaks the $O(N)$ symmetry. The symmetries respected by this system are: $\phi_i \leftrightarrow -\phi_i$ and $\phi_i \leftrightarrow \phi_j$.  

We will now study our bootstrap conditions to understand what happens in the cubic anisotropy case without referring to the lagrangian. Since $O(N)$ symmety is absent, we cannot use the form \eqref{s-ope}. To get the operator content let us look at the two point OPE $\phi_i \times \phi_j$. 
\be
\phi_i \times \phi_j \supset \{1,S,T_{(ij)},V_{ij},A_{[ij]}\}\,.
\ee
Here schematically $S \equiv \phi_i\partial^\ell\phi_i$ denotes singlet operators, $T_{ij}\equiv \phi_{(i}\partial^\ell\phi_{j)}-\d_{ijkl}\phi_k\partial^\ell\phi_l$  \ is a symmetric operator with no diagonal element, $V_{ij}=\d_{ijkl}\phi_k\partial^\ell\phi_l-\frac{\d_{ij}}{N}\phi_k\partial^\ell\phi_k$ is a traceless diagonal operator (i.e. 0 when $i\ne j$) and $A_{[ij]}=\phi_{[i}\partial^\ell\phi_{j]}$ denotes the antisymmetric operator. The traceless symmetric operator of the $O(N)$ case has now broken up into the two pieces $T$ and $V$ which are different multiplets that do not mix with each other. In free theory, the two-point function of one operator with another in a different multiplet is 0, and we will take the same operator basis for the interacting theory.

Now let us take $\langle \phi_{i_1}\phi_{i_2}\phi_{i_3}\phi_{i_4}\rangle = (x_{12}^2x_{34}^2)^{-\D_\phi}\mathcal{A}(u,v)$. We will use the above OPE to write the conformal blocks and the associated irreducible tensor structures. We have,
\begin{align}\label{cubic-ope}
\mathcal{A}(u,v) &= \sum_{S^+}C_{\D,\ell}\,(\delta_{{i_1}{i_2}}\delta_{{i_3}{i_4}})\,g_{\D,\ell}(u,v)\nonumber \\ &
+\sum_{T^+}C_{\D,\ell}\,(\delta_{{i_1}{i_3}}\delta_{{i_2}{i_4}}+\delta_{{i_1}{i_4}}\delta_{{i_2}{i_3}}-2\,\delta_{{i_1}{i_2},{i_3}{i_4}})\,g_{\D,\ell}(u,v)
+\sum_{V^+}C_{\D,\ell}(\d_{i_1,i_2,i_3,i_4}-\frac{1}{N}\d_{i_1 i_2}\d_{i_3i_4})g_{\D,\ell}(u,v)
\nonumber \\ & +\sum_{A^-}C_{\D,\ell}\,(\delta_{{i_1}{i_4}}\delta_{{i_2}{i_3}}-\delta_{{i_1}{i_3}}\delta_{{i_2}{i_4}})\,g_{\D,\ell}(u,v)
\end{align}
where $+$ and $-$ are to indicate even and odd spins respectively and the $C_{\D,\ell}$ appearing in a sector corresponds to operators in  that sector.

The corresponding Witten diagram expansion can be written as,
\begin{align}\label{schWitten}
\sum_{\D,\ell}c_{\D,\ell}W^{(s)}_{\D,\ell}(u,v) &= \int \frac{ds\, dt}{(2\pi i)^2}\, u^s\, v^t\, \G(-t)^2\,\G(s+t)^2\,\G(\D_\f-s)^2 \bigg( (\delta_{{i_1}{i_2}}\delta_{{i_3}{i_4}})\,M^{S, (s)}(s, t) \nonumber \\&+(\delta_{{i_1}{i_3}}\delta_{{i_2}{i_4}}+\delta_{{i_1}{i_4}}\delta_{{i_2}{i_3}}-2\,\delta_{{i_1}{i_2},{i_3}{i_4}})\,M^{T,(s)}(s, t) + (\d_{i_1,i_2,i_3,i_4}-\frac{1}{N}\d_{i_1 i_2}\d_{i_3i_4}) M^{V,(s)}\nonumber \\& + (\delta_{{i_1}{i_4}}\delta_{{i_2}{i_3}}-\delta_{{i_1}{i_3}}\delta_{{i_2}{i_4}})\,M^{A, (s)}(s, t) \bigg).
\end{align}
As before, the $M^{i,(s)}$ are given by \eqref{Mdef}. The $t$ channel is obtained by changing $s\to t+\D_\phi$, $t\to s-\D_\phi$ and $2 \leftrightarrow 4$. The $u$ channel is obtained by changing $s\to \D_\phi-s-t$, and $2 \leftrightarrow 3$. The rest of the analysis is similar to the $O(N)$ case. We sum over all the channels and rearrange them according to the tensor structures appearing in the $s$-channel \eqref{schWitten}. Now we expand the Mellin amplitudes in terms of the continuous Hahn polynomials as in \eqref{Mtoq}. Then  corresponding to $\d_{i_1i_2}\d_{i_3i_4}$ we get the equations,
\begin{align}\label{cubic1}
\sum_{\D}\bigg[2 \sum_{\ell'}c_{\D,\ell'}^Vq_{\Delta ,\ell |\ell '}^{V,(1,t)} (N-1)+N \Big[2 \sum_{\ell'}c_{\D,\ell'}^S q_{\Delta ,\ell |\ell '}^{S,(1,t)}+2 q_{\Delta =0,\ell |\ell '=0}^{(1,t)}+2 \sum_{\ell'}c_{\D,\ell'}^Tq_{\Delta ,\ell |\ell '}^{T,(1,t)} (N-1)&\nonumber\\ +c_{\D,\ell}^Sq_{\Delta ,\ell}^{S,(1,s)} N\Big]\bigg] &=0\nonumber\\
\sum_{\D} \bigg[2 \sum_{\ell'}c_{\D,\ell'}^V q_{\Delta ,\ell |\ell '}^{V,(2,t)} (N-1)+N \Big[2 \sum_{\ell'}c_{\D,\ell'}^S q_{\Delta ,\ell |\ell '}^{S,(2,t)}+2\sum_{\ell'}c_{\D,\ell'}^T q_{\Delta ,\ell |\ell '}^{T,(2,t)} (N-1) &\nonumber\\   +c_{\D,\ell}^Sq_{\Delta ,\ell }^{S,(2,s)} N\Big]\bigg] &= 0 \,.
\end{align}
Corresponding to $\d_{i_1i_3}\d_{i_2i_4}+\d_{i_1i_4}\d_{i_2i_3}-2\d_{i_1i_2i_3i_4}$ we get the equations,
\begin{align}\label{cubic2}
\sum_{\D} \bigg[c_{\D,\ell}^Tq_{\Delta ,\ell }^{T,(1,s)}+\sum_{\ell'}c_{\D,\ell'}^Sq_{\Delta ,\ell |\ell '}^{S,(1,t)}+\sum_{\ell'}c_{\D,\ell'}^Tq_{\Delta ,\ell |\ell '}^{T,(1,t)}+q_{\Delta =0,\ell |\ell '=0}^{(1,t)}-\frac{1}{N}\sum_{\ell'}c_{\D,\ell'}^Vq_{\Delta ,\ell |\ell '}^{V,(1,t)}\bigg]&=0\nonumber\\
\sum_{\D} \bigg[c_{\D,\ell}^T q_{\Delta ,\ell }^{T,(2,s)}+ \sum_{\ell'}c_{\D,\ell'}^S q_{\Delta ,\ell |\ell '}^{S,(2,t)}+ \sum_{\ell'}c_{\D,\ell'}^T q_{\Delta ,\ell |\ell '}^{T,(2,t)}-\frac{1}{N}\sum_{\ell'}c_{\D,\ell'}^V q_{\Delta ,\ell |\ell '}^{V,(2,t)}\bigg]&=0
\end{align}
The part $\d_{i_1i_2i_3i_4}-\d_{i_1i_2}\d_{i_3i_4}/N$ gives,
\begin{align}\label{cubic3}
\sum_{\D} \bigg[ c_{\D,\ell}^Vq_{\Delta ,\ell}^{V,(1,s)}+2 \sum_{\ell'}\Big[\sum_{\ell'}c_{\D,\ell'}^Sq_{\Delta ,\ell |\ell '}^{S,(1,t)}- \sum_{\ell'}c_{\D,\ell'}^Tq_{\Delta ,\ell |\ell '}^{T,(1,t)}+q_{\Delta =0,\ell |\ell '=0}^{(1,t)}+\sum_{\ell'}c_{\D,\ell'}^V q_{\Delta ,\ell |\ell '}^{V,(1,t)}\big(1-\frac{1}{N}\big)\Big]\bigg]&=0\nonumber\\
\sum_{\D} \bigg[ c_{\D,\ell}^V q_{\Delta ,\ell }^{V,(2,s)}+2 \Big[ \sum_{\ell'}c_{\D,\ell'}^Sq_{\Delta ,\ell |\ell '}^{S,(2,t)}-\sum_{\ell'}c_{\D,\ell'}^T q_{\Delta ,\ell |\ell '}^{T,(2,t)}+ \sum_{\ell'}c_{\D,\ell'}^Vq_{\Delta ,\ell |\ell '}^{V,(2,t)}\big(1-\frac{1}{N}\big)\Big]\bigg]&=0\,.
\end{align}
Finally we have the antisymmetric $\d_{i_1i_4}\d_{i_2i_3}-\d_{i_1i_3}\d_{i_2i_4}$ sector  equations, which are
\begin{align}\label{cubic4}
\sum_{\D} \bigg[c_{\D,\ell}^A q_{\Delta ,\ell }^{A,(1,s)}+\sum_{\ell'}c_{\D,\ell'}^Sq_{\Delta ,\ell |\ell '}^{S,(1,t)}-\sum_{\ell'}c_{\D,\ell'}^Tq_{\Delta ,\ell |\ell '}^{T,(1,t)}+q_{\Delta =0,\ell |\ell '=0}^{(1,t)}-\frac{1}{N}\sum_{\ell'}c_{\D,\ell'}^Vq_{\Delta ,\ell |\ell '}^{V,(1,t)}\bigg]&=0\nonumber\\
\sum_{\D} \bigg[c_{\D,\ell}^A q_{\Delta ,\ell}^{A,(2,s)}+\sum_{\ell'}c_{\D,\ell'}^Sq_{\Delta ,\ell |\ell '}^{S,(2,t)}-\sum_{\ell'}c_{\D,\ell'}^Tq_{\Delta ,\ell |\ell '}^{T,(2,t)}-\frac{1}{N}\sum_{\ell'}c_{\D,\ell'}^Vq_{\Delta ,\ell |\ell '}^{V,(2,t)}\bigg]&=0
\end{align}
In writing the above equations we have used the relation \eqref{ueqtot}.

\subsection{Solutions}

In this subsection we will solve the above equations to find the anomalous dimensions and OPE coefficients of the  operators in the spectrum. Once again, we will use the conservation of stress tensor (i.e. $\D_{\ell=2}=4-\e$) as the input. This will give the dimension of $\phi$. Let us write it as,
\be
\D_\phi=1+\d_\phi^{(1)}\e+\d_\phi^{(2)}\e^2+\d_\phi^{(3)}\e^3+O(\e^4)
\ee
To determine $\D_\phi$ we will have to solve the above equations simultaneously order by order in $\e$. This will require the crossed channels too. For the same reason as described in \ref{tch-cancel} only the scalar operators of lowest dimension will contribute to the $t$-channel  till the $O(\e^3)$ order. So, in order to solve $\D_\phi$ till $O(\e^3)$ we also have to know the dimensions and OPE coefficients of the $\ell=0$ operators and  OPE coefficient of the spin 2 operator.,
\begin{align}
\D_{i,0}&=2+\d_{i,0}^{(1)}\e+\d_{i,0}^{(2)}\e^2+O(\e^3)\,, \nonumber\\C_{i,0}&=C_{i,0}^{(0)}+C_{i,0}^{(1)}\e+C_{i,0}^{(2)}\e^2+O(\e^3)\,,\nonumber\\
C_{2h,\ell=2}&\equiv C_{S,2}=C_{S,2}^{(0)}+C_{S,2}^{(1)}\e+C_{S,2}^{(2)}\e^2+C_{S,2}^{(3)}\e^3+O(\e^4)
\end{align}
where $i=S,T,V$. There is no spin 0 antisymmetric operator. Now using this parametrisation we solve the equations \eqref{cubic1} for $\ell=2$ and \eqref{cubic1},\eqref{cubic2} and \eqref{cubic3} for $\ell=0$ simultaneously. This gives,
\be\label{Dphicubic}
\D_\phi=1-\frac{\epsilon }{2}+\frac{(N-1) (2+N) \epsilon ^2}{108 N^2}+\frac{(N-1)\left(1728 N-222 N^2+109 N^3-1696\right) \epsilon ^3}{11664 N^4}\,.
\ee
The spin 0 dimensions and OPE coefficients are given by,
\begin{align}\label{Dim0}
\D_{S,0}&=2- \frac{2+N}{3 N} \epsilon-\frac{(N-1) (424+N (-326+19 N))}{162 N^3}\epsilon ^2\,,
\\\label{ope1}
C_{S,0}&=\frac{2}{N}-\frac{4 (N-1) }{3 N^2}\epsilon +\frac{4 \left(N-1) (106 - 89 N + N^2\right) }{81 N^4}\epsilon ^2\,,
\\\label{dimT}
\D_{T,0}&=2+\frac{2-3 N}{3 N}\epsilon +\frac{-424+N (530+N (-127+3 N))}{162 N^3}\epsilon ^2\,,
\\\label{ope2}
C_{T,0}&=1-\frac{2  }{3 N}\epsilon+\frac{2 (N-1) (19 N-106) }{81 N^3}\epsilon ^2\,,
\\\label{dimV}
\D_{V,0}&=2-\frac{2 (1+N)}{3 N}\epsilon +\frac{424+N (-538+N (131+19 N))}{162 N^3}\epsilon ^2\,,
\\\label{Dimlast}
C_{V,0}&=2-\frac{2 (-2+N)  }{3 N}\epsilon-\frac{2 (N-2) (N-1) (106+17 N)}{81 N^3} \epsilon ^2\,.
\end{align}
and the spin 2 OPE coefficient which is given by,
\be
C_{S,2}=\frac{2}{3 N}-\frac{11  }{18 N}\epsilon+\frac{(-22+N (11+74 N))  }{486 N^3}\epsilon^2+\frac{18656+N (-37664+N (22206+N (-4019+902 N)))}{52488 N^5}\e^3\,.
\ee
The quantities $\D_\phi$ and $\D_{S,0}$ are known in literature \cite{phi4}, and our results agree with them.

Now let us turn our attention to higher spin operators. Using the information obtained above we can determine their anomalous dimensions and OPE coefficients order by order in $\e$. Let us denote them as $\D_{i,\ell}$ and $C_{i,\ell}$. Here $i$ stands for $S,T,V,A$. Now we solve \eqref{cubic1}-\eqref{cubic4} and use \eqref{Dphicubic}-\eqref{Dimlast} order by order, to determine the above unknowns. For $i=S,T,V$ we have only even spins, and for $i=A$ we have only odd spins. We obtain,
\begin{align}\label{dimspinS}
\D_{S,\ell}&=2-\e+\ell  +\frac{(N+2)(N-1)(N-2)(N+3) }{54 N^2 \ell  (1+\ell )}\epsilon ^2+\frac{(N-1) }{5832 N^4 \ell ^2 (1+\ell )^2} ( (\ell -2) \ell  (1+\ell ) (3+\ell )\times\nonumber\\&
(1728 N -1696)-6 N^2 (\ell  (\ell  (37 \ell  (2+\ell )-491)-384)+36)+N^3 (\ell  (\ell  (109 \ell  (2+\ell )+373)\nonumber\\& -168)-108)
\left.+216 N^2 (N+2) \ell  (1+\ell ) \left(H_{\ell -1}-3 H_{\ell }\right)\right)\epsilon ^3
\end{align}
\begin{align}\label{dimspinT}
\D_{T,\ell}&=2-\epsilon+\ell +\frac{(12-18 N+(-1+N) (2+N) \ell  (1+\ell )) }{54 N^2 \ell  (1+\ell )}\epsilon ^2+\frac{1}{5832 N^4 \ell ^2 (1+\ell )^2} \left(109 N^4 \ell ^2 (1+\ell )^2\right.\nonumber\\&
+1696 (-2+\ell ) \ell  (1+\ell ) (3+\ell )-32 N (1+\ell ) (-108+\ell  (-636+107 \ell  (1+\ell )))+N^3 (540+\ell  (2952 \nonumber\\&
+\ell  (3053-331 \ell  (2+\ell ))))-6 N^2 (540+\ell  (2656+\ell  (1899-325 \ell  (2+\ell ))))+216 N \ell  (1+\ell ) \times \nonumber\\&
\left.\left(\left(16-18 N+7 N^2\right) H_{-1+\ell }+3 (2-3 N) N H_{\ell }\right)\right)\epsilon ^3
\end{align}
\begin{align}\label{dimspinV}
\D_{V,\ell}&=2-\epsilon+\ell  +\frac{\left(-2 \left(\ell +\ell ^2-6\right)+N^2 \left(\ell +\ell ^2-6\right)+N \left(6+\ell +\ell ^2\right)\right) }{54 N^2 \ell  (1+\ell )}\epsilon ^2+\frac{1}{5832 N^4 \ell ^2 (1+\ell )^2} \nonumber\\&
\left(1696 (\ell -2) \ell  (1+\ell ) (3+\ell )-32 N \ell  (1+\ell ) (107 \ell  (1+\ell )-324)+6 N^2 (-108+\ell  (1+\ell ) (-112\right.\nonumber\\&
+325 \ell  (1+\ell )))+N^4 (-108+\ell  (-168+\ell  (373+109 \ell  (2+\ell ))))+N^3 (540-\ell  (648+\ell  (1843\nonumber\\&
\left.+331 \ell  (2+\ell ))))+216 N^2 \ell  (1+\ell ) \left(\left(N+N^2-6\right) H_{\ell -1}-3 (-2+N) (1+N) H_{\ell }\right)\right)\epsilon ^3
\end{align}
\begin{align}\label{dimspinA}
\D_{A,\ell}&=2-\epsilon+\ell +\frac{(-18 (N-2)+(N-1) (2+N) \ell  (1+\ell )) }{54 N^2 \ell  (1+\ell )}\epsilon ^2+\frac{1}{5832 N^4 \ell ^2 (1+\ell )^2} \nonumber\\&
\big(109 N^4 \ell ^2 (1+\ell )^2+1696 \ell  (1+\ell ) \left(-18+\ell +\ell ^2\right)-6 N^2 (324+\ell  (1+\ell ) (3696-325 \ell  (1+\ell )))\nonumber\\&
-32 N (1+\ell ) (-54+\ell  (-1440+107 \ell  (1+\ell )))+N^3 (540+\ell  (2952+\ell  (3053-331 \ell  (2+\ell ))))\nonumber\\&
+216 (-2+N) N \ell  (1+\ell ) \left((7 N-4) H_{\ell -1}-9 N H_{\ell }\right)\big)\epsilon ^3\,.
\end{align}
Note that the anomalous dimension of spin-1 current does not vanish. This is expected since the rotational symmetry of $O(N)$ is no longer present, implying $J_{\mu}$ is not conserved. 

Now let us come to the OPE coefficients. We write them as,
\be\label{cubicope}
\frac{C_{i,\ell}}{C_{i,\ell}^{free}}=1+c_{i,\ell}^{(2)}\e^2+c_{i,\ell}^{(3)}\e^3\,.
\ee
Here we define the free field OPE coefficients by $C_{S,\ell}^{free}=C_{O(1),\ell}^{free}/N$, $C_{T,\ell}^{free}=C_{O(1),\ell}^{free}/2$, $C_{V,\ell}^{free}=C_{O(1),\ell}^{free}$ and $C_{A,\ell}^{free}=C_{O(1),\ell}^{free}/2$\,. Here $C_{O(1),\ell}^{free}$ is the free field OPE coefficient for any spin in the $O(1)$ theory, given by,
\be\label{Cfree}
C_{O(1),\ell}^{free}=\frac{2 \Gamma\left(\ell+h-1\right)^2\Gamma\left(\ell+2h-3\right)}{\ell!\Gamma\left(h-1\right)^2\Gamma\left(2h+2\ell-3\right)}\,.
\ee

\begin{table}
\begin{center}
\renewcommand{\arraystretch}{1.5}
\begin{tabular}{|c|c|c|c|c|c|}\hline
  & $\text{$\ell $=2}$ & $\text{$\ell $=4}$ & $\text{$\ell $=6}$ & $\text{$\ell $=8}$ & $\text{$\ell $=10}$ \\ \hline 
 $N\text{=2}$ & $\frac{11}{324} $& $\frac{2039}{64800}$ &$ \frac{30893}{873180}$ & $\frac{1666423}{42456960}$ & $\frac{1133429981}{26592073200}$ \\ \hline
$N \text{=3}$ & $\frac{55}{1458}$ & $\frac{2039}{58320}$ & $\frac{30893}{785862}$ & $\frac{1666423}{38211264}$ & $\frac{1133429981}{23932865880}$ \\[0.2cm] \hline
 $N\text{=10}$ & $\frac{11}{300}$ & $\frac{2039}{60000}$ & $\frac{30893}{808500}$ & $\frac{1666423}{39312000}$ & $\frac{1133429981}{24622290000}$ \\ \hline
\end{tabular}\caption{$c_{S,\ell}^{(2)} $ as given in \eqref{cubic-ope}}\label{cS2}
\end{center}
\end{table}
\begin{table}
\begin{center}
\renewcommand{\arraystretch}{1.5}
\begin{tabular}{|c|c|c|c|c|c|} \hline
   & $ \text{$\ell $=2}$ & $ \text{$\ell $=4} $ & $ \text{$\ell $=6} $ & $ \text{$\ell $=8} $ & $ \text{$\ell $=10}$ \\ \hline
$  N\text{=2} $ & $ \frac{955}{17496} $ & $ \frac{33071}{699840} $ & $ \frac{3665024719}{72613648800} $ & $ \frac{45230019647}{834534005760} $ & $ \frac{16447155548067179}{285712467504710400}$ \\ \hline
 $ N\text{=3} $ & $ \frac{40379}{708588} $ & $ \frac{6959771}{141717600} $ & $ \frac{30739672087}{588170555280} $ & $ \frac{1892198482723}{33798627233280} $ & $ \frac{686900153698555567}{11571354933940771200}$ \\ \hline
 $ N\text{=10} $ & $ \frac{2291}{45000} $ & $ \frac{130753}{3000000} $ & $ \frac{1723185619}{37352700000} $ & $ \frac{317067803821}{6439305600000} $ & $ \frac{38269562772181723}{734857169508000000}$  \\ \hline
\end{tabular}\caption{$c_{S,\ell}^{(3)}$ as given in \eqref{cubic-ope}}\label{cS3}
\end{center}
\end{table}
\begin{table}
\begin{center}
\renewcommand{\arraystretch}{1.5}
\begin{tabular}{|c|c|c|c|c|c|} \hline
  & $\text{$\ell $=2}$ &$ \text{$\ell $=4} $&$ \text{$\ell $=6}$ &$ \text{$\ell $=8} $& $\text{$\ell $=10}$ \\ \hline
$N \text{=2} $&$ \frac{11}{324} $  &$ \frac{2039}{64800} $  &$ \frac{30893}{873180} $ &$ \frac{1666423}{42456960} $&$ \frac{1133429981}{26592073200}$ \\ \hline
$ N\text{=3} $&$ \frac{187}{5832} $&$ \frac{136421}{4082400} $&$ \frac{57793}{1496880}$ &$ \frac{36326611}{840647808} $&$ \frac{293100163357}{6222545128800} $\\ \hline
$N \text{=10} $&$ \frac{187}{8100} $&$ \frac{114437}{3780000} $&$ \frac{170857}{4677750} $&$ \frac{483771131}{11675664000} $&$ \frac{130842828107}{2880807930000}$ \\ \hline
 \end{tabular}\caption{$c_{T,\ell}^{(2)}$ as in \eqref{cubic-ope}}\label{cT2}
\end{center}
\end{table}
\begin{table}
\begin{center}
\renewcommand{\arraystretch}{1.5}
\begin{tabular}{|c|c|c|c|c|c|} \hline
  &$ \text{$\ell $=2} $&$ \text{$\ell $=4} $&$ \text{$\ell $=6} $&$ \text{$\ell $=8} $&$ \text{$\ell $=10}$ \\ \hline$
N \text{=2} $&$ \frac{955}{17496} $&$ \frac{33071}{699840} $&$ \frac{3665024719}{72613648800} $&$ \frac{45230019647}{834534005760} $&$ \frac{16447155548067179}{285712467504710400} $\\ \hline$
 N\text{=3} $&$ \frac{64489}{1417176} $&$ \frac{1263227303}{27776649600}$ &$ \frac{11300769391}{224064973440} $&$ \frac{320561250041821}{5842334136038400} $&$ \frac{2292070375098731839349}{39111179676719806656000} $\\ \hline$
N \text{=10} $&$ \frac{29837}{1215000} $&$ \frac{420441947}{11907000000} $&$ \frac{27216198529}{648336150000} $&$ \frac{140366883740173}{3005315913600000} $&$ \frac{506565599576907257701}{10059459793395012000000}$ \\ \hline
\end{tabular}\caption{$c_{T,\ell}^{(3)}$ as in \eqref{cubic-ope}}\label{cT3}
\end{center}
\end{table}
\begin{table} 
\begin{center}
\renewcommand{\arraystretch}{1.5}
\begin{tabular}{|c|c|c|c|c|c|}
\hline
  & $\text{$\ell $=2} $&$ \text{$\ell $=4} $&$ \text{$\ell $=6} $&$ \text{$\ell $=8} $&$ \text{$\ell $=10} $\\ \hline$
 N\text{=2} $&$ \frac{11}{648} $&$ \frac{1217}{45360} $&$ \frac{49807}{1496880} $&$ \frac{1481069}{38918880} $&$ \frac{526157011}{12570798240} $\\ \hline
$ N\text{=3} $&$ \frac{77}{2916} $&$ \frac{4066}{127575}$ &$ \frac{54163}{1428840} $&$ \frac{8997979}{210161952} $&$ \frac{145754265827}{3111272564400} $\\\hline$
 N\text{=10} $&$ \frac{539}{16200} $&$ \frac{15619}{472500} $&$ \frac{9898997}{261954000} $&$ \frac{246069253}{5837832000} $&$ \frac{24030761429}{523783260000}$ \\ \hline
\end{tabular}\caption{$c_{V,\ell}^{(2)}$ as in \eqref{cubic-ope}}\label{cV2}
\end{center}
\end{table}
\begin{table}
\begin{center}
\renewcommand{\arraystretch}{1.5}
\begin{tabular}{|c|c|c|c|c|c|}\hline
  & $\text{$\ell $=2}$ &$ \text{$\ell $=4} $&$ \text{$\ell $=6} $&$ \text{$\ell $=8} $&$ \text{$\ell $=10} $\\ \hline$
 N\text{=2} $&$ \frac{223}{8748} $&$ \frac{13146719}{342921600} $&$ \frac{5729638637}{124480540800} $&$ \frac{866670614587}{16829769116160} $&$ \frac{97858846018004947}{1755832982119856640} $\\ \hline$
 N\text{=3} $&$ \frac{29213}{708588} $&$ \frac{121937213}{2777664960} $&$ \frac{53003131961}{1069401009600} $&$ \frac{17350445830481}{319502648064600} $&$ \frac{1138230092181377071783}{19555589838359903328000} $\\ \hline
$ N\text{=10} $&$ \frac{7624}{151875} $&$ \frac{1024214479}{23814000000} $&$ \frac{1660750757513}{36306824400000}$ &$ \frac{257502902121943}{5259302848800000}$ &$ \frac{94869301081115147749}{1828992689708184000000} $\\ \hline
\end{tabular}\caption{$c_{V,\ell}^{(3)}$ as in \eqref{cubic-ope}}\label{cV3}
\end{center}
\end{table}
\begin{table} 
\begin{center}
\renewcommand{\arraystretch}{1.5}
\begin{tabular}{|c|c|c|c|c|c|} \hline
  &$ \text{$\ell $=1} $&$ \text{$\ell $=3} $&$ \text{$\ell $=5} $&$ \text{$\ell $=7} $&$ \text{$\ell $=9} $\\  \hline$
 N\text{=2} $&$ \frac{1}{108}$ &$ \frac{73}{3240} $&$ \frac{4127}{136080} $&$ \frac{696991}{19459440} $&$ \frac{26499493}{661620960}$ \\  \hline$
 N\text{=3} $&$ \frac{7}{243} $&$ \frac{101}{3645} $&$ \frac{212491}{6123600} $&$ \frac{24712403}{612972360}$ &$ \frac{444787193}{9924314400} $\\  \hline
$ N\text{=10} $&$ \frac{7}{300} $&$ \frac{709}{27000} $&$ \frac{189809}{5670000} $&$ \frac{14778019}{378378000}$ &$ \frac{3596205413}{82702620000} $\\  \hline
\end{tabular}\caption{$c_{A,\ell}^{(2)}$ as in \eqref{cubic-ope}}\label{cA2}
\end{center}
\end{table}
\begin{table} 
\begin{center}
\renewcommand{\arraystretch}{1.5}
\begin{tabular}{|c|c|c|c|c|c|}
 \hline
  & $\text{$\ell $=1} $&$ \text{$\ell $=3} $&$ \text{$\ell $=5} $&$ \text{$\ell $=7} $&$ \text{$\ell $=9}$ \\ \hline$
 N\text{=2} $&$ \frac{163}{11664} $&$ \frac{28807}{874800} $&$ \frac{43789397}{1028764800}$ &$ \frac{1029950903453}{21037211395200}$ &$ \frac{261349538475643}{4863803274570240}$ \\ \hline
 $N=\text{=3}$ & $\frac{43633}{944784} $& $\frac{21601949}{566870400}$ &$ \frac{18974539573}{416649744000} $& $\frac{4899887619317521}{95424790888627200}$ & $\frac{550521620277178021}{9849201631004736000}$ \\ \hline
$N \text{=10}$ &$ \frac{1759}{90000} $&$ \frac{1483133}{48600000}$ &$ \frac{925929479}{23814000000} $&$ \frac{121213773268411}{2727045921600000} $&$ \frac{738499703203970219}{15199385233032000000}$ \\ \hline 
\end{tabular}\caption{$c_{A,\ell}^{(3)}$ as in \eqref{cubic-ope}}\label{cA3}
\end{center}
\end{table}

The quantities $c_{i,\ell}^{(2)}$ and $c_{i,\ell}^{(3)}$ have been obtained in a closed form for general $N$ and $\ell$. However the expression is too big to present here. So we give their values for specific $N$-s and $\ell$-s. The general form can be made available on request.

To the best of our knowledge, almost none of these results have been computed before. So apart from $\D_\phi$ and $\D_{S,0}$ all results presented in this section are new. The new ones also pass some consistency checks like giving free theory results at $N=1$ and Ising model results for $N=2$ \cite{phi4}.

\section{Discussion}
We have analyzed the Mellin space analytic bootstrap techniques to conformal field theories with $O(N)$ symmetry. Consistency with the OPE imposes non trivial constraints on the dimensions and the OPE coefficients of the operators appearing in the singlet, symmetric traceless and antisymmetric representations of $O(N)$. By considering the leading spurious pole $s=\D_\phi$, we looked at the $\epsilon$-expansion and the large-$N$ expansion and demonstrated that the consistency conditions lead to known results as well as new results for OPE coefficients. We also studied the case with cubic anisotropy and obtained new results. We list below some future directions.
	
	\begin{itemize}
\item It will be interesting to compare the new $O(\e^3)$ results we have for the OPE coefficients with what arises from numerical boostrap. In \cite{usprl,uslong} we compared the Ising case with the spin-4 OPE result in \cite{zohar} and found impressive agreement. 
	\item 
It will be desirable to develop and algorithm to compute systematically subleading corrections. We have not used all the equations. There are spurious poles of the form $s=\D_\phi+n$ and we just considered $n=0$. With a judicious choice, it should be possible to extract a lot more information from these equations. By exploiting these equations it should be possible \cite{RGAS} to extract more information about subleading terms as well as about other higher order operators in the spectrum for which some information is known \cite{kehrein}. It should also be possible to consider CFTs in higher dimensions as in \cite{klebanov}.

\item 
It will be important to develop numerical algorithms to solve these equations. As pointed out in \cite{uslong}, it may be easier to consider expanding these equations around some $t=t_0$ rather than in terms of the continuous Hahn polynomials. We go from one set to the other by taking an infinite linear combination. Hence, it is not apriori guaranteed that convergence (as a sum over the spectrum) in one case will lead to convergence in the other. It appears to us that expanding around a special point in $t$ may be more suited for numerics. While the issue about convergence as a sum over the spectrum is solved in the conventional approach to numerics \cite{rychkovaa}, this question still needs to be resolved in our approach.

\item
It will be interesting to understand whether this approach can be extended to logarithmic conformal field theories \cite{Cardy:2013rqg}. For $N=0, -2, -4$, a logarithmic behavior arises in the correlation function. It is desirable to extend our analysis to physical systems exhibiting logarithmic behavior in appropriate limits.
\item 
To make contact with AdS/CFT it will be interesting to understand the large-N systematics in more detail. Our progress in this paper was quite modest but that was because we concentrated on the leading spurious pole. However, our methods should be useful for future studies along similar lines and also for extending them to supersymmetric theories \cite{susyboot}.
	
\end{itemize}
\section*{Acknowledgments}
We especially thank Rajesh Gopakumar for numerous discussions and comments on the manuscript. We acknowledge useful discussions  with  F. Alday, J. Cardy,  S. Giombi, T. Hartman, K. Jensen,  J. Kaplan, I. Klebanov, P Kraus, G. Mandal, S. Minwalla, H. Osborn, J. Penedones, D. Poland, S. Pufu,  L. Rastelli, S. Rychkov, K. Sen, M. Serone, S. Wadia and W. Witczak-Krempa.  PD thanks Johns Hopkins University, SFSU, UC Berkeley and UCLA for hospitality during the course of this work. AK thanks Princeton University, Cornell University, Johns Hopkins University, YITP Stony Brook, Yale University, EPFL Lausanne, SISSA and CERN for hospitality during  this work. A.S. acknowledges support from a DST SwarnaJayanti Fellowship Award DST/SJF/PSA-01/2013-14. 

\appendix

\section{Essential formulas}\label{formulas}
\subsection{The normalization}
When we expand the correlator $\langle \phi\phi\phi\phi \rangle$ in terms of Witten diagrams, we write the constant coefficients as $c_{\D,\ell}$. These constants are related to the OPE coefficients $C_{\D,\ell}$ through a normalization $\mathfrak{N}_{\D,\ell}$ which is given by,
\begin{align}\label{norm}
&c_{\D,\ell}= C_{\D,\ell}\mathfrak{N}_{\D,\ell} =C_{\D,\ell}\frac{(-2)^{\ell }  (\ell +\Delta-1 ) \Gamma (1-h+\Delta ) \Gamma ^2(\ell +\Delta-1 )}{ \Gamma (\Delta -1) \Gamma ^4\left(\frac{\ell +\Delta }{2}\right) } \Gamma^{-1} \left(\frac{\ell -\Delta +\Delta _1+\Delta _2}{2}\right) \nonumber\\
& \Gamma^{-1}  \left(\frac{\Delta +\Delta _1+\Delta _2-2 h+\ell }{2}\right) \Gamma^{-1}  \left(\frac{\ell -\Delta +\Delta _3+\Delta _4}{2}\right)\Gamma^{-1}  \left(\frac{\Delta +\Delta _3+\Delta _4-2 h+\ell }{2}\right)\,.
\end{align}
As explained in \cite{uslong}  this is obtained by computing the leading power law $u^{(\D-\ell)/2}(1-v)^\ell$ from the Witten diagram and comparing with the conformal blocks.
\subsection{Mack Polynomials}\label{mack}
The Mack polynomials $P^{(s)}_{\nu,\ell}(s,t)$, for identical external scalars, are given by \cite{mack, joao, dolanosborn2}
\begin{align}\label{sumt}
\begin{split}
& P^{(s)}_{\nu,\ell}(s,t)=\widetilde{\sum}\frac{\G^2(\l_1)\G^2(\bar{\l}_1)(\l_2-s)_k(\bar{\l}_2-s)_k(s+t)_\b(s+t)_\a(-t)_{m-\a}(-t)_{\ell-2k-m-\b}}{\prod_i\G(l_i)}\,,\\
& \text{where \ \  }\widetilde{\sum}\equiv \frac{\ell!}{2^{\ell}(h-1)_\ell}\sum_{k=0}^{[\frac{\ell}{2}]}\sum_{m=0}^{\ell-2k}\sum_{\a=0}^m\sum_{\b=0}^{\ell-2k-m}\frac{(-1)^{\ell-k-\a-\b}\G(\ell-k+h-1)}{\G(h-1)k!(\ell-2k)!}\binom{\ell-2k}{m}\binom{m}{\a}\\
& \ \ \ \ \ \ \ \times\binom{\ell-2k-m}{\b}\,.
\end{split}
\end{align}
The other notations are given by,
\be
 \l_1=\frac{h+\nu+\ell}{2} \,, \ \ \bar{\l}_1=\frac{h-\nu+\ell}{2}\,, \ \ \l_2=\frac{h+\nu-\ell}{2}\text{ \ \  and \ \  }\bar{\l}_2=\frac{h-\nu-\ell}{2}\,,
\ee
\be\label{li}
l_1=\l_2+\ell-k-m+\a-\b\,,\ l_2=\l_2+k+m-\a+\b\,, \ l_3=\bar{\l}_2+k+m\,,\ l_4=\bar{\l}_2+\ell-k-m\,.
\ee
\subsection{Continuous Hahn Polynomials}\label{A3}
We briefly summarize the key properties of the continuous Hahn polynomials. More details can be found in \cite{uslong}. It is given by,
\be\label{Qdefn}
Q^{2s+\ell}_{\ell,0}(t)=\frac{2^\ell (s)^2_\ell}{(2s+\ell-1)_\ell}\ {}_3F_2\bigg[\begin{matrix} -\ell,2s+\ell-1,s+t\\
\ \ s \ \ , \ \ \ \ \ \  s
\end{matrix};1\bigg]\,.
\ee
These polynomials have the orthogonality property \cite{AAR},
\be
\frac{1}{2\pi i}\int_{-i\infty}^{i\infty} dt \ \G(s+t)^2\G(-t)^2 Q^{2s+\ell}_{\ell,0}(t) Q^{2s+\ell '}_{\ell',0}(t)=(-1)^\ell\kappa_{\ell}(s)\d_{\ell,\ell'}\,,
\ee
where,
\be\label{kappadef}
\kappa_{\ell}(s)=\frac{4^\ell \ell!}{(2s+\ell-1)_\ell^2}\frac{\G^4(\ell+s)}{(2s+2\ell-1)\G(2s+\ell-1)}\,.
\ee
Further we have the identity,
\be\label{Qid}
Q^{2s+\ell}_{\ell,0}(t)=(-1)^\ell Q^{2s+\ell}_{\ell,0}(-s-t)\,.
\ee

Now one can use this identity \eqref{Qid} on the $t$-channel expression \eqref{ts} and $u$-channel expression \eqref{tu}, with which the two expressions become equal under the exchange $t \leftrightarrow -s-t$. Hence we get the equality \eqref{ueqtot}.
\subsection{$t$-channel integral}
The most general form of $q_{\D,\ell |\ell'}^{(t)}(s)$ in the $t$-channel for an exchange of  operator with spin $\ell'$, is given by,
\begin{align}\label{qtgen}
\begin{split}
\!\!\!\!\!\! \hspace{-1cm}&q_{\D,\ell |\ell'}^{(t)}(s)= \k_\ell(s)^{-1}\sum_{q=0}^\ell\widetilde{\sum}\int d\n  c_{\D,\ell'} \m^{(t)}_{\D,\ell'}(\n)\G^2(\l_1)\G^2(\bar{\l}_1)(\D_\phi-s)_{m-\a}(\D_\phi-s)_{\ell'-2k-m-\b}\frac{1}{\prod_i\G(l_i)}\frac{2^\ell((s)_\ell)^2}{(2s+\ell-1)_\ell}\\
&\times \frac{(-\ell)_q(2s+\ell-1)_q}{((s)_q)^2\ q!} \frac{\G(k+q+s+\a+\l_2-\D_\phi)\G(k+q+s+\a+\bar{\l}_2-\D_\phi)}{\G(q+2s+2k+\a+\b+\l_2+\bar{\l}_2-2\D_\phi)}\G(k+s+\b-\D_\phi+\l_2) \\&\times \G(k+s+\b-\D_\phi+\bar{\l}_2) {}_3F_2\bigg[\begin{matrix}-q,-\a,1-2k-q-2s-\a-\b+2\D_\phi-\l_2-\bar{\l}_2\\
1-k-q-s-\a-\l_2+\D_\phi,1-k-q-s-\a-\bar{\l}_2+\D_\phi\end{matrix};1\bigg]\,.
\end{split}
\end{align}
This general form is derived in \cite{uslong}.

\section{Obtaining the $c_T$ from symmetry}\label{hugh}
The central charge $c_T$ which is given by $c_T=\frac{d^2 \Delta_\phi ^2}{(d-1)^2 C_{2h,2}}$ can be obtained using symmetries of the problem and known large $N$ results--the argument for $c_T$ in this section is due to Hugh Osborn. $C_{2h,2}$ is related to the square of the three point function $\langle \phi_i \phi_j T\ra$. Here $T$ is the stress tensor singlet operator schematically given by $\phi_k \partial_\mu \partial_\nu \phi_k$. To obtain the three point function let us look at the general function $\langle\phi_i(x) \phi_j(y) (\phi_k\partial_\mu\partial_\nu\phi_r(z))\ra$. For the stress tensor we will have to contract this with $\delta_{kr}$ first. Then we will contract the whole 3-point function with itself. 

Now assume a generalised interacting term given by $\frac{1}{24}\lambda_{ijkl}\phi_i\phi_j\phi_k\phi_l$. The Feynman diagrams relevant  to $\langle\phi_i \phi_j (\phi_k\partial_\mu\partial_\nu\phi_r)\ra$ at the 2-loop and 3-loop orders are shown below in Figure \ref{feynmandiag}. Other diagrams go to 0 upon the action of derivatives in $\phi_k\partial_\mu\partial_\nu\phi_r$.
\begin{figure}[!htpb]
  \centering
        \includegraphics[width=\textwidth]{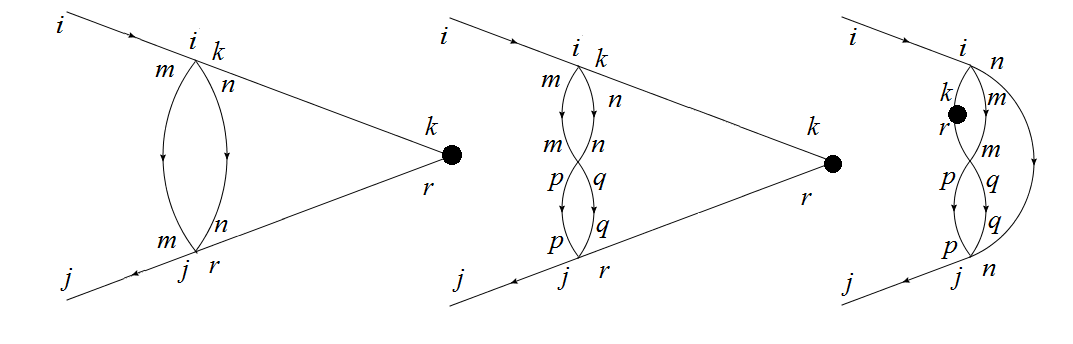}
    \caption{Feynman diagrams with $\lambda_{ijkl}$ parameter. The first diagram on left is a two loop diagram. The second and third on the right are three loop diagrams. The colour indices are indicated at each vertex. The blobs denote a composite operator ($T$ or $J$) insertion.}
    \label{feynmandiag}
\end{figure}

Let us rescale the interaction $\lambda_{ijkl} \to 16\pi^2\la_{ijkl}$. The general term from these two processes can be written as 
\be\label{struc1}
\langle\phi_i \phi_j (\phi_k\partial_\mu\partial_\nu\phi_r)\ra= O(1)+a \la_{ikmn}\la_{mnjr}+b \la_{ikmn}\la_{mnpq}\la_{pqjr}+c\la_{inkm}\la_{rmpq}\la_{pqjn}\,.
\ee
The $O(\la^0)$ term can be anything of the form $x\d_{ij}\d_{kl}+y\d_{ik}\d_{jl}+z\d_{il}\d_{jk}$. Now the contracting the above with $\d_{kr}$, we get the form,
\be
\langle\phi_i \phi_j (\phi_k\partial_\mu\partial_\nu\phi_k)\ra= \g_1\d_{ij}+\a_1 \la_{ikmn}\la_{jkmn}+\b_1 \la_{ikmn}\la_{mnpq}\la_{pqkj}+\b_2\la_{inkm}\la_{kmpq}\la_{pqjn}\,.
\ee
This contracted with itself should give us the OPE coefficient $C_{2h,2}$. Hence we can correctly guess the form,
\be
c_T/c_{T,\text{scalar}}=N+\a \la_{ijkl}\la_{ijkl}+\b \la_{ijkl}\la_{klmn}\la_{mnij}\,.
\ee
Here $c_{T,\text{scalar}}$ is the central charge for $N=1$ theory. The $O(\la^0)$ is just $N$ which follows from free theory. Now for the $O(N)$ case we have, 
\be
\la_{ijkl}=\la(\d_{ij}\d_{kl}+\d_{ik}\d_{jl}+\d_{il}\d_{jk})\,.
\ee
Also at the fixed point in $d=4-\e$ we have,
\be
\lambda =\frac{\epsilon }{N+8}+3\frac{(3N+14)}{(N+8)^3}\epsilon ^2\,.
\ee
This gives,
\be
c_T/c_{T,free}=1+\frac{3 \a \epsilon ^2+\b \epsilon ^3}{N}\,.
\ee
Now the large $N$ expansion of $c_T/c_{T,free}$ can be found in \cite{klebanov2,petkou}, and it is given by,
\be
c_T/c_{T,free}=1-\frac{5 \epsilon ^2}{12}-\frac{7 \epsilon ^3}{36}\,.
\ee
Using this we get, $\a=-5/36$ and $\b=-7/36$. So we obtain,
\be
{c_T/c_T}_{free} = 1-\frac{5\,(N+2)}{12\,(N+8)^2}\,\e^2-\,\frac{(N+2)(7N^2+382N+1708)}{36\,(N+8)^4}\,\e^3 +\,O(\e^4)\,,
\ee
which exactly matches with our result \eqref{cT}.

Even though $c_T$ was obtained this way, it is not possible to do the same for $c_J$--in terms of the OPE coefficients $c_J/c_{J_{free}}=C_{A,1}^{free}/C_{A,1}$. This is because although $c_J$ is known up to the $1/N$ order, it has a more complicated structure in terms of the perturbative parameter $\la_{ijkl}$. So instead of \eqref{struc1} we have the form,
\be
\langle\phi_i \phi_jJ_{[kr]}\ra= a_0(\d_{ik}\d_{jl}-\d_{il}\d_{jk})+a \la_{ikmn}\la_{mnjr}+b \la_{ikmn}\la_{mnpq}\la_{pqjr}+c\la_{inkm}\la_{rmpq}\la_{pqjn}\,.
\ee
Here $J_{[kr]}$ is the spin 1 antisymmetric current. The first term in the rhs above comes from antisymmetrization. We get the OPE coefficient of $J$ by contracting $\langle\phi_i \phi_jJ_{[kr]}\ra$ with itself. This gives the form,
\be
c_J/c_{J,free}=1+\a \la_{ijkl}\la_{ijkl}+\b \la_{ijkl}\la_{klmn}\la_{mnij}+\g\la_{iikl}\la_{kmnp}\la_{lmnp}\,.
\ee
Since there are three undetermined coefficients $\a,\b,\g$ to fix, we would not be able to do it from the $\e^2$ and $\e^3$ terms of $1/N$ expansion alone.

\section{Obtaining the large $N$ corrections}\label{largeNexact}

The correction to the 4-point function $\langle\phi_i\phi_j\phi_k\phi_l\rangle$ has been computed exactly at the $1/N$ order \cite{Lang:1991kp}. It is can be written in a compact way as \cite{alday, Alday2},
\begin{align}
\langle\phi_i\phi_j\phi_j\phi_l\rangle&=\frac{\d_{ij}\d_{kl}}{(x_{12}x_{34})^{\D_\phi}}+\frac{\d_{il}\d_{jk}}{(x_{14}x_{23})^{\D_\phi}}+\frac{\d_{ik}\d_{jl}}{(x_{13}x_{24})^{\D_\phi}}  +\frac{\d_\phi^{(1)}h\G(h)}{(h-2)\G^2(h-1)N}\frac{f_{ijkl}}{(x_{13}^2 x_{24}^2)^{\D_\phi}}
\end{align}
where,
\be
f_{ijkl}=\d_{ij}\d_{kl}\bar{D}_{1,1,\mu-1,\mu-1}(u,v)+\d_{il}\d_{jk}\bar{D}_{\mu-1,1,1,\mu-1}(u,v)+\d_{ik}\d_{jl}\bar{D}_{\mu-1,1,\mu-1,1}(u,v)\,.
\ee 
The $\bar{D}$ functions are defined in \cite{dolanosborn2}. We can rearrange the 4-point function into singlet, symmetric traceless and antisymmetric parts. With the overall factor of $(x_{12}^2 x_{34}^2)^{-\D_\phi}$, the $1/N$ correction coefficients of the latter two sectors are given by (upper sign for symmetric traceless and lower for antisymmetric),
\begin{align}\label{exactN}
&\sum _{n,m}u^{h-1}\frac{h \delta _{\phi }^{(1)}}{h-2}\frac{ u^n(1-v)^m}{2n!m!}\left(\pm\frac{(h-1)_n(h-1)_{n+m}(n+m)!}{(h)_{2n+m}}(-\log  u+\psi (n+1)-\psi (n+m+1)+2\psi (h+2n+m)\right.\nonumber\\&
\left.-\psi (h-1+n)-\psi (h-1+n+m))+\frac{(h-1)_{n+m}{}^2n!}{(h)_{2n+m}}(-\log  u+2\psi (h+2n+m)-2\psi (h-1+n+m))\right) \nonumber\\&
 \pm \frac{\delta_{\phi }^{(1)}}{2}u^{h-1} \log  u+\sum _m\left[u^{h-1}(1-v)^m\delta _{\phi }{}^{(1)}\frac{(-1)^m}{2}\bigg[
\binom{1-h}{m}
\log u+\sum _{q=1}^m\left(
\binom{1-h}{q-1}\frac{(-1)^{m-q+1}}{m-q+1}\right)\bigg]\right] 
\end{align}
The last line comes from the disconnected piece. The singlet sector coefficient is simply,
\be\label{exactNsing}
\frac{\d_\phi^{(1)}h\G(h)}{(h-2)\G^2(h-1)}u^{h-1}\bar{D}_{1,1,\mu-1,\mu-1}\,.
\ee
To read off the anomalous dimension and OPE coefficient corrections, we have to identify the above with the conformal blocks. The conformal block in the small $u$ limit reads,
\be
g_{\D,\ell}(u,v)=u^{(\D-\ell)/2}(1-v)^\ell {}_2F_1\big(\frac{\D-\ell}{2},\frac{\D-\ell}{2},\D-\ell,1-v\big)\,.
\ee
Consider the corrections $C_{\D,\ell}=C_\ell^{(0)}+\frac{C_\ell^{(1)}}{N}$ and $\D=\D^{(0)}+\frac{\d_\ell^{(1)}}{N}$. We have $\D^{(0)}=2$ for the singlet scalar and $\D_0=2h-2+\ell$ for all other operators. Thus we get for general $\ell$,
\be
C_{\D,\ell}g_{\D,\ell}(u,v)=C^{(0)}g_{2h-2+\ell,\ell}(u,v)+\frac{u^{h-1}(1-v)^\ell}{N}\big(\frac{\d_\ell}{2}\log u + C_\ell^{(1)}\big)+O(u,1-v)\,.
\ee
So the coefficient of $u^{h-1}(1-v)^\ell \log u$ gives the anomalous dimension $\d_\ell$ directly. The coefficient of the nonlog term $u^{h-1}(1-v)^\ell$ is associated with the OPE coefficient correction. However all the conformal blocks $C_{\D,\tilde{\ell}}g_{\D,\tilde{\ell}}$ with $\tilde{\ell} < \ell$ give a contribution to the nonlog term, which is of the form,
\begin{align}
\sum_{\tilde{\ell} \le \ell}& \frac{\text{  }\Gamma (2 h-2) \Gamma ^2\big(h+\ell -\tilde{\ell }-1\big) }{(\ell -\tilde{\ell })! \Gamma ^2(h-1) \Gamma (2 h+\ell -\tilde{\ell }-2)}\times \nonumber\\& \left(C^{(1)}_{\tilde{\ell }}+C^{(0)}_{\tilde{\ell }} \delta _{\tilde{\ell }} \big(\psi (2 h-2)+\psi \big(h+\ell -\tilde{\ell }-1\big)-\psi \big(2 h+\ell -\tilde{\ell }-2\big)-\psi (h-1)\big)\right)\,.
\end{align}
The above is then compared to the $u^{h-1}(1-v)^\ell\log u$ and $u^{h-1}(1-v)^\ell$ terms from \eqref{exactN} and \eqref{exactNsing} to read off the anomalous dimensions and OPE coefficients which match exactly with our results \eqref{largen1}-\eqref{largen4}.

\section{Higher spin OPE in $\e$-expansion}\label{e3}
The OPE coefficients of the higher spin operators in $d=4-\e$ can be written as 
\be\label{clcfr}
\frac{C_{\ell}}{C_{\ell}^{free}}=1 \ + \  c^{(2)}_{i,\ell}\e^3 \ + \ c^{(3)}_{i,\ell}\e^3 \ + \ O(\e^4)\,.
\ee
Here as usual $i$ indicates the singlet, traceless symmetric and antisymmetric sectors. In the above formula $c^{(2)}_{S,\ell}$, $c^{(2)}_{T,\ell}$ and $c^{(2)}_{A,\ell}$ were given in \eqref{opesing}, \eqref{opetraceless} and \eqref{opeanti}. Here we give the $O(\e^3)$ orders for the first few spins.
\subsection{Singlet sector}
\begin{align}
c^{(3)}_{S,\ell=4}&=\frac{(2+N) (405848+N (89228+989 N)) }{4800 (8+N)^4}\nonumber\\
c^{(3)}_{S,\ell=6}&=\frac{(2+N) (27035046944+N (5902407776+47767751 N)) }{298821600 (8+N)^4}\nonumber\\
c^{(3)}_{S,\ell=8}&=\frac{(2+N) (1002110534752+N (217772423200+1327572517 N))}{10302888960 (8+N)^4}\nonumber\\
c^{(3)}_{S,\ell=10}&=\frac{(2+N) (121568105958318592+N (26332733153306704+123560820979315 N))}{1175771471212800 (8+N)^4}\,.
\end{align}
\subsection{Traceless symmetric}
\begin{align}
c^{(3)}_{T,\ell=2}&=\frac{42096+N (22504+N (2878+13 N))}{216 (8+N)^4}\nonumber\\
c^{(3)}_{T,\ell=4}&=\frac{59659656+N (38917600+N (5670765+34133 N))}{352800 (8+N)^4}\nonumber\\
c^{(3)}_{T,\ell=6}&=\frac{2919785069952+N (2004861573920+N (299681375684+1572151439 N))}{16136366400 (8+N)^4}\nonumber\\
c^{(3)}_{T,\ell=8}&=\frac{3637661241149760+N (2543976284881184+N (382765405683350+1644002316149 N))}{18699743462400 (8+N)^4}\nonumber\\
c^{(3)}_{T,\ell=10}&=(1490290339762224000 (8+N)^4)^{-1}(308175148604337630720+N (217302237938493487024\nonumber\\
& \ \ +N (32745609281101869405+113160657172438904 N)))\,.
\end{align}
\subsection{Antisymmetric}
\begin{align}
c^{(3)}_{A,\ell=3}&=\frac{(2+N) (229376+N (48672+295 N))}{3456 (8+N)^4}\nonumber\\
c^{(3)}_{A,\ell=5}&=\frac{(2+N) (77087104+N (16575112+96553 N))}{972000 (8+N)^4}\nonumber\\
c^{(3)}_{A,\ell=7}&=\frac{(2+N) (144671572069952+N (31180232920640+150688639187 N))}{1616027212800 (8+N)^4}\nonumber\\
c^{(3)}_{A,\ell=9}&=\frac{(2+N) (5447552685503360+N (1173953131219392+4579923671359 N))}{55828779552000 (8+N)^4}\,.
\end{align}

The singlet $O(\e^3)$ OPE coefficients are found to obey the following general $\ell$ formula,
\begin{align}
c^{(3)}_{S,\ell}&=\frac{(2+N)}{8 (8+N)^4 \ell ^3 (1+\ell )^3} \Big[-2 N^2 \left(14+\ell ^2 (-25+\ell  (3+\ell  (3+\ell )))\right)+16 N (-28\nonumber\\&
+\ell  (-27+\ell  (32+7 \ell  (3+\ell  (3+\ell )))))+32 (-56+\ell  (-63+\ell  (58+17 \ell  (3+\ell  (3+\ell )))))\nonumber\\&
-\ell  (1+\ell ) \big(16 (8+N)^2 \ell  (1+\ell ) H_{-1+\ell }^2+\left(-224 (8+9 \ell )+272 \ell ^2 (3+\ell  (2+\ell ))-N^2 \left(28+\ell ^2 (-39\right.\right.\nonumber\\&
+\ell  (2+\ell )))+8 N (-56+(-1+\ell ) \ell  (54+7 \ell  (3+\ell )))) H_{2 \ell }+2 H_{-1+\ell } \left(N^2 \left(22+\ell ^2 (-19+\ell  (2+\ell ))\right)\right.\nonumber\\&
-16 (-88+\ell  (-63+17 \ell  (2+\ell  (2+\ell ))))-8 N (-44+\ell  (-27+\ell  (20+7 \ell  (2+\ell ))))\left.-8 (8+N)^2 \ell  (1+\ell ) H_{2 \ell }\right)\nonumber\\&
+2 (8+N)^2 \ell  (1+\ell ) \big((8-3 \ell  (1+\ell )) H^{(2)}_{\ell }+2 \left(-6+\ell +\ell ^2\right)H^{(2)}_{2 \ell }\big)\big)\Big]\,.
\end{align}

The traceless symmetric $O(\e^3)$ OPE coeficients are found to obey the following formula,
\begin{align}
c^{(3)}_{T,\ell}&=\frac{1}{8 (8+N)^4 \ell ^2 (1+\ell )^3}\left(-4 (8+N)^2 \ell  (1+\ell )^2 (-2 (6+N)+(2+N) \ell  (1+\ell ))H^{(2)}_{2 \ell }\right.\nonumber\\&
-2 (1+\ell ) \left(2 (6+N) (8+N)^2+8 \left(-4+N^2\right) \ell -(1088+N (640+3 N (34+N))) \ell ^2+2 (2+N)\right.\nonumber\\&
\left. (-272+(-56+N) N) \ell ^3+(2+N) (-272+(-56+N) N) \ell ^4-32 (4+N) (8+N) \ell  (1+\ell )\text{  }H_{2 \ell }\right) H_{\ell }\nonumber\\&
+(1+\ell ) \left(4 (6+N) (8+N)^2+16 (124+N (48+5 N)) \ell -(4+N) (408+N (122+7 N)) \ell ^2+\right.\nonumber\\&
\left.2 (2+N) (-272+(-56+N) N) \ell ^3+(2+N) (-272+(-56+N) N) \ell ^4\right) H_{2 \ell }+2 (8+N) \ell  (1+\ell )^2\nonumber\\&
\left. \left(-4 (32+N (10+N))+3 (2+N) (8+N) \ell +3 (2+N) (8+N) \ell ^2\right) H^{(2)}_{\ell }\right)\,.
\end{align}

The antisymmetric part is given by,
\begin{align}
c^{(3)}_{A,\ell}&=-\frac{1}{8 (8+N)^4 \ell ^2 (1+\ell )^3}(2+N) \left(-(1+\ell ) \left(4 (8+N)^2+16 (26+7 N) \ell -\left(368+72 N+7 N^2\right) \ell ^2\right.\right.\nonumber\\&
\left.+2 \left(-272-56 N+N^2\right) \ell ^3+\left(-272-56 N+N^2\right) \ell ^4\right) H_{\ell }+32 (8+N) \ell  (1+\ell )^2 H_{\ell }^2+2 (1+\ell ) H_{\ell } \nonumber\\&
\left(2 (8+N)^2+40 (2+N) \ell -\left(320+64 N+3 N^2\right) \ell ^2+2 \left(-272-56 N+N^2\right) \ell ^3+\left(-272-56 N+N^2\right) \ell ^4\right.\nonumber\\&
\left.-16 (8+N) \ell  (1+\ell ) H_{\ell }\right)-2 \big((8+N) \ell  (1+\ell )^2 \left(-4 (6+N)+3 (8+N) \ell +3 (8+N) \ell ^2\right) H^{(2)}_{\ell }\nonumber\\&
\left.-2 \big(24 \ell ^2-64-\! 16 N-N^2-168 \ell\! -44 N \ell -N^2 \ell +4 N \ell ^2+2 N^2 \ell ^2+(8+N)^2 \ell  (1+\ell )^2 \left(\ell +\ell ^2-2\right) \! H^{(2)}_{\ell }\big)\big)\right)\,.
\end{align}

\begin{thebibliography}{99}
	
	\bibitem{wilsonkogut}
	K. G. Wilson and J. B. Kogut, Phys.Rept. 12 (1974) 75-200
	
\bibitem{Migdal:1972tk} 
  A.~A.~Migdal,
  ``Conformal invariance and bootstrap,''
  Phys.\ Lett.\  {\bf 37B}, 386 (1971).

\bibitem{Ferrara:1973vz} 
  S.~Ferrara, A.~F.~Grillo, G.~Parisi and R.~Gatto,
  ``Covariant expansion of the conformal four-point function,''
  Nucl.\ Phys.\ B {\bf 49}, 77 (1972)
  Erratum: [Nucl.\ Phys.\ B {\bf 53}, 643 (1973)].\\
  S.~Ferrara, A.~F.~Grillo and R.~Gatto,
  ``Tensor representations of conformal algebra and conformally covariant operator product expansion,''
  Annals Phys.\  {\bf 76}, 161 (1973).

\bibitem{Polyakov}
A.~M.~Polyakov,
  ``Nonhamiltonian approach to conformal quantum field theory,''
  Zh.\ Eksp.\ Teor.\ Fiz.\  {\bf 66}, 23 (1974).
  




 \bibitem{bpz} 
  A.~A.~Belavin, A.~M.~Polyakov and A.~B.~Zamolodchikov,
  ``Infinite Conformal Symmetry in Two-Dimensional Quantum Field Theory,''
  Nucl.\ Phys.\ B {\bf 241}, 333 (1984). 

\bibitem{rrtv}
R.~Rattazzi, V.~S.~Rychkov, E.~Tonni and A.~Vichi,
``Bounding Scalar operator dimensions in 4D CFT,"
JHEP \textbf{0812} (2008) 031.
arXiv: 0807.0004[hep-th].

\bibitem{dolanosborn} 
F.~A.~Dolan and H.~Osborn,
``Conformal partial waves and operator product expansion,"
Nucl. Phys. {\bf B678}(2004) 491-507.
arXiv: hep-th/0309180.
\bibitem{do2}
F.~A.~Dolan and H.~Osborn,
``Conformal four point functions and the operator product expansion,"
Nucl. Phys. {\bf B599}(2001) 459-496.
arXiv: hep-th/0011040.

\bibitem{reviews}
S.~Rychkov,
  ``EPFL Lectures on Conformal Field Theory in $D\geq 3$  Dimensions,''
  arXiv:1601.05000 [hep-th].\\
  D.~Simmons-Duffin,
  ``TASI Lectures on the Conformal Bootstrap,''
  arXiv:1602.07982 [hep-th].\\
  J.~D.~Qualls,
  ``Lectures on Conformal Field Theory,''
  arXiv:1511.04074 [hep-th].
 
\bibitem{bootstrap}
V.~S.~Rychkov and A.~Vichi,
``Universal constraints on conformal operator dimensions,"
Phys. Rev. \textbf{D80} (2009) 045006,
arXiv: 0905.0211[hep-th].\\
F.~Caracciolo and V.~S.~Rychkov,
``Rigorous limits on the Interaction strength in Quantim field theory,"
Phys. Rev. \textbf{D81} (2010) 085037,
arXiv: 0912.2726[hep-th].\\
D.~Poland and D.~Simmons-Duffin,
``Bounds on 4D Conformal and Superconformal field theories,''
JHEP \textbf{1105} (2011) 017,
arXiv: 1009.2087.\\
R.~Rattazzi, V.~S.~Rychkov and A.~Vichi,
``Central charge bounds in 4D conformal field theory,"
 Phys. Rev \textbf{D83} (2011) 046011,
arXiv: 1009.2725[hep-th].\\
R.~Rattazzi, S.~Rychkov and A.~Vichi,
``Bounds on 4D conformal field theories and global symmetry,"
J. Phys \textbf{A44} (2011) 035402,
arXiv:1009.5985[hep-th]\\
D.~Poland, D.~Simmons-Duffin and A.~Vichi,
``Carving out a space of 4D CFTs,"
JHEP \textbf{1205} (2012) 110,
arXiv: 1109.5176[hep-th].\\
 F.~Gliozzi,
  ``More constraining conformal bootstrap,''
  Phys.\ Rev.\ Lett.\  {\bf 111}, 161602 (2013)
  [arXiv:1307.3111].\\
   F.~Gliozzi and A.~Rago,
  ``Critical exponents of the 3d Ising and related models from Conformal Bootstrap,''
  JHEP {\bf 1410}, 042 (2014)
  [arXiv:1403.6003 [hep-th]].\\
  Y.~Nakayama and T.~Ohtsuki,
  ``Five dimensional $O(N)$-symmetric CFTs from conformal bootstrap,''
  Phys.\ Lett.\ B {\bf 734}, 193 (2014)
  [arXiv:1404.5201 [hep-th]].\\
   F.~Kos, D.~Poland, D.~Simmons-Duffin and A.~Vichi,
  ``Bootstrapping the O(N) Archipelago,''
  JHEP {\bf 1511}, 106 (2015)
  [arXiv:1504.07997 [hep-th]].\\
  J.~D.~Qualls,
  ``Universal Bounds on Operator Dimensions in General 2D Conformal Field Theories,''
  arXiv:1508.00548 [hep-th].\\
  M.~Hogervorst,
  ``Dimensional Reduction for Conformal Blocks,''
  arXiv:1604.08913 [hep-th].

\bibitem{3dising}
 S.~El-Showk, M.~F.~Paulos, D.~Poland, S.~Rychkov, D.~Simmons-Duffin and A.~Vichi,
  ``Solving the 3D Ising Model with the Conformal Bootstrap,''
  Phys.\ Rev.\ D {\bf 86}, 025022 (2012)
  [arXiv:1203.6064 [hep-th]].\\
  S.~El-Showk, M.~F.~Paulos, D.~Poland, S.~Rychkov, D.~Simmons-Duffin and A.~Vichi,
  ``Solving the 3d Ising Model with the Conformal Bootstrap II. c-Minimization and Precise Critical Exponents,''
  J.\ Stat.\ Phys.\  {\bf 157}, 869 (2014)
  [arXiv:1403.4545 [hep-th]].

\bibitem{mostprecise}
F.~Kos, D.~Poland, D.~Simmons-Duffin and A.~Vichi,
  ``Precision islands in the Ising and O(N) models,''
  JHEP {\bf 1608}, 036 (2016)
  [arXiv:1603.04436 [hep-th]].

 \bibitem{dlce} 
  A.~L.~Fitzpatrick, J.~Kaplan, D.~Poland and D.~Simmons-Duffin,
 ``The Analytic Bootstrap and AdS Superhorizon Locality,''
  JHEP {\bf 1312}, 004 (2013) 
  [arXiv:1212.3616 [hep-th]]. \\
Z.~Komargodski and A.~Zhiboedov,
  ``Convexity and Liberation at Large Spin,''
  JHEP {\bf 1311}, 140 (2013)
  [arXiv:1212.4103 [hep-th]].

\bibitem{others}
A.~Liam Fitzpatrick, J.~Kaplan, M.~T.~Walters and J.~Wang,
``Eikonalization of conformal blocks,"
JHEP {\bf1509} (2015) 019 
arXiv:1504.01737[hep-th].\\
S.~Hellerman, D.~Orlando, S.~Reffert and M.~Watanabe,
``On the CFT operator spectrum at large global charge,"
arXiv: 1505.01537[hep-th].\\
T.~Hartman, S.~Jain and S.~Kundu,
``Causality Constraints in Conformal Field Theory,"
arXiv: 1509.00014[hep-th].\\
D.~M.~Hofman, D.~Li, D.~Meltzer, D.~Poland and F.~Rejon-Barrera,
  ``A Proof of the Conformal Collider Bounds,''
  JHEP {\bf 1606}, 111 (2016)
  [arXiv:1603.03771 [hep-th]].\\   
   L.~Alvarez-Gaume, O.~Loukas, D.~Orlando and S.~Reffert,
  ``Compensating strong coupling with large charge,''
  arXiv:1610.04495 [hep-th].

\bibitem{kss}
L.~F.~Alday, A.~Bissi and T.~Lukowski,
  ``Large spin systematics in CFT,''
  arXiv:1502.07707 [hep-th].\\
 A.~Kaviraj, K.~Sen and A.~Sinha,
  ``Analytic bootstrap at large spin,''
  JHEP {\bf 1511}, 083 (2015)
  [arXiv:1502.01437 [hep-th]].\\
A.~Kaviraj, K.~Sen and A.~Sinha,
``Universal anomalous dimensions at large spin and large twist,"
 JHEP \textbf{1507} (2015) 026,
arXiv:1504.00772 [hep-th].\\
 P.~Dey, A.~Kaviraj and K.~Sen,
  ``More on analytic bootstrap for O(N) models,''
  JHEP {\bf 1606}, 136 (2016)
  [arXiv:1602.04928 [hep-th]].\\
  G.~Vos,
  ``Generalized Additivity in Unitary Conformal Field Theories,''
  Nucl.\ Phys.\ B {\bf 899}, 91 (2015)
  [arXiv:1411.7941 [hep-th]].\\
   D.~Li, D.~Meltzer and D.~Poland,
 ``Non-Abelian Binding Energies from the Lightcone Bootstrap,''
 arXiv:1510.07044 [hep-th].

\bibitem{spins}
M.~S.~Costa, J.Penedones, D.~Poland and S.~Rychkov,
``Spinning conformal correlators,"
JHEP \textbf{1111} (2011) 071,
arXiv: 1107.3554[hep-th].\\
M.~S.~Costa, J.Penedones, D.~Poland and S.~Rychkov,
``Spinning conformal blocks,"
JHEP \textbf{1111} (2011) 154,
arXiv: 1109.6321[hep-th].\\
D.~Simmons-Duffin,
``Projectors, shadows and Conformal blocks,"
JHEP \textbf{1404} (2014) 146,
arXiv: 1204.3894[hep-th].\\
A.~Castedo Echeverri, E.~Elkhidir, D.~Karateev and M.~Serone,
  ``Deconstructing Conformal Blocks in 4D CFT,''
  JHEP {\bf 1508}, 101 (2015)
  [arXiv:1505.03750 [hep-th]].\\
  L.~Iliesiu, F.~Kos, D.~Poland, S.~S.~Pufu, D.~Simmons-Duffin and R.~Yacoby,
  ``Bootstrapping 3D Fermions,''
  JHEP {\bf 1603}, 120 (2016)
  [arXiv:1508.00012 [hep-th]].\\
    L.~Iliesiu, F.~Kos, D.~Poland, S.~S.~Pufu, D.~Simmons-Duffin and R.~Yacoby,
  ``Fermion-Scalar Conformal Blocks,''
  JHEP {\bf 1604}, 074 (2016)
  [arXiv:1511.01497 [hep-th]].\\
  A.~Castedo Echeverri, E.~Elkhidir, D.~Karateev and M.~Serone,
  ``Seed Conformal Blocks in 4D CFT,''
  JHEP {\bf 1602}, 183 (2016)
  [arXiv:1601.05325 [hep-th]].\\
  L.~Iliesiu, F.~Kos, D.~Poland, S.~S.~Pufu, D.~Simmons-Duffin and R.~Yacoby,
  ``Fermion-Scalar Conformal Blocks,''
  JHEP {\bf 1604}, 074 (2016)
  [arXiv:1511.01497 [hep-th]].

 

\bibitem{giombii}
S.~Giombi and V.~Kirilin,
  ``Anomalous Dimensions in CFT with Weakly Broken Higher Spin Symmetry,''
  arXiv:1601.01310 [hep-th].
    
 \bibitem{skvortt} 
  E.~D.~Skvortsov,
  ``On (Un)Broken Higher-Spin Symmetry in Vector Models,''
  arXiv:1512.05994 [hep-th].\\
   S.~Giombi, V.~Gurucharan, V.~Kirilin, S.~Prakash and E.~Skvortsov,
  ``On the Higher-Spin Spectrum in Large N Chern-Simons Vector Models,''
  arXiv:1610.08472 [hep-th].












\bibitem{alday}
L.~F.~Alday and A.~Zhiboedov,
``Conformal bootstrap with slightly broken higher spin symmetry,"
 arXiv:1506.04659 [hep-th]. 
 
 





\bibitem{Alday2} 
  L.~F.~Alday and A.~Zhiboedov,
  ``An Algebraic Approach to the Analytic Bootstrap,''
  arXiv:1510.08091 [hep-th].

\bibitem{Alday4} 
  L.~F.~Alday,
  ``Large Spin Perturbation Theory,''
  arXiv:1611.01500 [hep-th].

\bibitem{Alday3} 
  L.~F.~Alday,
  ``Solving CFTs with Weakly Broken Higher Spin Symmetry,''
  arXiv:1612.00696 [hep-th].
 \bibitem{rychkovtan} 
  S.~Rychkov and Z.~M.~Tan,
  ``The $\epsilon$-expansion from conformal field theory,''
  J.\ Phys.\ A {\bf 48}, no. 29, 29FT01 (2015)
  [arXiv:1505.00963 [hep-th]].
 
  
\bibitem{rtothers}
P.~Basu and C.~Krishnan,
  ``$\epsilon$-Expansions Near Three Dimensions from Conformal Field Theory,''
  arXiv:1506.06616 [hep-th].\\
S.~Ghosh, R.~K.~Gupta, K.~Jaswin and A.~A.~Nizami,
  ``$\epsilon$-Expansion in the Gross-Neveu Model from Conformal Field Theory,''
  arXiv:1510.04887 [hep-th]; A.~Raju,
  ``$\epsilon$-Expansion in the Gross-Neveu CFT,''
  arXiv:1510.05287 [hep-th]\\
   S.~Yamaguchi,
  ``The $\epsilon$-expansion of the codimension two twist defect from conformal field theory,''
  arXiv:1607.05551 [hep-th]; K.~Nii,
  ``Classical equation of motion and Anomalous dimensions at leading order,''
  JHEP {\bf 1607}, 107 (2016)
  [arXiv:1605.08868 [hep-th]].
    F.~Gliozzi, A.~Guerrieri, A.~C.~Petkou and C.~Wen,
    ``Generalized Wilson-Fisher critical points from the conformal OPE,''
    arXiv:1611.10344 [hep-th].

\bibitem{epsnum}
 S.~El-Showk, M.~Paulos, D.~Poland, S.~Rychkov, D.~Simmons-Duffin and A.~Vichi,
  ``Conformal Field Theories in Fractional Dimensions,''
  Phys.\ Rev.\ Lett.\  {\bf 112}, 141601 (2014)
  [arXiv:1309.5089 [hep-th]].

 \bibitem{sensinha} 
  K.~Sen and A.~Sinha,
    ``On critical exponents without Feynman diagrams,''
    J.\ Phys.\ A {\bf 49}, no. 44, 445401 (2016)
     [arXiv:1510.07770 [hep-th]].
 
  
  
\bibitem{usprl}
R.~Gopakumar, A.~Kaviraj, K.~Sen and A.~Sinha,
  ``Conformal Bootstrap in Mellin Space,''
  arXiv:1609.00572 [hep-th].
  
\bibitem{uslong} 
  R.~Gopakumar, A.~Kaviraj, K.~Sen and A.~Sinha,
  ``A Mellin space approach to the conformal bootstrap,''
  arXiv:1611.08407 [hep-th].


\bibitem{mack} 
  G.~Mack,
  ``D-independent representation of Conformal Field Theories in D dimensions via transformation to auxiliary Dual Resonance Models. Scalar amplitudes,''
  arXiv:0907.2407 [hep-th].

 \bibitem{pene} 
  J.~Penedones,
  ``Writing CFT correlation functions as AdS scattering amplitudes,''
  JHEP {\bf 1103}, 025 (2011) 
  [arXiv:1011.1485 [hep-th]]. \\
  A.~L.~Fitzpatrick, J.~Kaplan, J.~Penedones, S.~Raju and B.~C.~van Rees,
 ``A Natural Language for AdS/CFT Correlators,''
  JHEP {\bf 1111}, 095 (2011)
  [arXiv:1107.1499 [hep-th]].
 
 \bibitem{mig} 
  M.~F.~Paulos,
  ``Towards Feynman rules for Mellin amplitudes,''
  JHEP {\bf 1110}, 074 (2011)
  [arXiv:1107.1504 [hep-th]].

\bibitem{dolanosborn2} 
F.~A.~Dolan and H.~Osborn,
  ``Conformal Partial Waves: Further Mathematical Results,''
  arXiv:1108.6194 [hep-th].

\bibitem{fitzpatrick}
A.~L.~Fitzpatrick and J.~Kaplan,
``AdS Field theory from conformal field theory,"
JHEP {\bf 1302} (2013) 054.
arXiv: 1208.0337[hep-th].



\bibitem{joao} 
  M.~S.~Costa, V.~Goncalves and J.~Penedones,
  ``Conformal Regge theory,''
  JHEP {\bf 1212}, 091 (2012)
  [arXiv:1209.4355 [hep-th]].

\bibitem{costa}
M.~S.~Costa, V.~Goncalves, J.~Penedones,
``Spinning AdS Propagators,"
JHEP {\bf 1409}(2014) 064.
arXiv: 1404.5625[hep-th].


\bibitem{joaoreview} 
 J.~Penedones,
  ``TASI lectures on AdS/CFT,''
  arXiv:1608.04948 [hep-th].
  
\bibitem{aldaybissi2}
  L.~F.~Alday and A.~Bissi,
  ``Unitarity and positivity constraints for CFT at large central charge,''
  arXiv:1606.09593 [hep-th].
  
\bibitem{rastelli} 
  L.~Rastelli and X.~Zhou,
  ``Mellin amplitudes for $AdS_5\times S^5$,''
  arXiv:1608.06624 [hep-th].

\bibitem{Aharony:2016dwx} 
    O.~Aharony, L.~F.~Alday, A.~Bissi and E.~Perlmutter,
    arXiv:1612.03891 [hep-th].


\bibitem{AAR} G.~E.~Andrews, R.~Askey and R~Roy, ``Special functions,'' Cambridge University Press, 1999.
 
\bibitem{Luke}
  Y.L. Luke, ``The Special Functions and Their Approximations,'' Vol. I–II', Academic Press, New York, 1969.
  

	\bibitem{petkoulots} 
	H.~Osborn and A.~C.~Petkou,
	``Implications of conformal invariance in field theories for general dimensions,''
	Annals Phys.\  {\bf 231}, 311 (1994)
	[hep-th/9307010].
	
	A.~C.~Petkou,
	``Evaluating the AdS dual of the critical O(N) vector model,''
	JHEP {\bf 0303} (2003) 049
	[hep-th/0302063].
	
	
	A.~C.~Petkou,
	``Operator product expansions and consistency relations in a O(N) invariant fermionic CFT for $2 < d < 4$,''
	Phys.\ Lett.\ B {\bf 389}, 18 (1996)
	[hep-th/9602054].
	
	
	
	A.~C.~Petkou and N.~D.~Vlachos,
	``Finite size and finite temperature effects in the conformally invariant O(N) vector model for 2 less than d less than 4,''
	hep-th/9809096.
	
	
	
	A.~C.~Petkou and N.~D.~Vlachos,
	``Finite size effects and operator product expansions in a CFT for $d > 2$,''
	Phys.\ Lett.\ B {\bf 446}, 306 (1999)
	[hep-th/9803149].
	
	
	A.~C.~Petkou,
	``C(T) and C(J) up to next-to-leading order in 1/N in the conformally invariant 0(N) vector model for $2 < d < 4$,''
	Phys.\ Lett.\ B {\bf 359}, 101 (1995)
	[hep-th/9506116].
	
	
	

	
	
	\bibitem{Lang:1992zw} 
	K.~Lang and W.~Ruhl,
	Nucl.\ Phys.\ B {\bf 400}, 597 (1993).
	
	
	
 \bibitem{phi4} 
  H.~Kleinert and V.~Schulte-Frohlinde,
  ``Critical Properties of Phi4 Theories,''
  World Scientific, Singapore, 2004 

 \bibitem{Braun} 
   V.~M.~Braun and A.~N.~Manashov,
   Eur.\ Phys.\ J.\ C {\bf 73}, 2544 (2013)
   [arXiv:1306.5644 [hep-th]].

\bibitem{kleinert}
H.~Kleinert, J.~Neu, V.~Schulte-Frohlinde, K.~G.~Chetrykin and S.~A.~Larin,
``Five loop renormalization group functions of O(n) symmetric phi**4 theory and epsilon expansions of critical exponents up to epsilon**5,''
Phys.Lett. \textbf{B272} (1991) 39-44, Phys.Lett. \textbf{B319} (1993) 545.\\
H.~Kleinert and V.~Schulte-Frohlinde,
``Exact five loop renormalization group functions of phi**4 theory with O(N) symmetric and cubic interactions: Critical exponents up to epsilon**5 ,"
 Phys.Lett. \textbf{B342} (1995) 284-296.\\
H.~Kleinert and V.~Schulte-Frohlinde,
``Critical exponents from five-loop strong coupling phi**4 theory in 4 - epsilon dimensions,"
J.Phys. \textbf{A34} (2001) 1037-1050.


\bibitem{gracey}
S.~E.~Derkachov, J.~A.~Gracey and A.~N.~Manashov,
  ``Four loop anomalous dimensions of gradient operators in phi**4 theory,''
  Eur.\ Phys.\ J.\ C {\bf 2}, 569 (1998). 
 
\bibitem{Witczak-Krempa:2013nua} 
  W.~Witczak-Krempa, E.~Sorensen and S.~Sachdev,
  ``The dynamics of quantum criticality via Quantum Monte Carlo and holography,''
  Nature Phys.\  {\bf 10}, 361 (2014)
  [arXiv:1309.2941 [cond-mat.str-el]].

\bibitem{prlcj}
S.~Gazit, D.~Podolsky and A.~Auerbach, ``Critical Capacitance and Charge-Vortex Duality Near the Superfluid-to-Insulator Transition," Phys.\ Rev.\ Lett.\ {\bf 113} 240601.

\bibitem{Kos:2013tga}
F.~Kos, D.~Poland and D.~Simmons-Duffin,
``Bootstrapping the $O(N)$ vector models,''
JHEP {\bf 1406}, 091 (2014)
[arXiv:1307.6856 [hep-th]].

\bibitem{Kosarchie}
F.~Kos, D.~Poland, D.~Simmons-Duffin and A.~Vichi,
``Bootstrapping the $O(N)$ archipelago,"
arXiv: 1504.07997[hep-th].


\bibitem{petkou}
A.~Petkou,
	``Conserved currents, consistency relations and operator product expansions in the conformally invariant O(N) vector model,''
	Annals Phys.\  {\bf 249}, 180 (1996)
	[hep-th/9410093].

\bibitem{klebanov2}
K.~Diab, L.~Fei, S.~Giombi, I.~R.~Klebanov and G.~Tarnopolsky,
  ``On $C_J$ and $C_T$ in the Gross-Neveu and $O(N)$ Models,''
  J.\ Phys.\ A {\bf 49}, no. 40, 405402 (2016)
  [arXiv:1601.07198 [hep-th]].

\bibitem{zohar} 
  Z.~Komargodski and D.~Simmons-Duffin,
  arXiv:1603.04444 [hep-th].

\bibitem{RGAS}
R.~Gopakumar and A.~Sinha, to appear.
 
\bibitem{kehrein}
S.~K.~Kehrein, F.~Wegner and Y.~Pismak,
``	Conformal symmetry and the spectrum of anomalous dimensions in the N vector model in four epsilon dimensions,"
Nucl.Phys. \textbf{B402} (1993) 669-692.\\
S.~K.~Kehrein and F.~Wegner,
``The Structure of the spectrum of anomalous dimensions in the N vector model in (4-epsilon)-dimensions,"
Nucl.Phys. \textbf{B424} (1994) 521-546\\ 
S.~K.~Kehrein,
``The structure of the spectrum of critical exponents of $(\phi^2)^2$ in two dimensions in D=4-epsilon dimensions: Resolution of degenracies and hierarchical structures,"
 Nucl.Phys. \textbf{B453} (1995) 777-806. 



\bibitem{klebanov}
L.~Fei, S.~Giombi and I.~Klebanov,
``Critical $O(N)$ models in $6-\e$ dimensions,"
Phys.Rev. {\bf D90} (2014) 2, 025018
arXiv:1404.1094.\\
L.~Fei, S.~Giombi, I.~Klebanov and G.~Tarnopolsky,
``Three loop analysis of the critical $O(N)$ models in $6-\e$ dimensions,"
Phys.Rev. {\bf D91} (2015) 4, 045011
arXiv: 1411.1099.\\
L.~Di Pietro, Z.~Komargodski, I.~Shamir and E.~Stamou,
``Quantum electrodynamics in d=3 from the epsilon-expansion,"
arXiv: 1508.06278[hep-th].

\bibitem{rychkovaa}
D.~Pappadopulo, S.~Rychkov, J.~Espin and R.~Rattazzi,
  ``OPE Convergence in Conformal Field Theory,''
  Phys.\ Rev.\ D {\bf 86}, 105043 (2012)
  [arXiv:1208.6449 [hep-th]].\\
S.~Rychkov and P.~Yvernay,
  ``Remarks on the Convergence Properties of the Conformal Block Expansion,''
  Phys.\ Lett.\ B {\bf 753}, 682 (2016)
  [arXiv:1510.08486 [hep-th]].

\bibitem{Cardy:2013rqg} 
J.~Cardy,
``Logarithmic conformal field theories as limits of ordinary CFTs and some physical applications,''
J.\ Phys.\ A {\bf 46}, 494001 (2013)
[arXiv:1302.4279 [cond-mat.stat-mech]].
\bibitem{susyboot}
C.~Beem, L.~Rastelli and B.~C.~van Rees,
  ``The $\mathcal N=4$ Superconformal Bootstrap,''
  Phys.\ Rev.\ Lett.\  {\bf 111}, 071601 (2013)
  [arXiv:1304.1803 [hep-th]].\\
   C.~Beem, M.~Lemos, P.~Liendo, L.~Rastelli and B.~C.~van Rees,
  ``The $ \mathcal{N}=2 $ superconformal bootstrap,''
  JHEP {\bf 1603}, 183 (2016)
  [arXiv:1412.7541 [hep-th]].\\
  M.~Lemos and P.~Liendo,
  ``Bootstrapping $ \mathcal{N}=2 $ chiral correlators,''
  JHEP {\bf 1601}, 025 (2016)
  [arXiv:1510.03866 [hep-th]].

 

	\bibitem{Lang:1991kp} 
	  K.~Lang and W.~Ruhl,
	  Nucl.\ Phys.\ B {\bf 377}, 371 (1992).
	  


 

\end{thebibliography}
\end{document}